\documentclass[review]{elsarticle}
\usepackage{subfigure}
\usepackage{lineno,hyperref}
\usepackage{indentfirst}
\usepackage{bm}
\usepackage{threeparttable}
\usepackage{multirow}
\usepackage{amsmath}
\usepackage{array}
\usepackage{amssymb}
\usepackage{graphicx}
\usepackage{url}
\usepackage{hyperref}
\usepackage{epstopdf}
\usepackage{booktabs}
\usepackage{color}
\usepackage{lscape}
\usepackage{geometry}
\usepackage{fancyhdr}

\journal{Elsevier}

\begin{document}

\begin{frontmatter}

\title{Multiple solutions of nonlinear coupled constitutive relation model and its rectification in non-equilibrium flow computation}
\author{Junzhe Cao$^a$}
\ead{caojunzhe@mail.nwpu.edu.cn}

\author[]{Sha Liu$^a$$^b$\corref{mycorrespondingauthor}}
\cortext[mycorrespondingauthor]{Corresponding author}
\ead{shaliu.@nwpu.edu.cn}

\author[]{Chengwen Zhong$^a$$^b$}
\ead{zhongcw@nwpu.edu.cn}

\author[]{Congshan Zhuo$^a$$^b$}
\ead{zhuocs@nwpu.edu.cn}

\author[]{Kun Xu$^c$$^d$}
\ead{makxu@ust.hk}

\address{$^a$School of Aeronautics, Northwestern Polytechnical University, Xi'an, Shaanxi 710072, China\\
$^b$Institute of Extreme Mechanics, Northwestern Polytechnical University, Xi'an, Shaanxi 710072, China\\
$^c$Department of Mathematics, Hong Kong University of Science and Technology, Hong Kong, China\\
$^d$Shenzhen Research Institute, Hong Kong University of Science and Technology, Shenzhen, China}

\begin{abstract}
In this study, the multiple solutions of Nonlinear Coupled Constitutive Relation (NCCR) model are firstly observed and a way for identifying the physical solution is proposed.  The NCCR model proposed by Myong is constructed from the generalized hydrodynamic equations of Eu, and aims to describe rarefied flows. The NCCR model is a complicated nonlinear system. Many assumptions have been used in the schemes for solving the NCCR equations. The corresponding numerical methods may be associated with unphysical solution and instability. At the same time, it is hard to analyze the physical accuracy and stability of NCCR model due to the uncertainties in the numerical discretization. In this study, a new numerical method for solving NCCR equations is proposed and used to analyze the properties of NCCR equations. More specifically, the nonlinear equations are converted into the solutions of an objective function of a single variable. Under this formulation, the multiple solutions of the NCCR system are identified and the criteria for picking up the physical solution are proposed. Therefore, a numerical scheme for solving NCCR equations is constructed. A series of flow problems in the near continuum and low transition regimes with a large variation of Mach numbers are conducted to validate the numerical performance of proposed method and the physical accuracy of NCCR model.
\end{abstract}


\end{frontmatter}

\section{Introduction}
With the development of hypersonic vehicle, spacecraft and micro-electromechanical system, the non-equilibrium flow receives more attention and developing corresponding numerical methods is a challenging topic. In recent years, both the stochastic particle method \cite{dsmcre,pbgkre,dsmcpbgkre} and Discrete Velocity Method (DVM) \cite{dvmre,imexre,ugksre,dugksre,idvmre,gsisre} are successfully developed. Meanwhile, a unified gas-kinetic wave-particle method is proposed \cite{ugkwp1} on the framework of Unified Gas-Kinetic Scheme (UGKS). It is a multiscale multi-efficiency preserving method, and is rapidly developed \cite{ugkwp5,suwp,suwp2}. All of above methods can accurately simulate flows from the continuum flow regime to the free molecular flow regime. However, efficiency problem exists in them. For example, the time step in the Direct Simulation Monte Carlo (DSMC) method is seriously restricted in the near-continuum flow regime. Generally, the number of numerical particles should be prohibitively large when using stochastic particle methods to simulate near-continuum flows. On the other hand, the DVM meets the curse of dimensionality, making the discrete velocity space expensive. The root cause of efficiency problem is that when the non-equilibrium feature becomes stronger, there are rather more degrees of freedom to be concerned about. It is worth noting that in macroscopic methods, less degrees of freedom are used to describe the non-equilibrium feature, which results in that the Knudsen number (Kn, the ratio of the molecular mean free path to the characteristic physical length scale) should not be too large when using these methods, or the accuracy problem exists. From another point of view, in macroscopic methods, the accuracy is sacrificed for efficiency. As a consequence, macroscopic methods are suitable for simulating near-continuum flows. Macroscopic methods, such as the Chapman-Enskog method \cite{ce} and the method of moments \cite{grad1}, are well developed\cite{grad2,grad3,grad4,grad5} and extended to polyatomic gases and multi-component mixture gases\cite{grad6,grad7}.

As a kind of macroscopic methods, the Nonlinear Coupled Constitutive Relation (NCCR) model is proposed by Myong \cite{myong1,myong2,myong3}, based on the generalized hydrodynamic equations of Eu \cite{Eu}. In recent years, the property of this model is studied from many points of view and is implemented to many flow mechanism studies and engineering applications. For example, the topology of the NCCR model is studied by Myong \cite{topology} and the linear stability is proved by Jiang \cite{stability}. The model is utilized to study complicated flows like shock-vortex interaction \cite{vortex1, vortex2}, dusty and granular flows \cite{dusty}. The model is applied in the framework of discontinuous Galerkin by Xiao \cite{dg} and in the framework of extended gas-kinetic scheme by Liu \cite{egks}, where this model is further improved according to one of recent conclusions in the non-equilibrium thermodynamics that the relaxation time ratio between the real nonlinear one and the linear (near equilibrium) one is bounded\cite{villani1,villani2}. Furthermore, the NCCR model is extended to the multi-species gas flow \cite{multispecies}, thermodynamic non-equilibrium flow \cite{temperature} and hypersonic reaction flow \cite{reaction}. The ability of NCCR model to describe the hypersonic near-continuum flow is systematically tested by Myong \cite{jcp2022}. Different from other kinds of macroscopic methods, there are no high order spatial derivatives in the NCCR equations, which reduces the difficulty in designing numerical methods and maintaining stability. However, this also results in that the stress and heat flux in NCCR equations are not expressed explicitly, and they can only be obtained by solving this complicated nonlinear system.

When trying to solve these complicated nonlinear equations, because of their complexity, the convergency property of classical iteration methods is hard to be analyzed, such as the fixed point iteration method and Newton's method. And actually they do not even converge. In Ref. \cite{myong2,myong3}, a decomposed solving method is proposed by Myong, which is the most popular method for solving the NCCR equations currently. In Myong's method, the three-dimensional problem is simplified approximately into three one-dimensional non-interfering problems in x, y, z directions. Though accurate solution on each direction can be obtained, coupled terms between different directions are not taken into consideration, which results in deviation. Aiming at Myong's method, Jiang comments that the most unsatisfied feature is the computational instability induced directly by an unphysical negative-density phenomenon, particularly in some expansion regions. In Ref. \cite{undecomposed}, the fixed point iteration method is modified, which brings about better stability. But the convergency property of this method is still hard to analyze and indeed its result is not accurate in tests. If the result of Modified Fixed Point Iteration (MFPI) method is taken as the precondition of Newton's method, the result is accurate and convergency property is better. This is the coupling solving method proposed by Jiang. However, the convergency property of this method is still hard to be analyzed. From another perspective, because these solving methods are approximate, it is a hard task to test the performance of NCCR model for describing non-equilibrium flows, whose development is hence restricted.

In this study, aiming at this problem, a new method is proposed to convert the problem of solving the multi-variable complicated nonlinear system into solving an objective function of a single variable. Under this foamulation, it is much more easy to analyze this complicated nonlinear system. The multiple solutions of NCCR equations are observed and the criteria for picking up the physical solution are proposed. An iterative solving method is then proposed without assumptions. A series of numerical test cases in the near continuum and low transition regimes with a large variation of Mach numbers are conducted. The numerical performance of proposed method and the physical accuracy of NCCR model are validated. In the test cases, results of experiments, DSMC method, UGKS method and Discrete UGKS (DUGKS) method are taken as references. Other macroscopic methods are also utilized for comparison, including the classical methods for solving Navier-Stokes(NS) equations (whether the bulk viscosity is considered or not and whether the slip boundary condition is implemented or not). The MFPI method for solving NCCR equations is implemented in these test cases as well, because this method has good stability and it is desirable to test its accuracy.

The remainder of this paper is organized as follows: the nondimensionalization system of this paper and the corresponding governing equations are introduced in Sec.\ref{sec:nccr}, along with a short analysis about the NCCR model. Sec.\ref{sec:solver} is the construction of the proposed solving method and multiple solutions are exhibited. The numerical test cases are conducted in Sec.\ref{sec:cases}. The conclusions are in Sec.\ref{sec:conclusion}.
\section{Nondimensionalization, governing equations and NCCR model}\label{sec:nccr}
This study is done under the framework of Stanford university unstructured (SU2) open source solver \cite{su1,su3}. The stress and heat flux in the viscous flux are calculated by NCCR equations, and the inviscid flux is computed by the Kinetic Inviscid Flux (KIF)\cite{kif} or AUSM+ -up scheme\cite{ausmplusup}. The nondimensionalization system should be introduced firstly,
\begin{equation}
\bm{\varphi} = \frac{\bm{\varphi}_{\rm{pre}}}{\bm{\varphi}_{\rm{ref}}},
\end{equation}
where $\bm{\varphi}$ denotes macroscopic values to be used in the nondimensionalized equations, $\bm{\varphi}_{\rm{pre}}$ denotes values before the nondimensionalization and $\bm{\varphi}_{\rm{ref}}$ denotes reference values, as Tab.\ref{tab:nondimen}.
\begin{table}[h]\label{tab:nondimen}
\centering
\caption{Reference values in the nondimensionalization}
\begin{tabular}{*{4}{|l|}}
\hline
	\textbf{Value name} &\textbf{Reference value} \\
\hline
	Length            & $l_{\rm{ref}}$                                            \\ \hline
	Pressure          & $p_{\rm{ref}}$                                            \\ \hline
	Density           & $\rho_{\rm{ref}}$                                         \\ \hline
    Temperature       & $T_{\rm{ref}}$                                            \\ \hline
    Velocity          & $U_{\rm{ref}}=\sqrt{p_{\rm{ref}}/\rho_{\rm{ref}}}$        \\ \hline
    Specific energy   & $e_{\rm{ref}}=U_{\rm{ref}}^2$                             \\ \hline
    Stress            & $\Pi_{\rm{ref}}=p_{\rm{ref}}$                             \\ \hline
    Heat flux         & $Q_{\rm{ref}}=\rho_{\rm{ref}}e_{\rm{ref}}U_{\rm{ref}}$    \\ \hline
    Gas constant      & $R_{\rm{ref}}=e_{\rm{ref}}/T_{\rm{ref}}$                  \\ \hline
    Heat capacity (constant pressure) & $C_{p,\rm{ref}}=R_{\rm{ref}}$             \\ \hline
    Dynamic viscosity & $\mu_{\rm{ref}}=\rho_{\rm{ref}}U_{\rm{ref}}l_{\rm{ref}}$  \\ \hline
    Heat conductivity & $k_{\rm{ref}}=C_{p,\rm{ref}}\mu_{\rm{ref}}$               \\
\hline
\end{tabular}
\end{table}

The governing equation used in this paper is
\begin{equation}
\partial_t\bm{\Psi} + \nabla\cdot\mathbf{F}^{\rm{c}} - \nabla\cdot\mathbf{F}^{\rm{v}} = \bm{0},
\end{equation}
where $\bm{\Psi}$ denotes the conserved value vector, $\mathbf{F}^{\rm{c}}$ denotes the inviscid flux vector and $\mathbf{F}^{\rm{v}}$ denotes the viscous flux vector. The equation can be expanded into
\begin{equation}
\left( \begin{array}{l}
\rho \\
\rho \mathbf{U}\\
\rho E
\end{array} \right)_t + \nabla  \cdot
\left( \begin{array}{l}
\rho \mathbf{U}\\
\rho \mathbf{UU} + p\mathbf{I}\\
(\rho E + p)\mathbf{U}
\end{array} \right) - \nabla  \cdot
\left( \begin{array}{l}
0\\
\bm{\Pi}  + \Delta \mathbf{I}\\
(\bm{\Pi}  + \Delta \mathbf{I}) \cdot \mathbf{U} + \mathbf{Q}
\end{array} \right) = \mathbf{0},
\end{equation}
where $\mathbf{I}$ is unit tenser and $\Delta$ is the excess normal stress, which is related to the bulk viscosity. It is perceived to have direct correlation with the rotational relaxation time of diatomic gas and be able to describe rotational relaxation effect to some extent \cite{bulk}. In NS equations, the linear constitutive relation is as follows:
\begin{equation}\label{eq:ns}
\begin{aligned}
{\Pi _{\rm{ij}}} &= 2\mu \frac{{\partial {U_{ \rm{< i}}}}}{{\partial {x_{\rm{j >} }}}},\\
{Q_{\rm{i}}} &= \frac{{\mu {C_p}}}{{\rm{Pr} }}\frac{{\partial T}}{{\partial {x_{\rm{i}}}}},\\
\Delta  &= {\mu _{\rm{b}}}\frac{{\partial {U_{\rm{i}}}}}{{\partial {x_{\rm{i}}}}},
\end{aligned}
\end{equation}
where $\rm{Pr}$ denotes the Prandtl number, $C_p=R\gamma/(\gamma-1)$, $\mu_{\rm{b}}$ is the bulk viscosity, and symbol $<\cdot>$ denotes the second-order symmetric and trace-free tensor:
\begin{equation}
A_{<\rm{ij}>}=\frac{1}{2}(A_{\rm{ij}}+A_{\rm{ji}})-\frac{1}{3}\delta_{\rm{ij}}A_{\rm{kk}},
\end{equation}
where $\bm{\delta}$ is the Kronecker tensor and the Einstein summation convention for repeated subscripts is used throughout this paper without special declaration. When using NS equations to simulate low speed continuum flows numerically, the bulk viscosity is always not considered, which means $\Delta=0$. In consideration of the ability of $\Delta$ to describe non-equilibrium flows, in this study, the performance is tested whether the $\Delta$ term is considered when using NS equations.
The first-order/linear constitutive relation can only describe continuum flows accurately. To extend the ability to describe rarefied flows, the second-order constitutive relation is derived from the Boltzmann-Curtiss kinetic equation \cite{BC1,BC2}, based on Eu's modified moment method \cite{Eu,Eu2} and Myong's closing-last balanced closure \cite{myong4}. It is known as the NCCR model,
\begin{equation}\label{eq:nccr1}
\begin{aligned}
{\Pi _{\rm{ij}}} &= \frac{1}{{q(\kappa )}}\frac{\mu }{p}\left\{ { - 2\frac{{\partial {U_{\rm{ < i}}}}}{{\partial {x_{\rm{k}}}}}{\Pi _{\rm{j > k}}} + 2(p - \Delta )\frac{{\partial {U_{\rm{ < i}}}}}{{\partial {x_{\rm{j > }}}}}} \right\},\\
{Q_{\rm{i}}} &= \frac{1}{{q(\kappa )}}\frac{\mu }{p}\frac{1}{{\Pr }}\left\{ {{C_p}(p - \Delta )\frac{{\partial T}}{{\partial {x_{\rm{i}}}}} - {C_p}{\Pi _{\rm{ik}}}\frac{{\partial T}}{{\partial {x_{\rm{k}}}}} - \frac{{\partial {U_{\rm{i}}}}}{{\partial {x_{\rm{k}}}}}{Q_{\rm{k}}}} \right\},\\
\Delta  &= \frac{1}{{q(\kappa )}}\left\{ {{\mu _{\rm{b}}}\frac{{\partial {U_{\rm{i}}}}}{{\partial {x_{\rm{i}}}}} - 3\frac{{{\mu _{\rm{b}}}}}{p}(\Delta {I_{\rm{ij}}} + {\Pi _{\rm{ij}}})\frac{{\partial {U_{\rm{i}}}}}{{\partial {x_{\rm{j}}}}}} \right\},
\end{aligned}
\end{equation}
and
\begin{equation}\label{eq:qkapa}
q(\kappa)=\frac{\rm{sinh}\kappa}{\kappa},
\end{equation}
\begin{equation}\label{eq:kapa}
\kappa  = \frac{{{\pi ^{\frac{1}{4}}}}}{{\sqrt {2\beta } }}{\left\{ {\frac{{{\Pi _{\rm{ij}}}{\Pi _{\rm{ji}}}}}{{2{p^2}}} + \frac{{5 - 3\gamma }}{{2{f_{\rm{b}}}{p^2}}}{\Delta ^2} + \frac{{{Q_{\rm{i}}}{Q_{\rm{i}}}}}{{T{p^2}}}\frac{\rm{Pr }}{{{C_p}}}} \right\}^{\frac{1}{2}}},
\end{equation}
where ``$\rm{sinh}$'' is the hyperbolic sine function, and $\kappa$ is related to the entropy relaxation rate. When the flow tends to equilibrium, $\kappa$ tends to the Rayleigh-Onsager dissipation function in non-equilibrium thermodynamics \cite{Onsager}. As to monatomic gas, the second term in Eq.\ref{eq:kapa} ($\Delta$ term) is zero. $f_{\rm{b}}=\mu_{\rm{b}}/\mu$, which is $0$ for monatomic gas and $0.8$ for nitrogen gas \cite{bulk}. According to the Variable Soft Sphere (VSS) model,
\begin{equation}\label{eq:vss}
\beta  = \frac{{5(\alpha  + 1)(\alpha  + 2)}}{{4\alpha (5 - 2\omega )(7 - 2\omega )}},
\end{equation}
where $\alpha$ donates the molecular scattering factor and $\omega$ donates the heat index.
\section{The multiple solutions of NCCR model and its solving method}\label{sec:solver}
As Eq.\ref{eq:nccr1}, the NCCR equations are always exhibited into a form where variables to be solved are lined explicitly on the left hand of equal sign . This form accords well with the classical fixed point iteration method. However, it does not converge when using classical methods like the fixed point iteration method and Newton's method. The convergency property is hard to analyze because of the complexity of these equations and the root of complexity is in Eq.\ref{eq:qkapa} and Eq.\ref{eq:kapa}. Due to the complexity of $q(\kappa)$, a new method is needed to analyze the NCCR equations.

At present, there are mainly two methods to solve NCCR equations, the coupling method of Myong \cite{myong2} and the undecomposed method of Jiang \cite{undecomposed}. The former is firstly proposed and used widely. In Myong's method, the equations are split into three series of equations corresponding to x, y, z directions. On each direction, an accurate solving method is proposed. However, the related terms between different directions are ignored, which means when this method is used, accurate solution of one-dimensional problem can be obtained but for multidimensional problems it can not. In Ref.\cite{undecomposed}, Jiang proposes that instability problem in some expansion regions exists when using this method.

Therefore, the coupling solving method is needed and Jiang's work focuses on this. In Ref.\cite{undecomposed}, the fixed point iteration method is modified. In this method, extra algorithm is designed to weaken the instability caused by $q(\kappa)$. However, after the modification, the accuracy becomes hard to analyze and actually accurate solution cannot be obtained through the MFPI method. From another point of view, in this method, accuracy is sacrificed for stability. Details of MFPI method can be referred to Appendix A. As it is simple and efficient, accuracy and stability of MFPI method are tested in this study. In particular, it is found that if taking the MFPI method as the preconditioner of Newton's method, the accuracy and stability become much better. This is the coupling method of Jiang, but its convergence property is still hard to analyze.

In this section, a new method is proposed, which makes it rather easy to analyze the NCCR equations. The multiple solutions of the equations are observed and a new numerical solving method is proposed. This method is to be introduced as follows. At the first stage, Eq.\ref{eq:nccr1} is exhibited in a new form:
\begin{equation}\label{eq:nccr2}
\mathbf{A}\cdot\mathbf{m}=\mathbf{b},
\end{equation}
where $\mathbf{m}$ is the vector of variables to be solved,
\begin{equation}\label{eq:mm}
\mathbf{m} = ({\Pi _{11}},{\Pi _{12}},{\Pi _{13}},{\Pi _{22}},{\Pi _{23}},{\Pi _{33}},{Q_1},{Q_2},{Q_3},\Delta )^{\rm{T}}.
\end{equation}
For clarity, every terms of the stress tensor and heat flux vector are listed. $\Pi_{\rm{ij}}=\Pi_{\rm{ji}}$ so there are ten variables to be solved. $\mathbf{A}$ is a $10\times10$ matrix,
\begin{landscape}
\tiny
\begin{equation}\label{eq:aa}
\begin{aligned}
&\mathbf{A}=
\\
&\frac{\mu}{p}
\begin{pmatrix}
-\frac{4}{3}\frac{\partial U_1}{\partial x_1}
&-\frac{4}{3}\frac{\partial U_1}{\partial x_2}+\frac{2}{3}\frac{\partial U_2}{\partial x_1}
&-\frac{4}{3}\frac{\partial U_1}{\partial x_3}+\frac{2}{3}\frac{\partial U_3}{\partial x_1}
&\frac{2}{3}\frac{\partial U_2}{\partial x_2}
&\frac{2}{3}\frac{\partial U_2}{\partial x_3}+\frac{2}{3}\frac{\partial U_3}{\partial x_2}
&\frac{2}{3}\frac{\partial U_3}{\partial x_3}
&0
&0
&0
&-2\left({ \frac{\partial U_1}{\partial x_1}-\frac{1}{3}\frac{\partial U_k}{\partial x_k} }\right)\\
-\frac{\partial U_2}{\partial x_1}
&-\left({ \frac{\partial U_1}{\partial x_1}+\frac{\partial U_2}{\partial x_2} }\right)
&-\frac{\partial U_2}{\partial x_3}
&-\frac{\partial U_1}{\partial x_2}
&-\frac{\partial U_1}{\partial x_3}
&0
&0
&0
&0
&-\left({ \frac{\partial U_1}{\partial x_2}+\frac{\partial U_2}{\partial x_1} }\right)\\
-\frac{\partial U_3}{\partial x_1}
&-\frac{\partial U_3}{\partial x_2}
&-\left({ \frac{\partial U_1}{\partial x_1}+\frac{\partial U_3}{\partial x_3} }\right)
&0
&-\frac{\partial U_1}{\partial x_2}
&-\frac{\partial U_1}{\partial x_3}
&0
&0
&0
&-\left({ \frac{\partial U_1}{\partial x_3}+\frac{\partial U_3}{\partial x_1} }\right)\\
\frac{2}{3}\frac{\partial U_1}{\partial x_1}
&\frac{2}{3}\frac{\partial U_1}{\partial x_2}-\frac{4}{3}\frac{\partial U_2}{\partial x_1}
&\frac{2}{3}\frac{\partial U_1}{\partial x_3}+\frac{2}{3}\frac{\partial U_3}{\partial x_1}
&-\frac{4}{3}\frac{\partial U_2}{\partial x_2}
&-\frac{4}{3}\frac{\partial U_2}{\partial x_3}+\frac{2}{3}\frac{\partial U_3}{\partial x_2}
&\frac{2}{3}\frac{\partial U_3}{\partial x_3}
&0
&0
&0
&-2\left({ \frac{\partial U_2}{\partial x_2}-\frac{1}{3}\frac{\partial U_k}{\partial x_k} }\right)\\
0
&-\frac{\partial U_3}{\partial x_1}
&-\frac{\partial U_2}{\partial x_1}
&-\frac{\partial U_3}{\partial x_2}
&-\left({ \frac{\partial U_2}{\partial x_2}+\frac{\partial U_3}{\partial x_3} }\right)
&-\frac{\partial U_2}{\partial x_3}
&0
&0
&0
&-\left({ \frac{\partial U_2}{\partial x_3}+\frac{\partial U_3}{\partial x_2} }\right)\\
\frac{2}{3}\frac{\partial U_1}{\partial x_1}
&\frac{2}{3}\frac{\partial U_1}{\partial x_2}+\frac{2}{3}\frac{\partial U_2}{\partial x_1}
&\frac{2}{3}\frac{\partial U_1}{\partial x_3}-\frac{4}{3}\frac{\partial U_3}{\partial x_1}
&\frac{2}{3}\frac{\partial U_2}{\partial x_2}
&\frac{2}{3}\frac{\partial U_2}{\partial x_3}-\frac{4}{3}\frac{\partial U_3}{\partial x_2}
&-\frac{4}{3}\frac{\partial U_3}{\partial x_3}
&0
&0
&0
&-2\left({ \frac{\partial U_3}{\partial x_3}-\frac{1}{3}\frac{\partial U_k}{\partial x_k }}\right)\\
-\frac{C_p}{\rm{Pr}}\frac{\partial T}{\partial x_1}
&-\frac{C_p}{\rm{Pr}}\frac{\partial T}{\partial x_2}
&-\frac{C_p}{\rm{Pr}}\frac{\partial T}{\partial x_3}
&0
&0
&0
&-\frac{1}{\rm{Pr}}\frac{\partial U_1}{\partial x_1}
&-\frac{1}{\rm{Pr}}\frac{\partial U_1}{\partial x_2}
&-\frac{1}{\rm{Pr}}\frac{\partial U_1}{\partial x_3}
&-\frac{C_p}{\rm{Pr}}\frac{\partial T}{\partial x_1}\\
0
&-\frac{C_p}{\rm{Pr}}\frac{\partial T}{\partial x_1}
&0
&-\frac{C_p}{\rm{Pr}}\frac{\partial T}{\partial x_2}
&-\frac{C_p}{\rm{Pr}}\frac{\partial T}{\partial x_3}
&0
&-\frac{1}{\rm{Pr}}\frac{\partial U_2}{\partial x_1}
&-\frac{1}{\rm{Pr}}\frac{\partial U_2}{\partial x_2}
&-\frac{1}{\rm{Pr}}\frac{\partial U_2}{\partial x_3}
&-\frac{C_p}{\rm{Pr}}\frac{\partial T}{\partial x_2}\\
0
&0
&-\frac{C_p}{\rm{Pr}}\frac{\partial T}{\partial x_1}
&0
&-\frac{C_p}{\rm{Pr}}\frac{\partial T}{\partial x_2}
&-\frac{C_p}{\rm{Pr}}\frac{\partial T}{\partial x_3}
&-\frac{1}{\rm{Pr}}\frac{\partial U_3}{\partial x_1}
&-\frac{1}{\rm{Pr}}\frac{\partial U_3}{\partial x_2}
&-\frac{1}{\rm{Pr}}\frac{\partial U_3}{\partial x_3}
&-\frac{C_p}{\rm{Pr}}\frac{\partial T}{\partial x_3}\\
-3f_{\rm{b}}\frac{\partial U_1}{\partial x_1}
&-3f_{\rm{b}}\left({ \frac{\partial U_1}{\partial x_2}+\frac{\partial U_2}{\partial x_1} }\right)
&-3f_{\rm{b}}\left({ \frac{\partial U_1}{\partial x_3}+\frac{\partial U_3}{\partial x_1} }\right)
&-3f_{\rm{b}}\frac{\partial U_2}{\partial x_2}
&-3f_{\rm{b}}\left({ \frac{\partial U_2}{\partial x_3}+\frac{\partial U_3}{\partial x_2} }\right)
&-3f_{\rm{b}}\frac{\partial U_3}{\partial x_3}
&0
&0
&0
&-3f_{\rm{b}}\frac{\partial U_k}{\partial x_k}
\end{pmatrix}
\\
&-q(\kappa)\mathbf{I},
\end{aligned}
\end{equation}
\begin{equation}\label{eq:bb}
\tiny
\mathbf{b} = -\left[{
2\mu\left({ \frac{\partial U_1}{\partial x_1}-\frac{1}{3}\frac{\partial U_k}{\partial x_k} }\right),
\mu\left({ \frac{\partial U_1}{\partial x_2}+\frac{\partial U_2}{\partial x_1} }\right),
\mu\left({ \frac{\partial U_1}{\partial x_3}+\frac{\partial U_3}{\partial x_1} }\right),
2\mu\left({ \frac{\partial U_2}{\partial x_2}-\frac{1}{3}\frac{\partial U_k}{\partial x_k} }\right),
\mu\left({ \frac{\partial U_2}{\partial x_3}+\frac{\partial U_3}{\partial x_2} }\right),
2\mu\left({ \frac{\partial U_3}{\partial x_3}-\frac{1}{3}\frac{\partial U_k}{\partial x_k} }\right),
\frac{\mu C_p}{\rm{Pr}}\frac{\partial T}{\partial x_1},
\frac{\mu C_p}{\rm{Pr}}\frac{\partial T}{\partial x_2},
\frac{\mu C_p}{\rm{Pr}}\frac{\partial T}{\partial x_3},
\mu_{\rm{b}}\frac{\partial U_k}{\partial x_k}
}\right]^{\rm{T}}.
\end{equation}
\end{landscape}
\restoregeometry
It is noticed that all values in $\mathbf{b}$ are known and that all values in $\mathbf{A}$ are known except for $q(\kappa)$. If the value of $q(\kappa)$ is given, $\mathbf{m}$ can be directly obtained by $\mathbf{A}^{-1}\cdot\mathbf{b}$. This means that $\mathbf{m}$ is a function of $q(\kappa)$, and here it is abbreviated to: (Subscript ``f'' denotes ``function'')
\begin{equation}\label{eq:mq}
\mathbf{m}_{\rm{f}}(\tilde{q}) = \mathbf{A(\tilde{q})}^{-1}\cdot\mathbf{b},
\end{equation}
where $\tilde{q}$ denotes a given $q(\kappa)$. Meanwhile, from Eq.\ref{eq:qkapa} and Eq.\ref{eq:kapa}, it is known that $q(\kappa)$ is also a function of $\mathbf{m}$. Here, it is abbreviated to: (The calculating procedure follows Eq.\ref{eq:qkapa} and Eq.\ref{eq:kapa})
\begin{equation}\label{eq:qm}
q_{\rm{f}}(\tilde{\mathbf{m}}),
\end{equation}
where $\tilde{\mathbf{m}}$ denotes a given $\mathbf{m}$. Assuming that $\tilde{q}$ is the one determined by the exact solution, then $\mathbf{m}_{\rm{f}}(\tilde{q})$ is the very solution, and certainly $q_{\rm{f}}(\mathbf{m}_{\rm{f}}(\tilde{q}))$ equals to $\tilde{q}$. Therefore, an objective function of a single variable is designed: (Referring to Eq.\ref{eq:qkapa} and Eq.\ref{eq:kapa}, $\tilde{q}\geq 1.0$)
\begin{equation}\label{eq:fq}
F_{\rm{ob}}(\tilde{q}) = q_{\rm{f}}(\mathbf{m}_{\rm{f}}(\tilde{q}))-\tilde{q}.
\end{equation}
Evidently, at the zero point of $F_{\rm{ob}}(\tilde{q})$, $\mathbf{m}_{\rm{f}}(\tilde{q})$ is the exact solution. Here, the problem of solving multi-variable nonlinear complicated system (Eq.\ref{eq:nccr1}) is converted into solving an objective function of a single variable (Eq.\ref{eq:fq}).

\begin{figure}
	\centering
	\subfigure[]{
			\label{fig1a}
			\includegraphics[width=0.45 \textwidth]{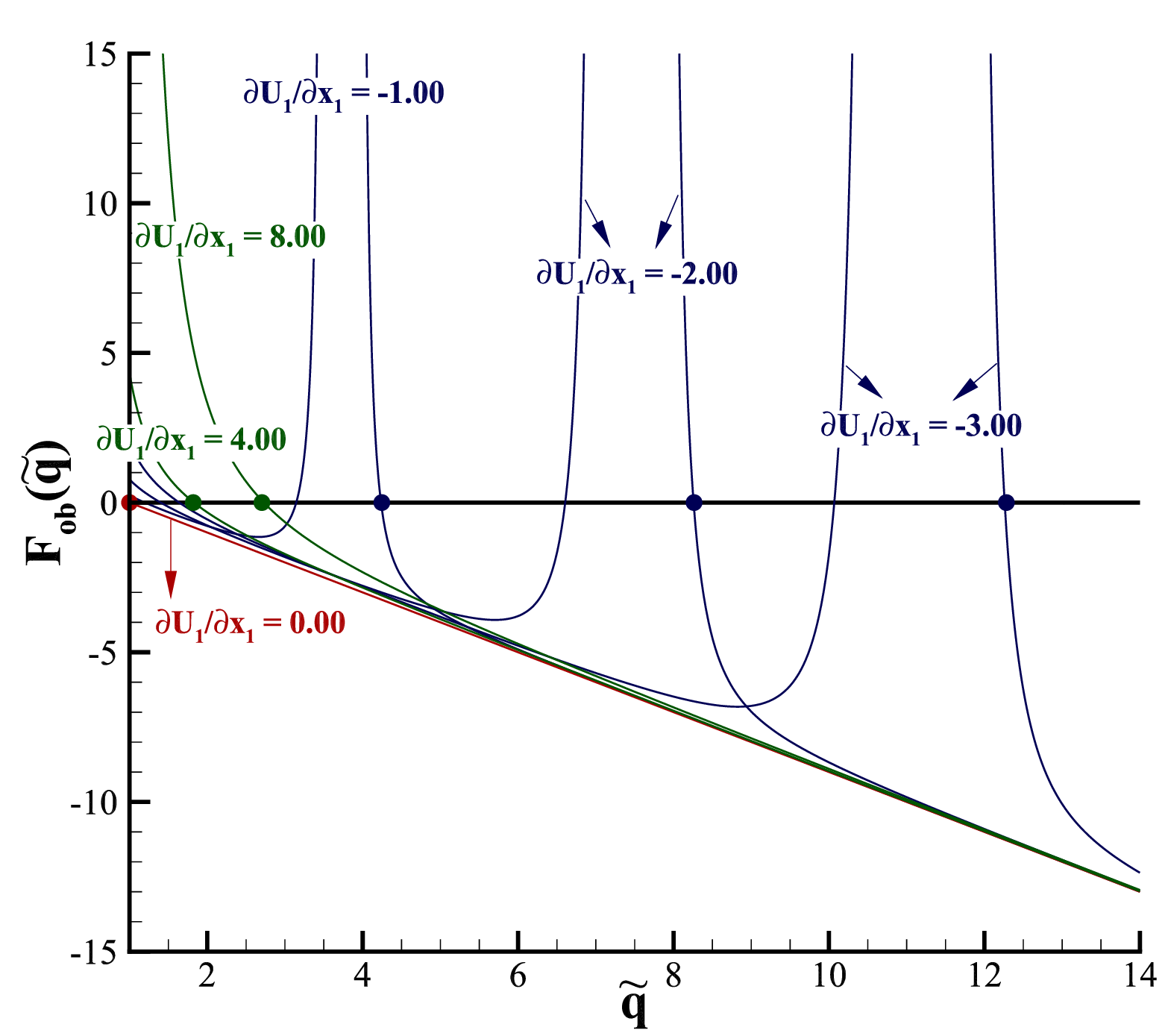}
		}
    \subfigure[]{
    		\label{fig1b}
    		\includegraphics[width=0.45 \textwidth]{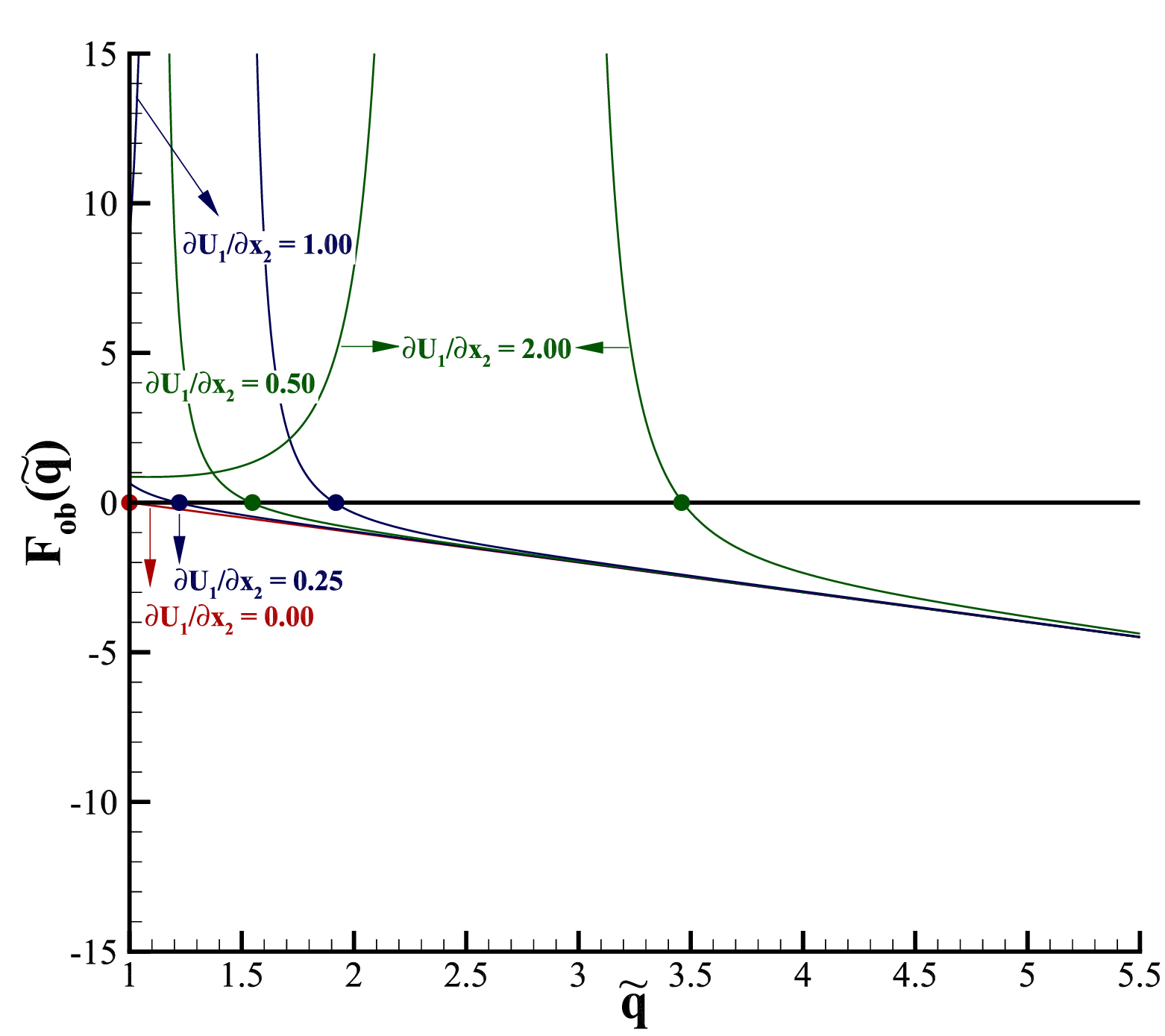}
    	}
	\caption{$F_{\rm{ob}}(\tilde{q})-\tilde{q}$ curves of simple cases: (a) Different $\frac{\partial U_1}{\partial x_1}$, (b) Different $\frac{\partial U_1}{\partial x_2}$.}
\end{figure}

\begin{figure}
	\centering
	\subfigure[]{
			\label{fig2a}
			\includegraphics[width=0.45 \textwidth]{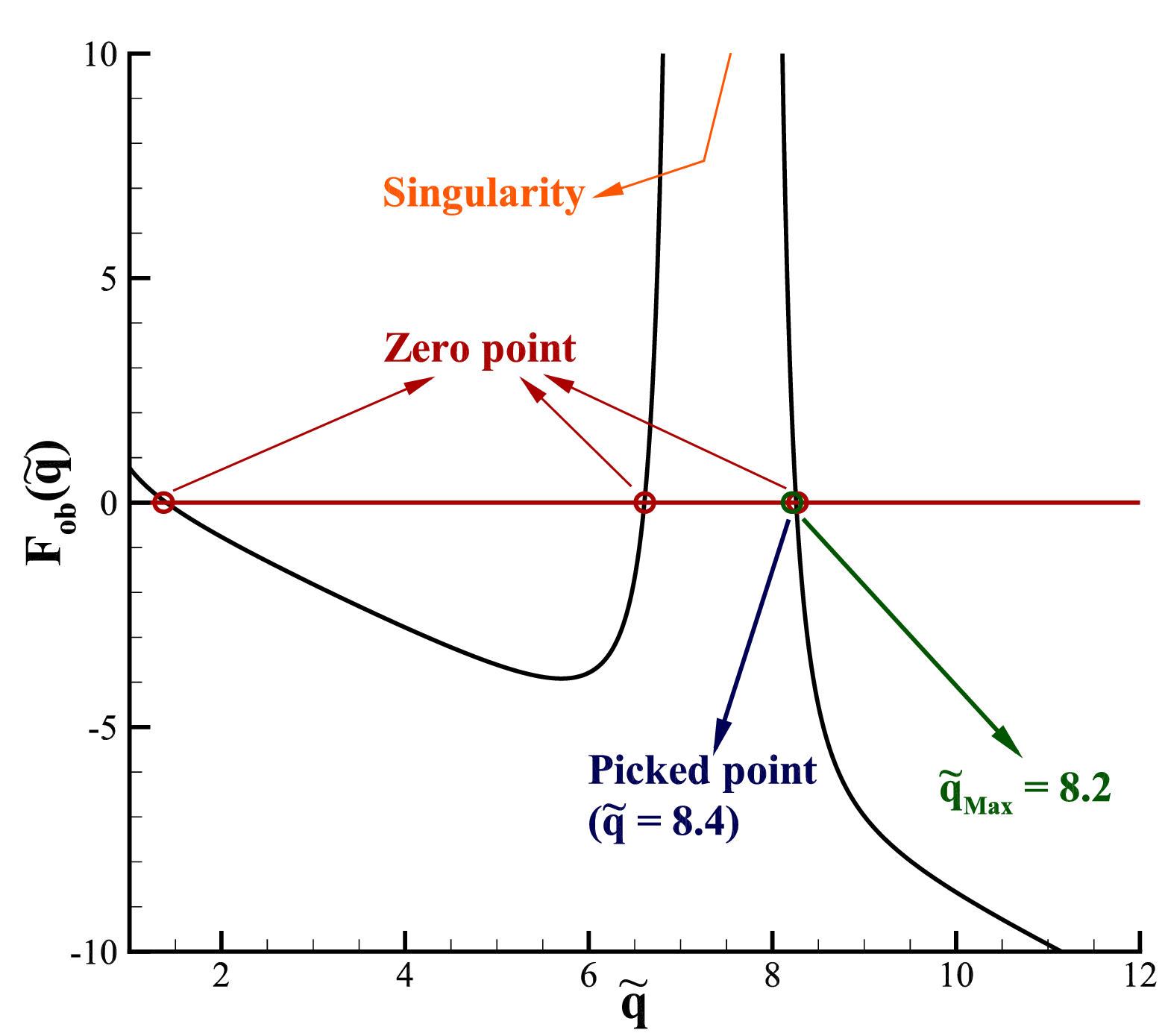}
		}
    \subfigure[]{
    		\label{fig2b}
    		\includegraphics[width=0.45 \textwidth]{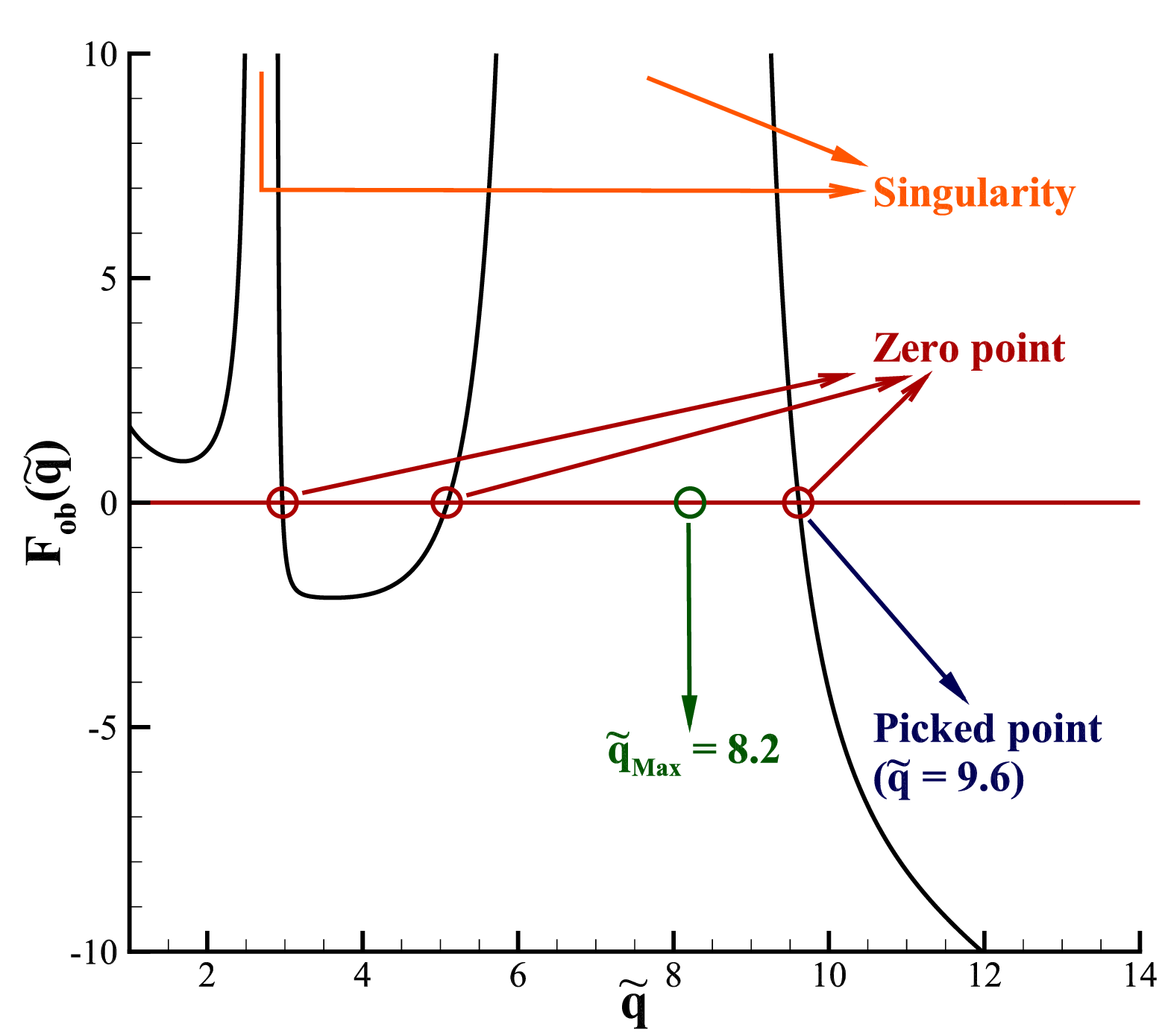}
    	}
	\caption{$F_{\rm{ob}}(\tilde{q})-\tilde{q}$ curves of certain cases: (a) $\frac{\partial U_1}{\partial x_1}=-2.0$, (b) $\frac{\partial U_1}{\partial x_1}=-2.0$ and $\frac{\partial T}{\partial x_1}=3.0$. ($\tilde{q}_{\rm{Max}}$ denotes the estimated maximum of the most right singularity and details are referred to Appendix B.)}
\end{figure}

To exhibit a clear picture of $F_{\rm{ob}}(\tilde{q})$, simple cases are taken as example (the value of $\frac{\pi^{0.25}}{\sqrt{2\beta}}$ in Eq.\ref{eq:kapa} is set to be $1.439528$). In Fig.\ref{fig1a}, all spatial derivatives of $\mathbf{U}$ and $T$ are set to zero except for $\frac{\partial U_1}{\partial x_1}$ (considering the normal stress), while in Fig.\ref{fig1b}, all spatial derivatives are set to zero except for $\frac{\partial U_1}{\partial x_2}$ (considering the shear stress). Surprisingly, when $\frac{\partial U_1}{\partial x_1}=-2.0$, there are three zero points and one singularity. This case is exhibited in detail in Fig.\ref{fig2a}. And in Fig.\ref{fig2b} ($\frac{\partial U_1}{\partial x_1}=-2.0$, $\frac{\partial T}{\partial x_1}=3.0$, other spatial derivatives are zero), there are three zero points and two singularities.

Several properties are summarized to get a more clear picture of $F_{\rm{ob}}(\tilde{q})$. These properties are easy to derive, as follows.
\begin{description}
    \item[Property (1)] When $\tilde{q}=1.0$, $F_{\rm{ob}}(\tilde{q}))\geq 0$.
    \item[Property (2)] When $\tilde{q}\rightarrow\infty$, $F_{\rm{ob}}(\tilde{q})\rightarrow -\infty$ and $\frac{dF_{\rm{ob}}(\tilde{q})}{d\tilde{q}}=-1$.
    \item[Property (3)] At every singularity, $F_{\rm{ob}}(\tilde{q})\rightarrow +\infty$.
\end{description}
From these properties, it can be inferred that there must be one or more than one zero point for $\tilde{q}$. Then, the problems become which is the physical one and how to find it. In this study, the largest zero point is regarded to be physical and here are reasons:
\begin{description}
    \item[Reason (1)] The $q$ corresponding to the $\mathbf{m}$ of NS equations (Eq.\ref{eq:ns}) through Eq.\ref{eq:qkapa} and Eq.\ref{eq:kapa} is always larger than all $\tilde{q}$ of zero points. Now that NCCR equations is considered to be a kind of extension of NS equations to non-equilibrium, the correct zero point should be nearest to the $q$ of NS equations.
    \item[Reason (2)] In zero-dimensional cases, if the largest zero point is chosen, the results are identical with the results of Myong. This is verified in Sec.\ref{sec:only}.
    \item[Reason (3)] If other zero points are chosen, discontinuity of solutions will happen. Fig.\ref{fig3} gives an example. There are three zero points when $\frac{\partial U_1}{\partial x_1}=-0.57,-0.60$ and there are one zero point when $\frac{\partial U_1}{\partial x_1}=-0.55,-0.50$. The solutions with the largest zero point is always continuous while others are not.
\end{description}

\begin{figure}
	\centering
	\subfigure{
			\includegraphics[width=0.5 \textwidth]{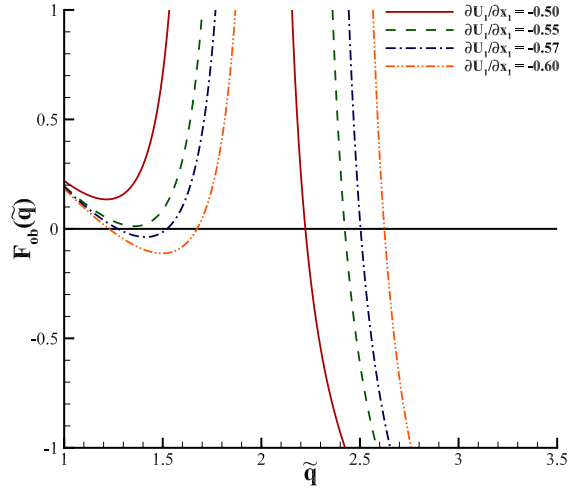}
		}
	\caption{\label{fig3} $F_{\rm{ob}}(\tilde{q})-\tilde{q}$ curves of $\frac{\partial U_1}{\partial x_1}=-0.50,-0.55,-0.56,-0.60$.}
\end{figure}

Currently, it is hard to evidence strictly which zero point is physical. While the second reason and the third reason above explain why other zero point can not be chosen and the first reason is a weak physical explanation. At this moment, it is time to design an algorithm to find the zero point at the most right side. Referring to the second property, the third property and examples above, the largest zero point is thought to be produced between the most right singularity and infinity. The former provides a positive $\mathbf{m}_{\rm{f}}(\tilde{q})$, and the latter provides the negative one. Then following the zero point theorem, the bisection method can be used for root finding. On the other hand, if there is no singularity, the bisection method is much easier to be used. In Appendix B, it is found that there is no singularity when:
\begin{equation}\label{eq:qmax}
\tilde{q}>\tilde{q}_{\rm{Max}},
\end{equation}
and a method to calculate $\tilde{q}_{\rm{Max}}$ is proposed there. In Fig.\ref{fig2a} and Fig.\ref{fig2b}, $\tilde{q}_{\rm{Max}}$ is marked for exhibition. Now that $\tilde{q}_{\rm{Max}}$ is calculated, it is complex but easy to find the two sides of the zero point theorem, then the bisection method can be used. The method for finding the exact solution of NCCR equations is shown as follows:
\begin{description}
    \item[Step (1)] Calculate $\tilde{q}_{\rm{Max}}$ in Eq.\ref{eq:qmax}. This is the first point of the zero point theorem.
    \item[Step (2)] Calculate $F_{\rm{ob}}(\tilde{q}_{\rm{Max}})$. If $F_{\rm{ob}}(\tilde{q}_{\rm{Max}})=0.0$, this is the correct zero point and go to step 4. If $F_{\rm{ob}}(\tilde{q}_{\rm{Max}})>0.0$, this is the most right singularity. Pick a $\tilde{q}$ large enough to be the second point, whose $F_{\rm{ob}}$ is negative. If $F_{\rm{ob}}(\tilde{q}_{\rm{Max}})<0.0$, the most right singularity is on the left side of $\tilde{q}_{\rm{Max}}$, so try different $\tilde{q}<\tilde{q}_{\rm{Max}}$ step by step until its $F_{\rm{ob}}$ is not negative, which is the second point. The interval for step-by-step method is always set $\frac{\tilde{q}_{\rm{Max}}}{40}$, which should be bounded in $\left[{0.02,1.5}\right]$ in experience.
    \item[Step (3)] Implement the bisection method between the two points to find the correct zero point.
    \item[Step (4)] Use Eq.\ref{eq:mq} to get the final solution.
\end{description}
The efficiency is tested in Sec.\ref{sec:arbb}. In that case, during the same iterations for the whole domain, the computation time of proposed method is $58.5$ times of method for NS equations, and $2.8$ times of MFPI for NCCR equations. Because MFPI is taken as the pretreatment of Jiang's coupling method, the efficiency of proposed method is as the same magnitude order of Jiang' method.

\section{Numerical experiments}\label{sec:cases}
In this section, a series of numerical test cases are conducted to validate the numerical performance of proposed method and the physical accuracy of NCCR model. The $u_{\rm{x/y}}$-only flows are simulated to show that in zero-dimensional cases, the results of proposed method are identical with results of Myong. Other kinds of flows are simulated to test the method and the model in different flow patterns, including shock structures in argon gas (Ma = $3.0$, $5.0$, $8.0$ and $20.0$), shock structures in nitrogen gas (Ma = $1.7$, $2.4$, $3.8$, $6.1$, $8.4$ and $10.0$), rarefied Couette flows, supersonic/hypersonic rarefied cylinder flows in nitrogen gas (Kn = 0.01, Ma = $3.0$, $6.0$ and $12.0$), the hypersonic rarefied cylinder flow in argon gas (Kn = $0.1$, Ma = $5.0$), the hypersonic rarefied flat flow in nitrogen gas (Kn = $0.016$, Ma = $20.2$), and supersonic rarefied sphere flows in nitrogen gas (Kn = $0.031$ and $0.121$, Ma = $4.25$). Other macroscopic methods are also tested for comparison, based on NS equations (no slip boundary condition), NS equations (slip boundary condition) and NS equations with the excess normal stress $\Delta$ of bulk viscosity (slip boundary condition). The MFPI method for solving NCCR equations is tested to verify its accuracy and robustness as well. Furthermore, in order to improve the stability of NCCR model at the wake (where the Kn number is too high and $q(\kappa)$ is too large), a remedy is proposed.

In one-dimensional cases, the Monotonic Upstream-centered Scheme for Conservation Laws (MUSCL) with van Leer limiter is used. To simulate two-dimensional flows, methods are imbed into SU2\cite{su1,su3}. SU2 is an open Source collection of software tools for the analysis of partial differential equations on unstructured meshes. The cell-vertex scheme with dual control volumes is used and the MUSCL with Venkatakrishnan limiter is used. Methods mentioned above can be referred to Ref.\cite{jiri,toro}. The slip boundary condition used in this study is the usual form proposed by Maxwell\cite{maxwell,lockerby}:
\begin{equation}
\begin{aligned}
U_{\rm{slip}} &= C_m\frac{2-\sigma_v}{\sigma_v}\lambda\frac{\partial U_1}{\partial x_2}+C_s\lambda\frac{\partial T}{\partial x_1},\\
T_{\rm{slip}} &= T_{\rm{wall}}+C_t\lambda\frac{\partial T}{\partial x_1},
\end{aligned}
\end{equation}
where,
\begin{equation}\label{eq:lambda}
\begin{aligned}
\lambda &= \frac{1}{\beta}\frac{\mu}{p}\sqrt{\frac{RT}{2\pi}},\\
C_m &= \pi\beta,\\
C_s &=\frac{3}{4}\pi\beta\sqrt{\frac{2R}{\pi T}},\\
C_t &=\frac{2-\sigma_T}{\sigma_T}\pi\beta\frac{2\gamma}{(\gamma+1)\rm{Pr}},
\end{aligned}
\end{equation}
where $\lambda$ denotes the molecular mean free path, $\beta$ is calculated through Eq.\ref{eq:vss}. Parameters $\sigma_v$ and $\sigma_T$ equal to $1$ for surfaces that reflect all incident molecules diffusely.

\subsection{$u_{\rm{x/y}}$-only flow}\label{sec:only}
This is a zero-dimensional test case. All spatial derivatives are set to zero except for $\frac{\partial U_1}{\partial x_1}$ or $\frac{\partial U_1}{\partial x_2}$. The gas is set to be nitrogen gas, and parameters are $\rm{Pr}=0.72$, $f_{\rm{b}}=0.8$, and $\frac{\pi^{0.25}}{\sqrt{2\beta}}=1.439528$ in Eq.\ref{eq:kapa}. In Myong's work\cite{myong1,myong2,myong3}, this case is used to exhibit difference among different constitutive relationship models. Because the decomposed method of Myong is accurate for one-dimensional flow, its results are used for comparison, to test the accuracy of proposed method and MFPI method.

\begin{figure}
	\centering
	\subfigure[]{
			\includegraphics[width=0.45 \textwidth]{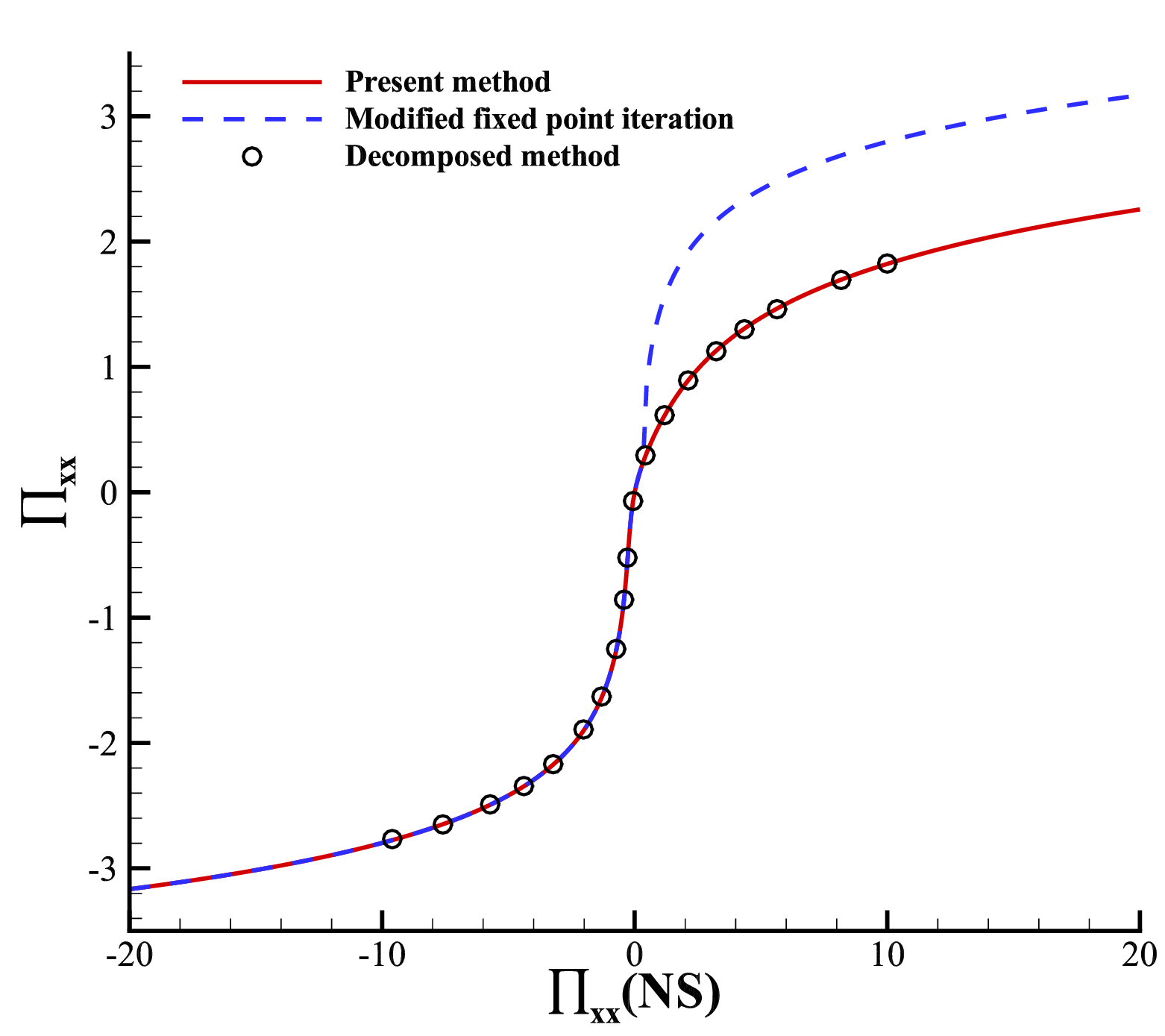}
		}
    \subfigure[]{
    		\includegraphics[width=0.45 \textwidth]{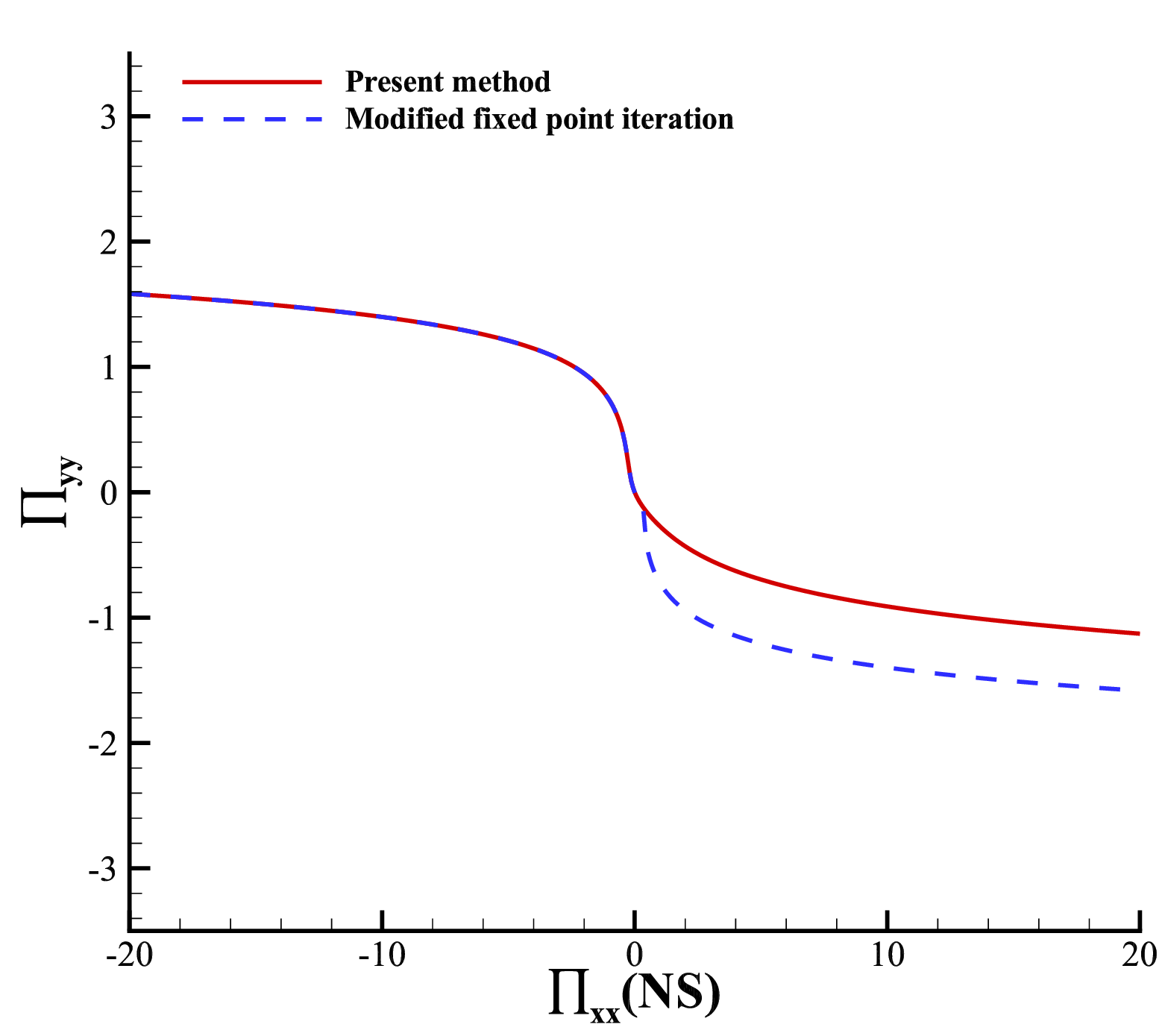}
    	}
    \subfigure[]{
            \label{fig3c}
			\includegraphics[width=0.45 \textwidth]{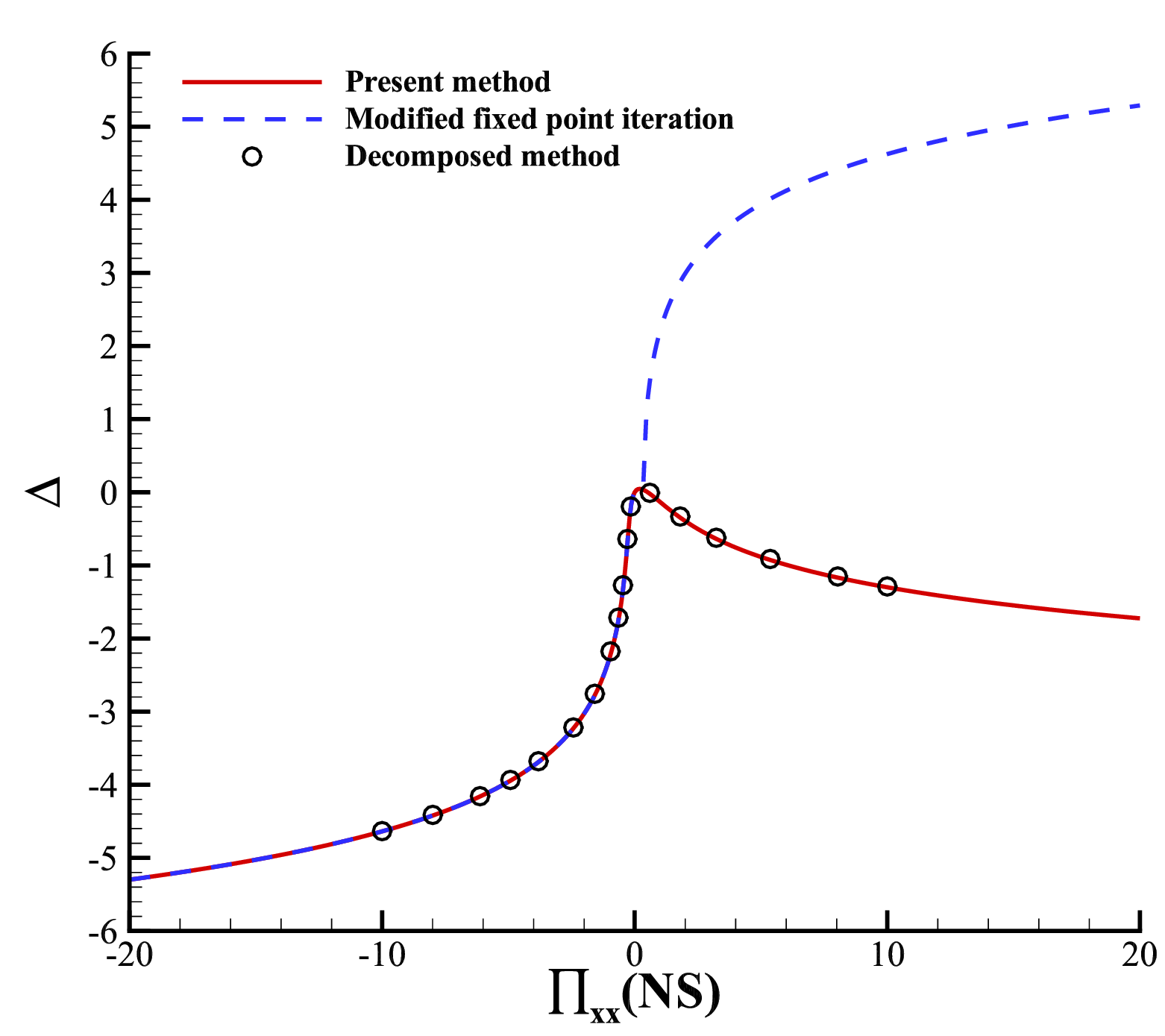}
		}
    \subfigure[]{
    		\includegraphics[width=0.45 \textwidth]{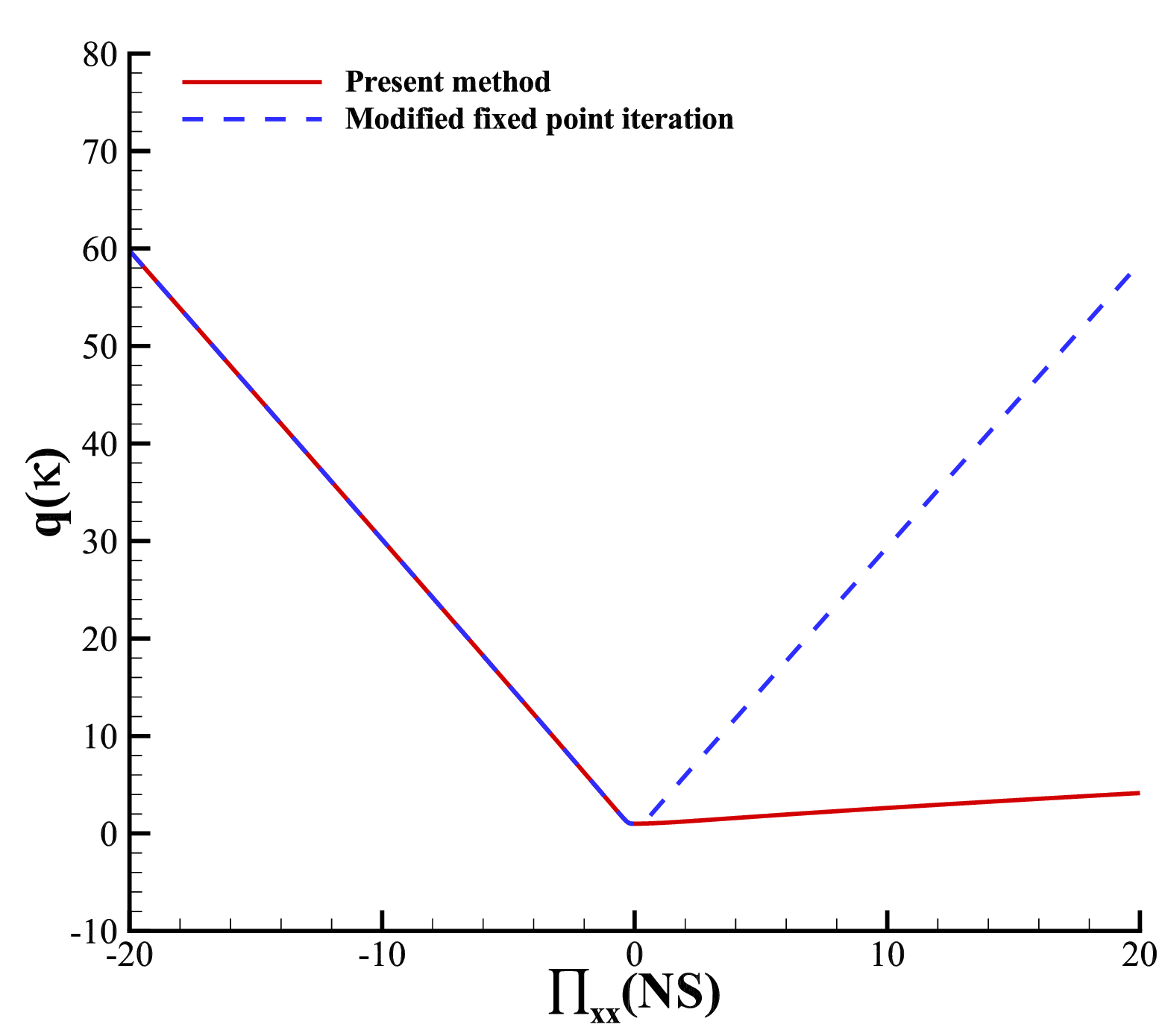}
    	}
	\caption{\label{fig3} Results of $u_{\rm{x}}$-only flow: (a) $\Pi_{\rm{xx}}$, (b) $\Pi_{\rm{yy}}$, (c) $\Delta$, (d) $q_{\kappa}$.}
\end{figure}

\begin{figure}
	\centering
	\subfigure[]{
			\includegraphics[width=0.45 \textwidth]{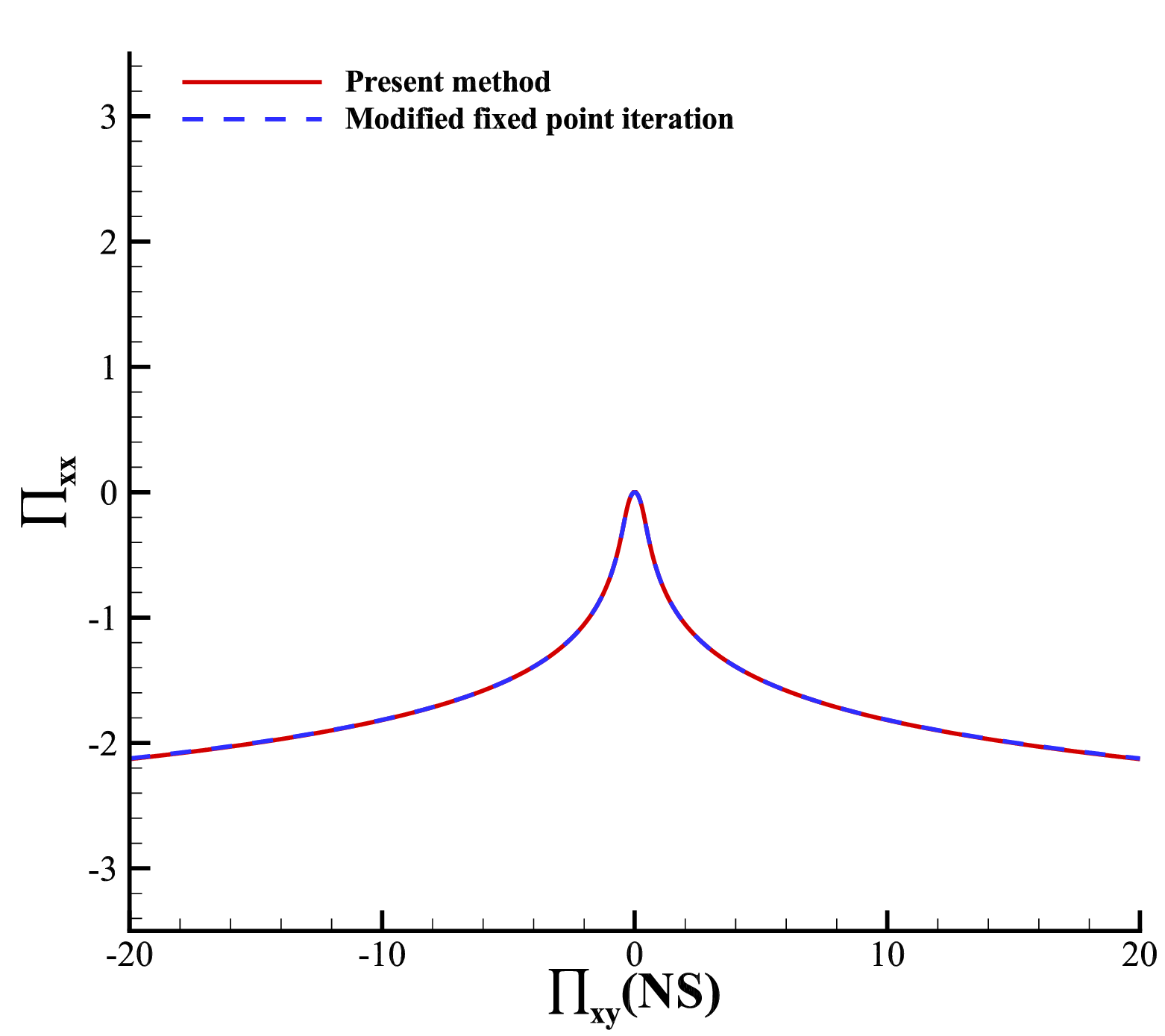}
		}
    \subfigure[]{
    		\includegraphics[width=0.45 \textwidth]{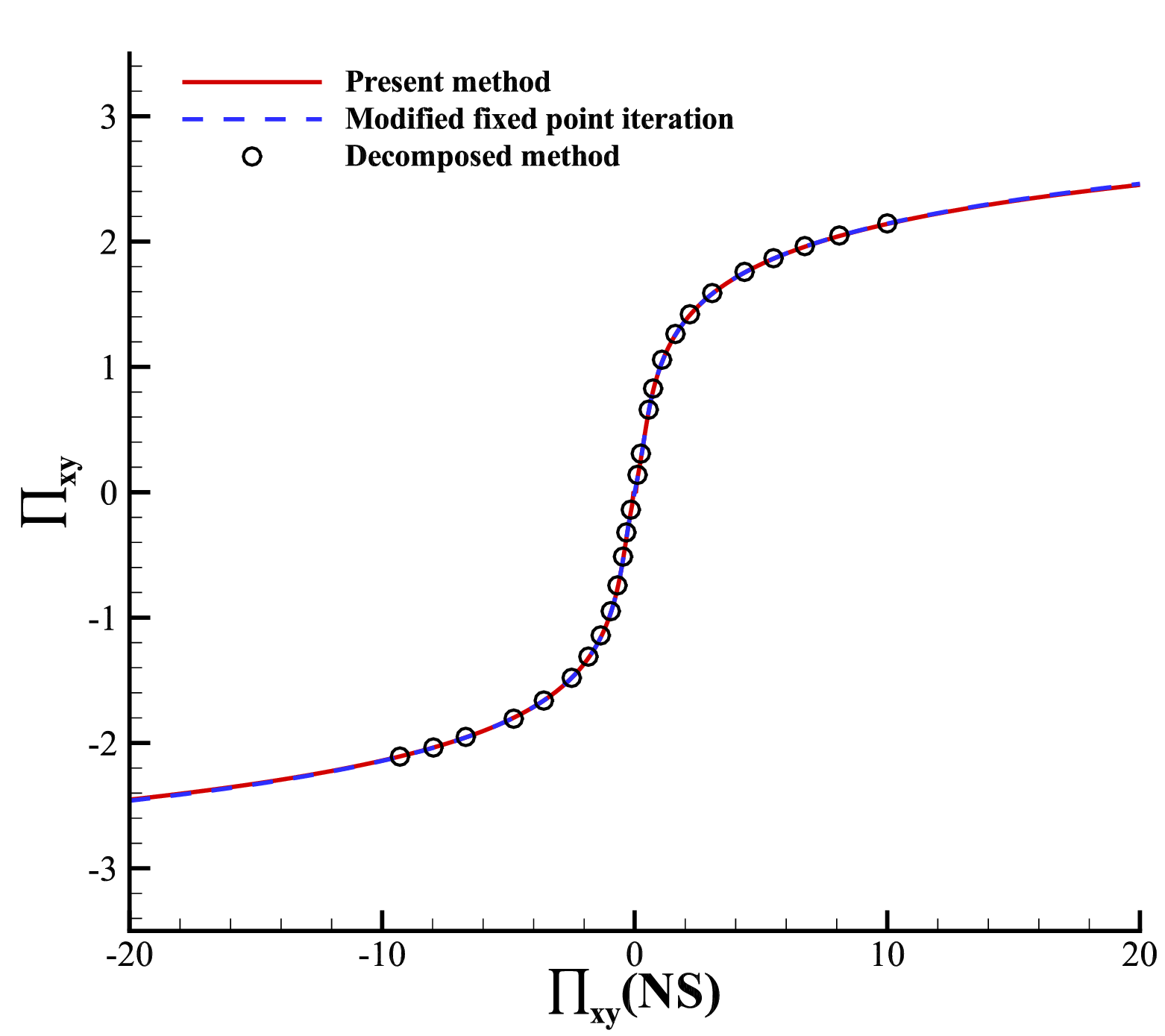}
    	}
    \subfigure[]{
			\includegraphics[width=0.45 \textwidth]{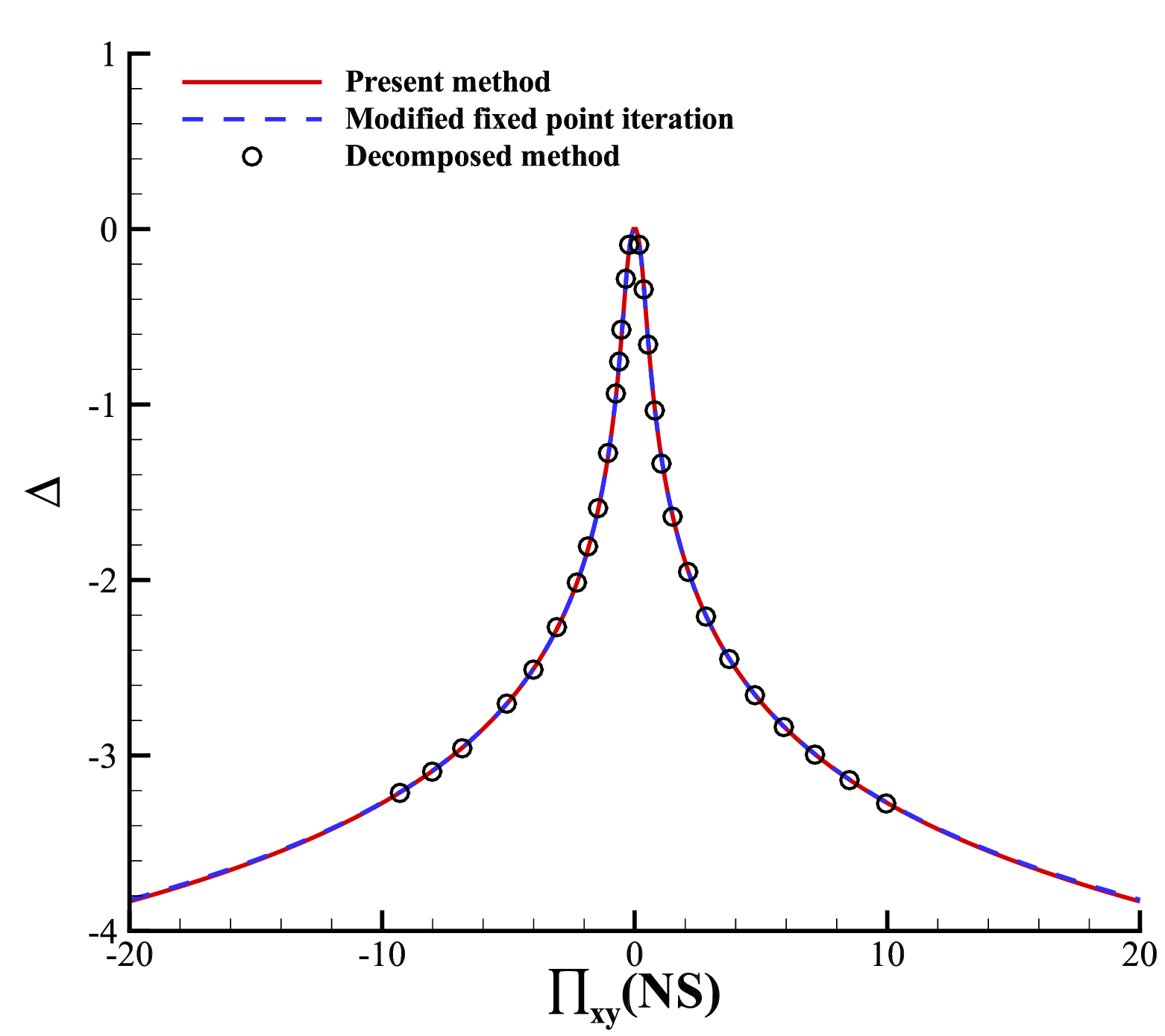}
		}
    \subfigure[]{
    		\includegraphics[width=0.45 \textwidth]{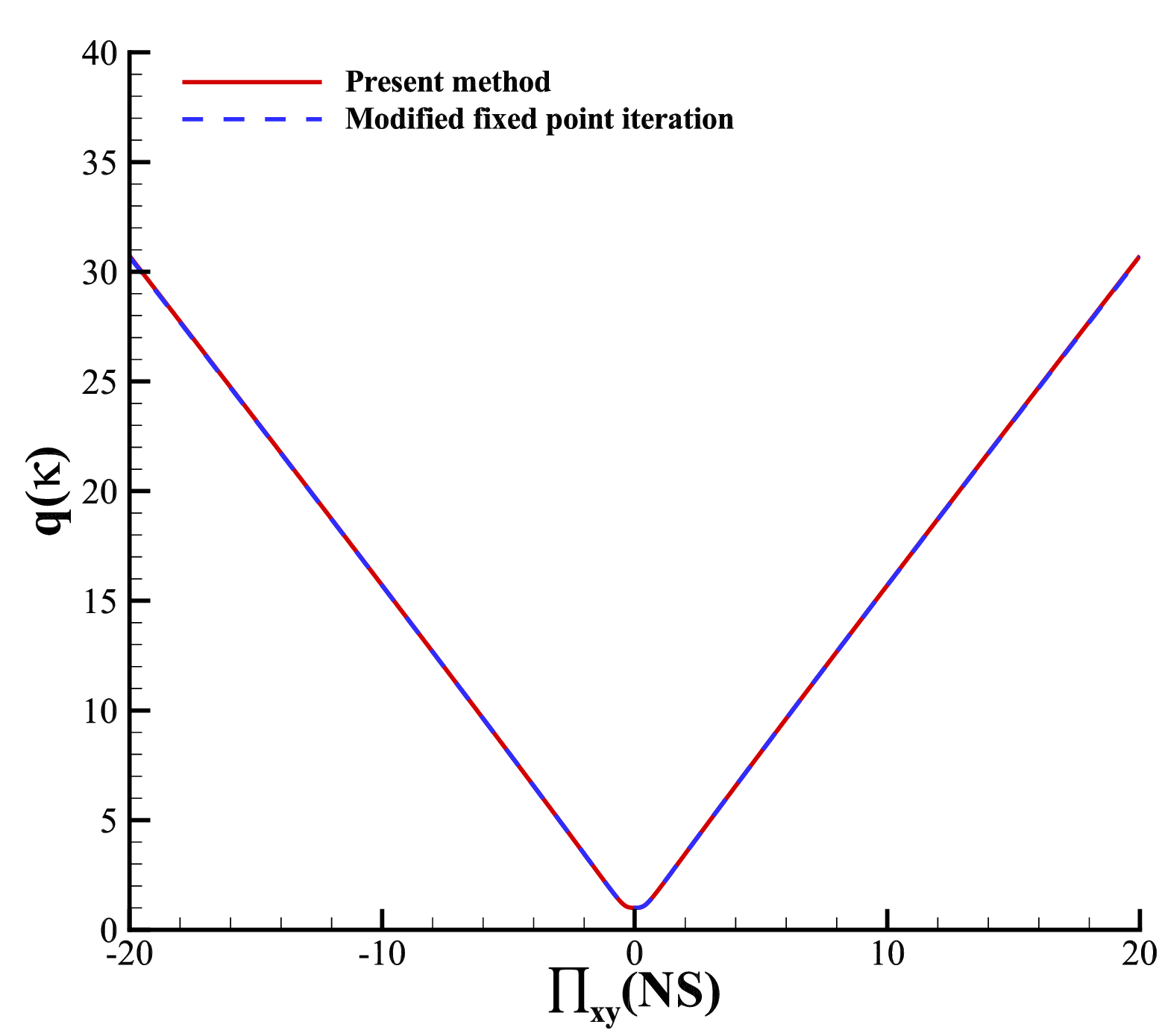}
    	}
	\caption{\label{fig4} Results of $u_{\rm{y}}$-only flow: (a) $\Pi_{\rm{xx}}$, (b) $\Pi_{\rm{xy}}$, (c) $\Delta$, (d) $q_{\kappa}$.}
\end{figure}

Results are shown in Fig.\ref{fig3} and Fig.\ref{fig4}. Results of the proposed method are in line well with ones of Myong's method. However, results of MFPI method do not match well. In the right part of Fig.\ref{fig3c}, the sign of excess normal stress $\Delta$ is wrong. Referring to Fig.\ref{fig1a}, there is only one zero point when $\frac{\partial U_1}{\partial x_1}>0$ ($\Pi_{\rm{xx},\rm{NS}}>0$), which indicates that $F_{\rm{ob}}(q_{\rm{MFPI}})\neq0.0$ and that the result of MFPI is not even any one solution of NCCR equations.

\subsection{Shock structures monatomic in argon gas at different Mach numbers}\label{sec:ssar}
Shock structure is a typical non-equilibrium flow. When Mach number is high, its nonequilibrium feature is strong. It is a classical case to test the ability of models or numerical methods to describe non-equilibrium flows. In order to avoid the influence of bulk viscosity, rotational relaxation and vibrational relaxation, monatomic argon gas flows are firstly simulated, at Mach numbers $3.0$, $5.0$, $8.0$, $20.0$. The inviscid flux used is KIF and parameters are: $\gamma=5/3$, $\rm{Pr}=2/3$. The dynamic viscosity is calculated through $\mu\sim T^{\omega}$. The size of mesh is set to be a half of molecular mean free path $\lambda$, which is calculated through Eq.\ref{eq:lambda}.

\begin{figure}
	\centering
	\subfigure[]{
			\includegraphics[width=0.45 \textwidth]{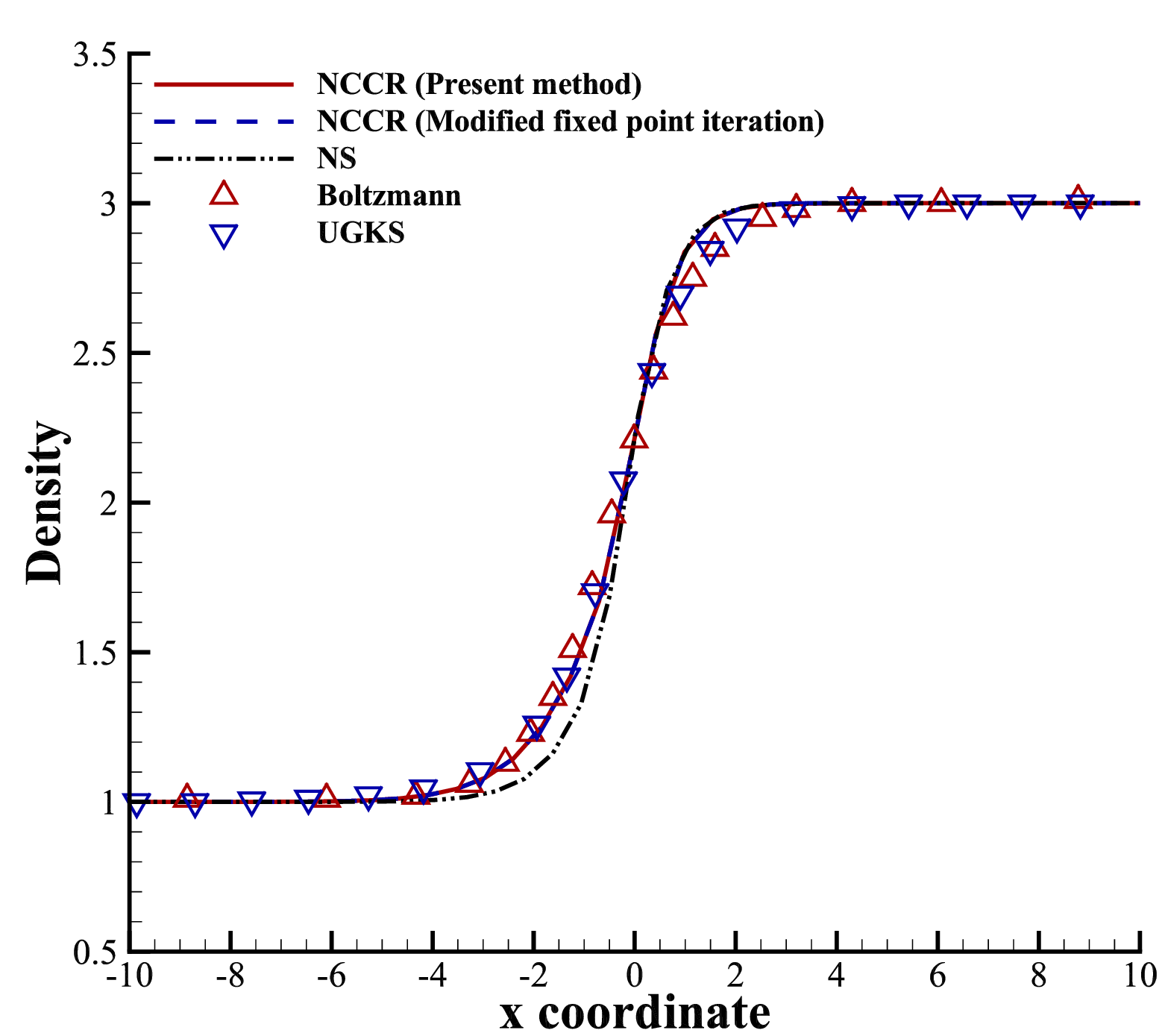}
		}
    \subfigure[]{
    		\includegraphics[width=0.45 \textwidth]{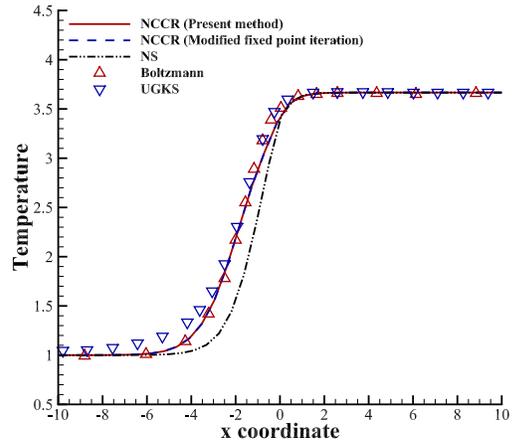}
    	}
    \subfigure[]{
			\includegraphics[width=0.45 \textwidth]{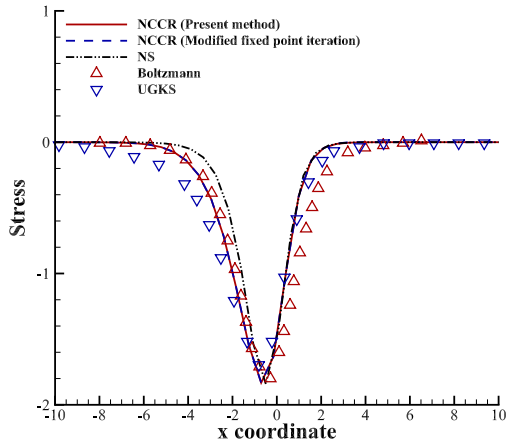}
		}
    \subfigure[]{
    		\includegraphics[width=0.45 \textwidth]{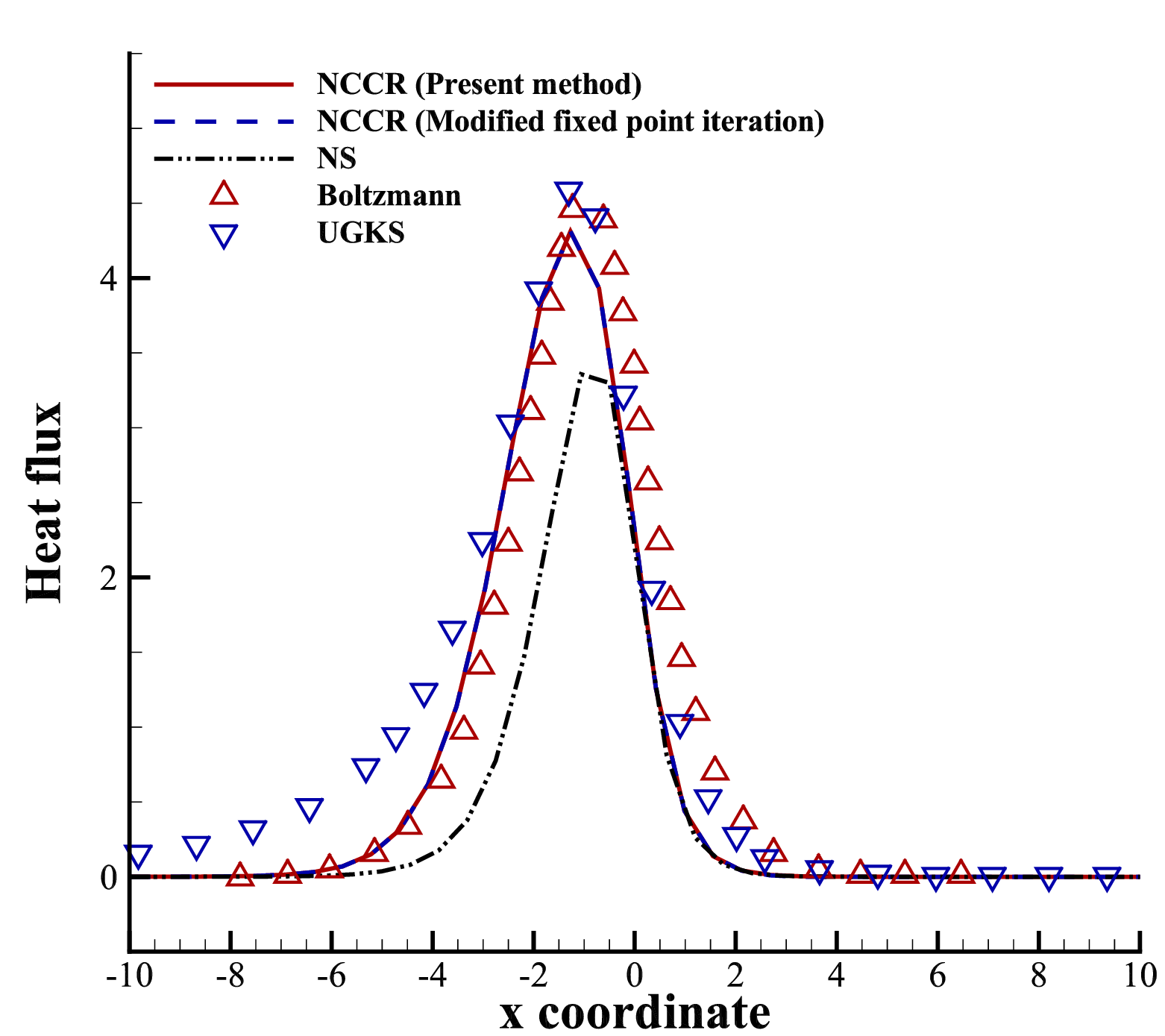}
    	}
	\caption{\label{fig5} Results of $\rm{Ma}=3.0$ argon gas shock structure: (a) Density, (b) temperature, (c) stress, (d) heat flux.}
\end{figure}

\begin{figure}
	\centering
	\subfigure[]{
			\includegraphics[width=0.45 \textwidth]{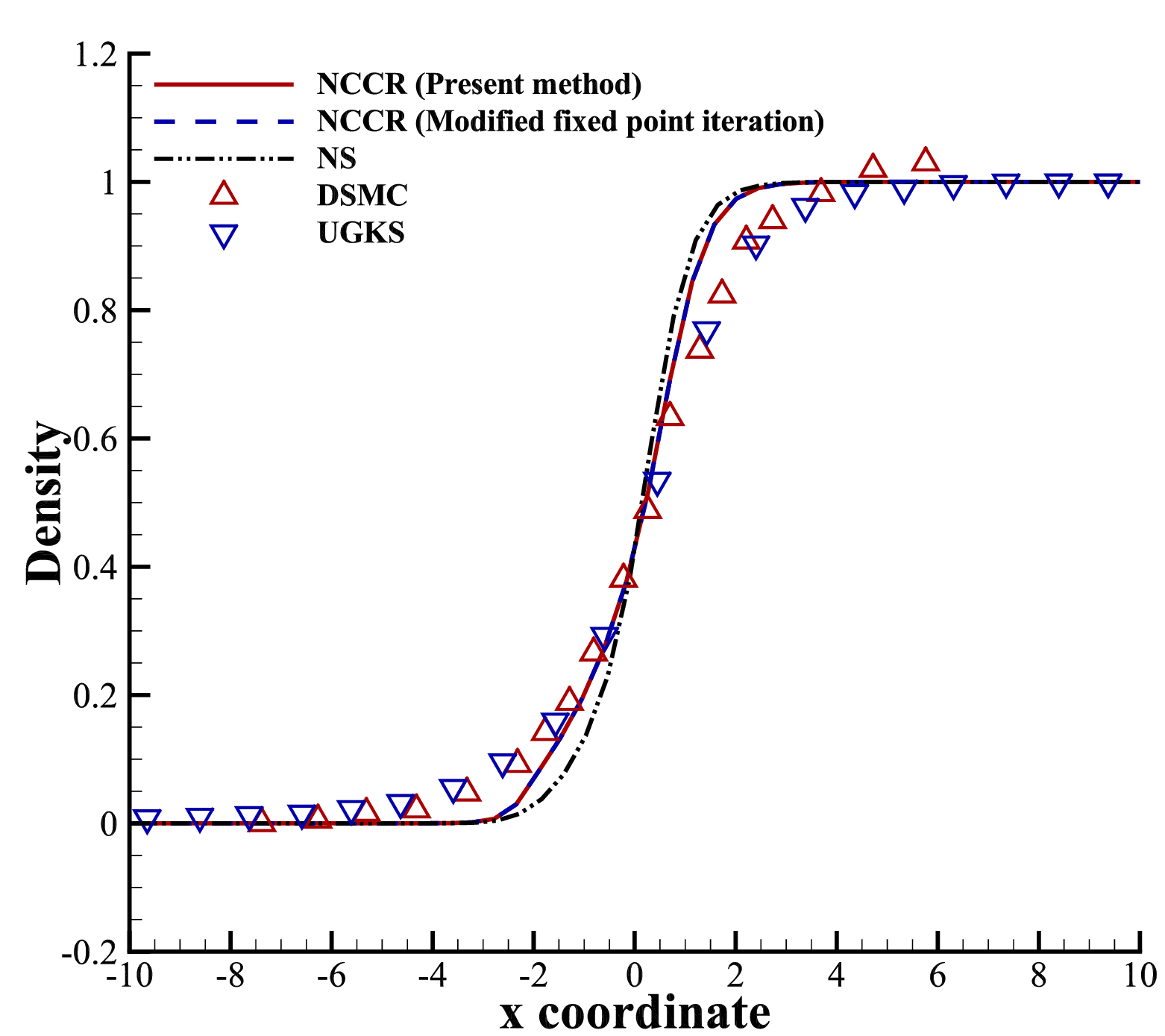}
		}
    \subfigure[]{
    		\includegraphics[width=0.45 \textwidth]{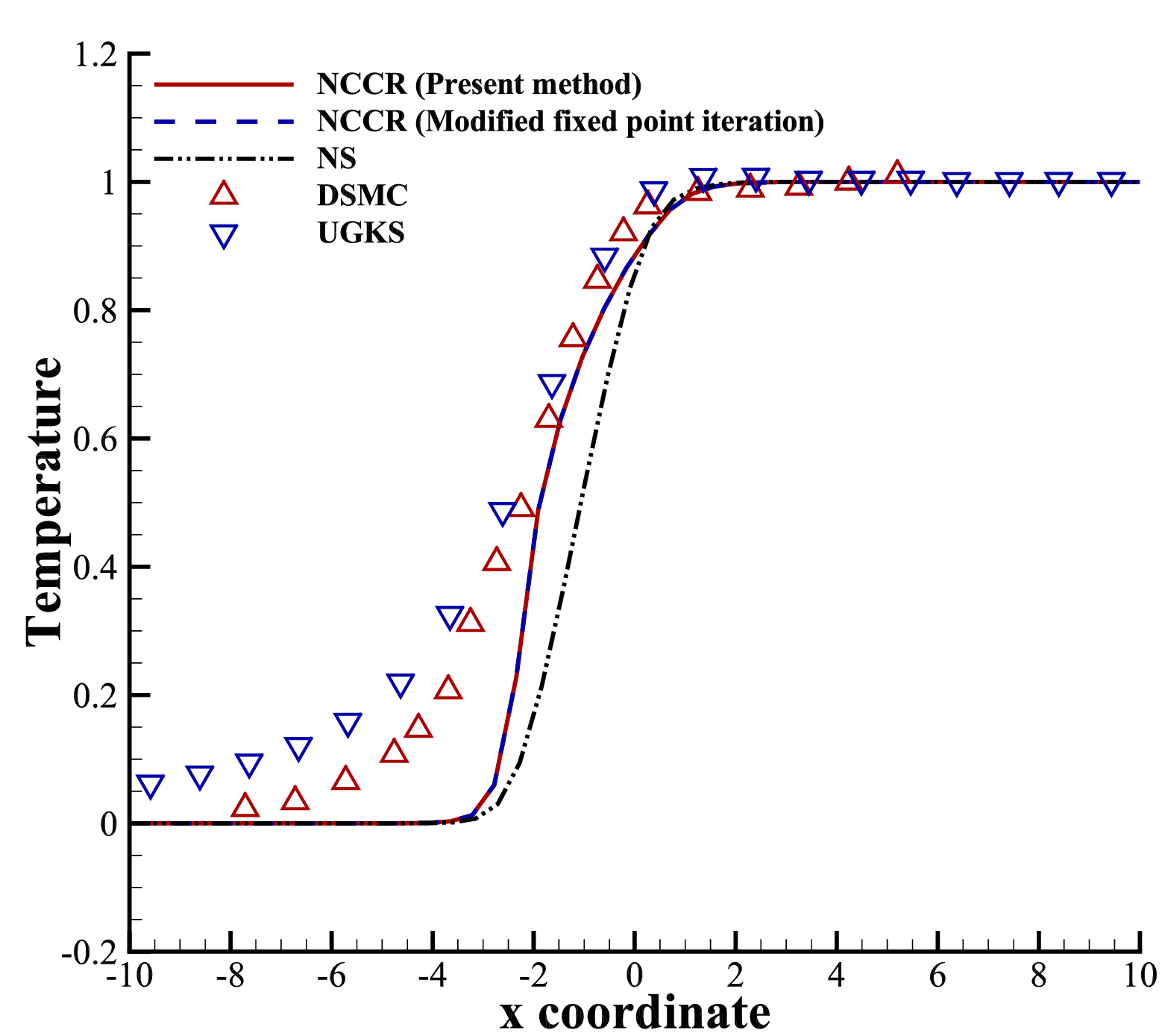}
    	}
    \subfigure[]{
			\includegraphics[width=0.45 \textwidth]{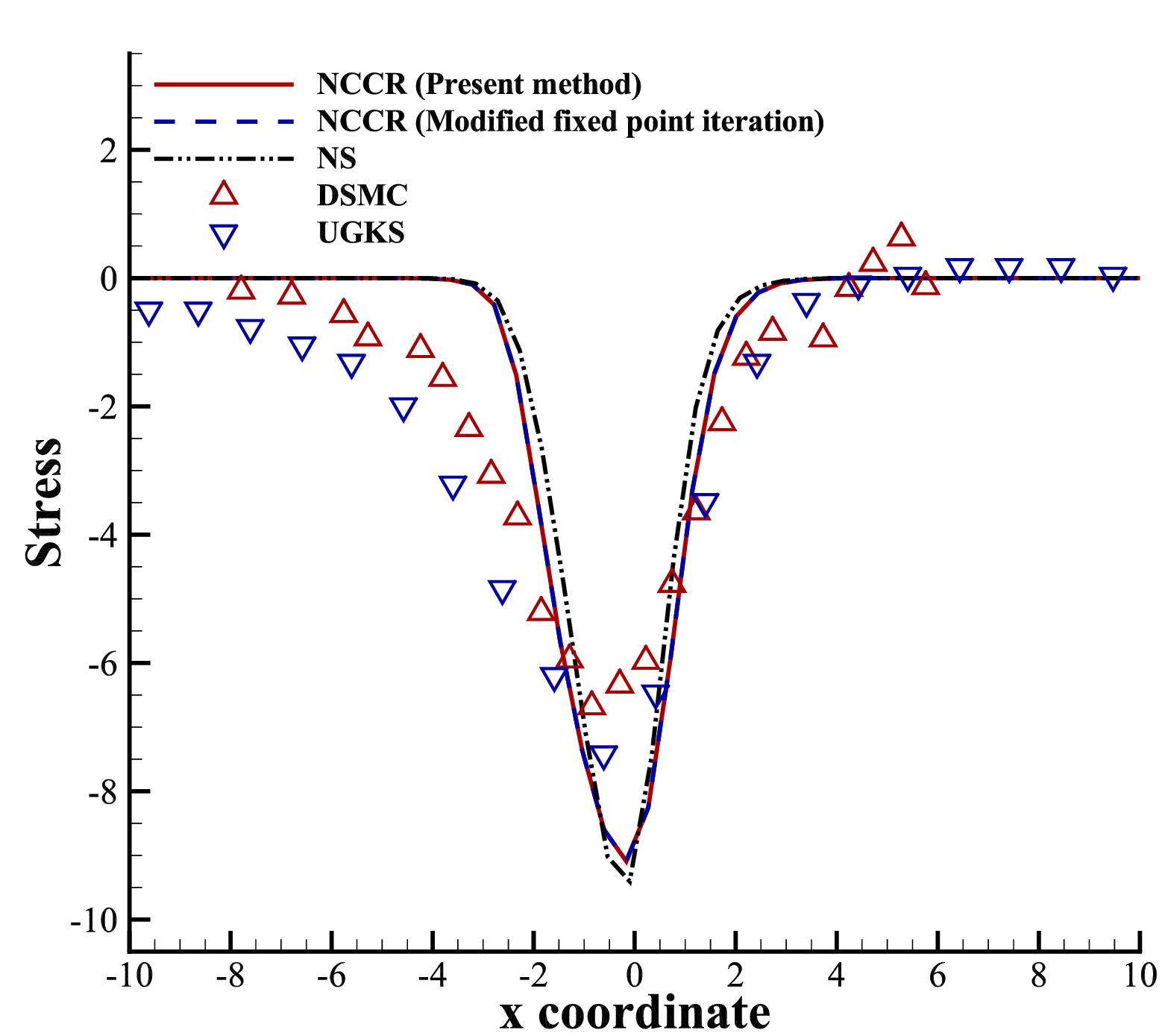}
		}
    \subfigure[]{
    		\includegraphics[width=0.45 \textwidth]{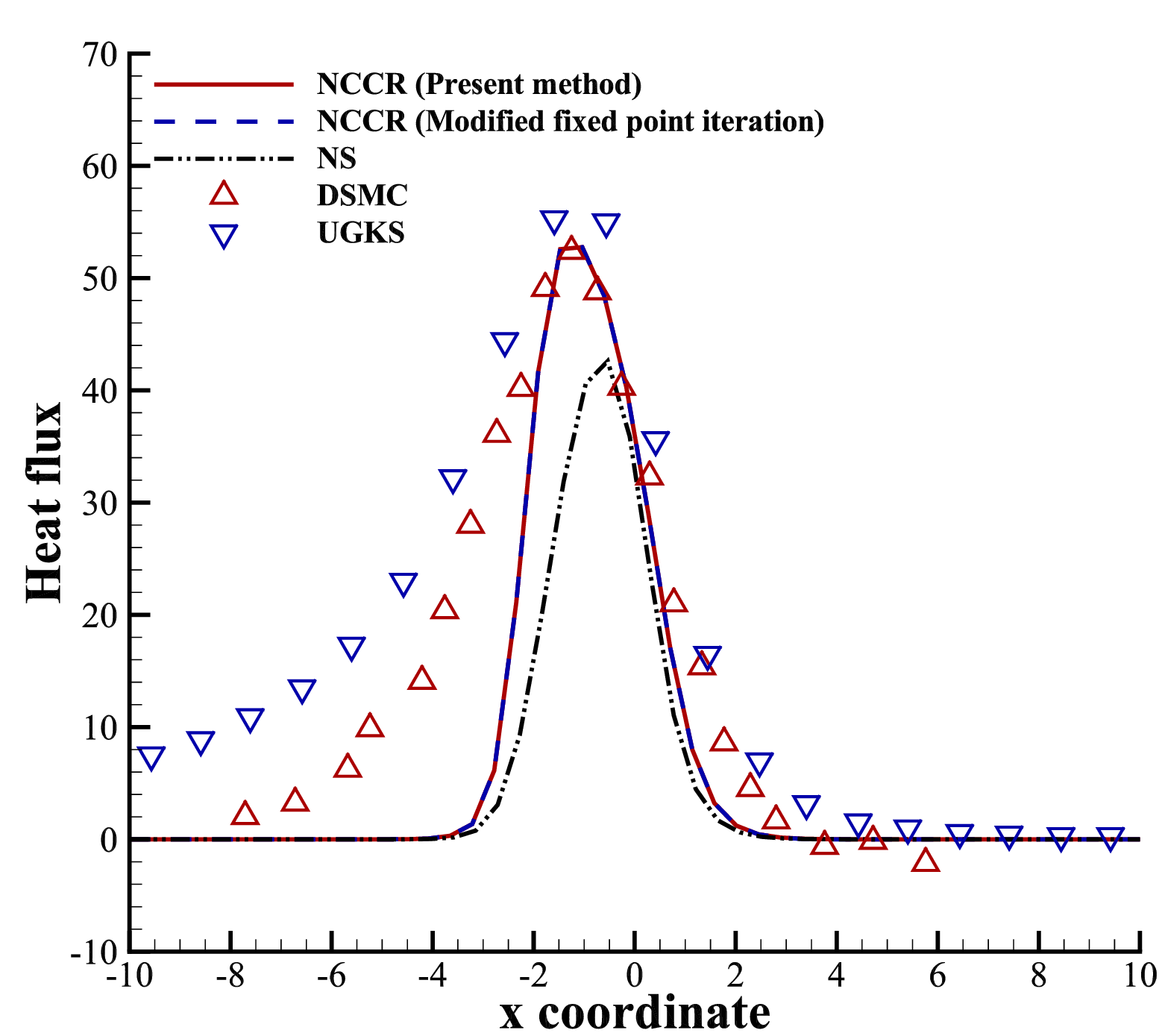}
    	}
	\caption{\label{fig6} Results of $\rm{Ma}=8.0$ argon gas shock structure: (a) Density, (b) temperature, (c) stress, (d) heat flux.}
\end{figure}

\begin{figure}
	\centering
	\subfigure[]{
			\includegraphics[width=0.45 \textwidth]{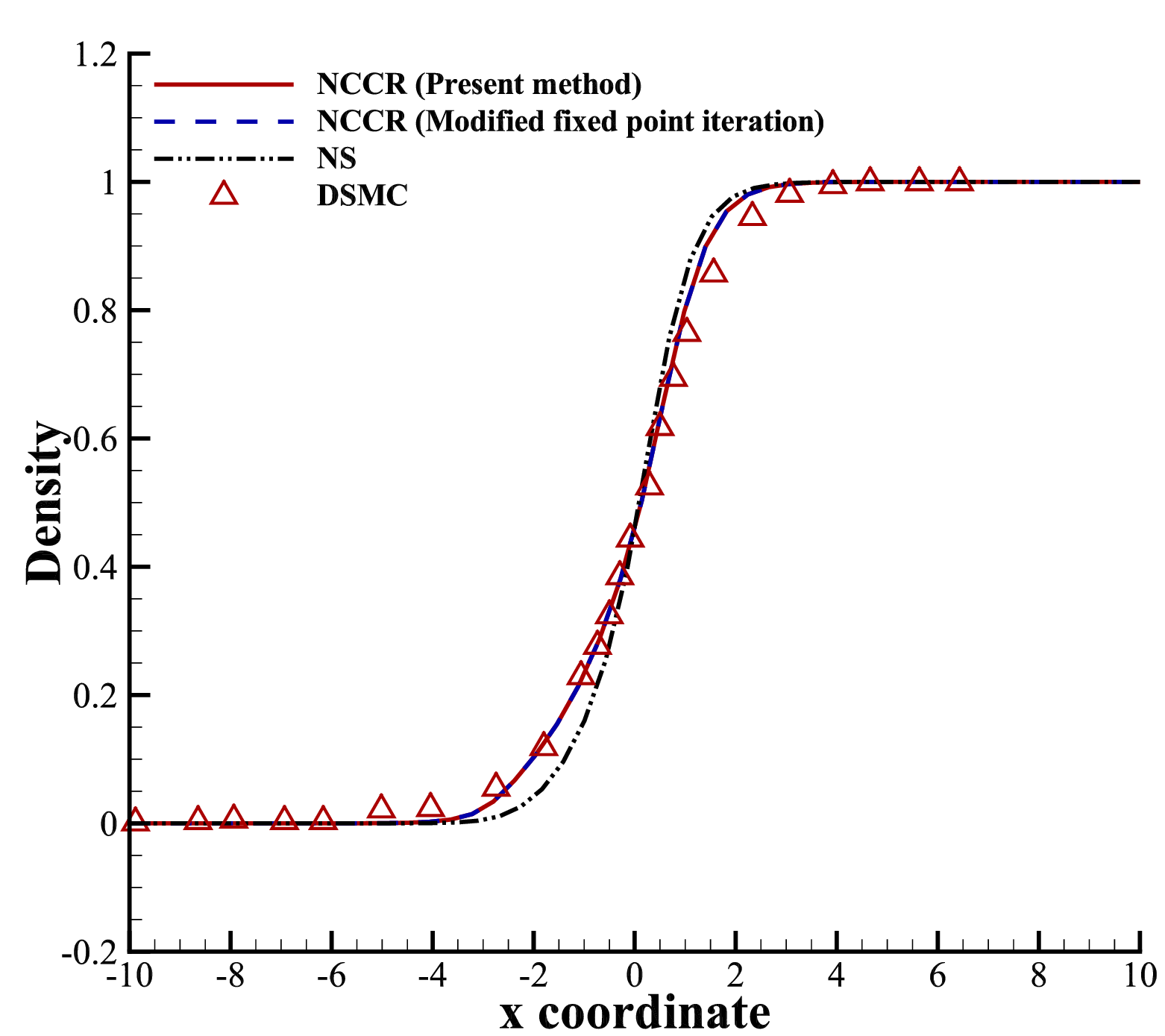}
		}
    \subfigure[]{
    		\includegraphics[width=0.45 \textwidth]{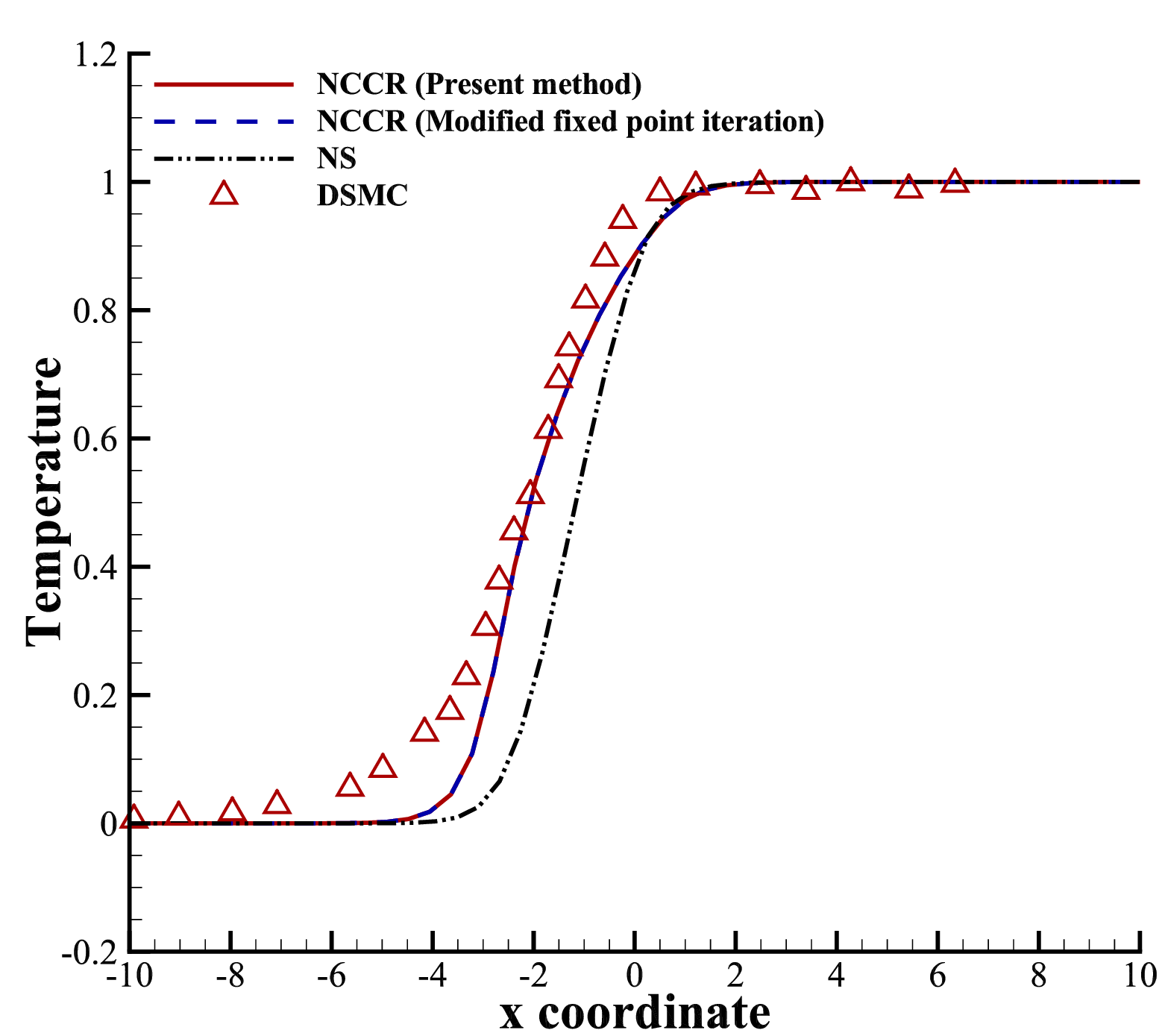}
    	}
	\caption{\label{fig7} Results of $\rm{Ma}=5.0$ argon gas shock structure: (a) Density, (b) temperature.}
\end{figure}

\begin{figure}
	\centering
	\subfigure[]{
			\includegraphics[width=0.45 \textwidth]{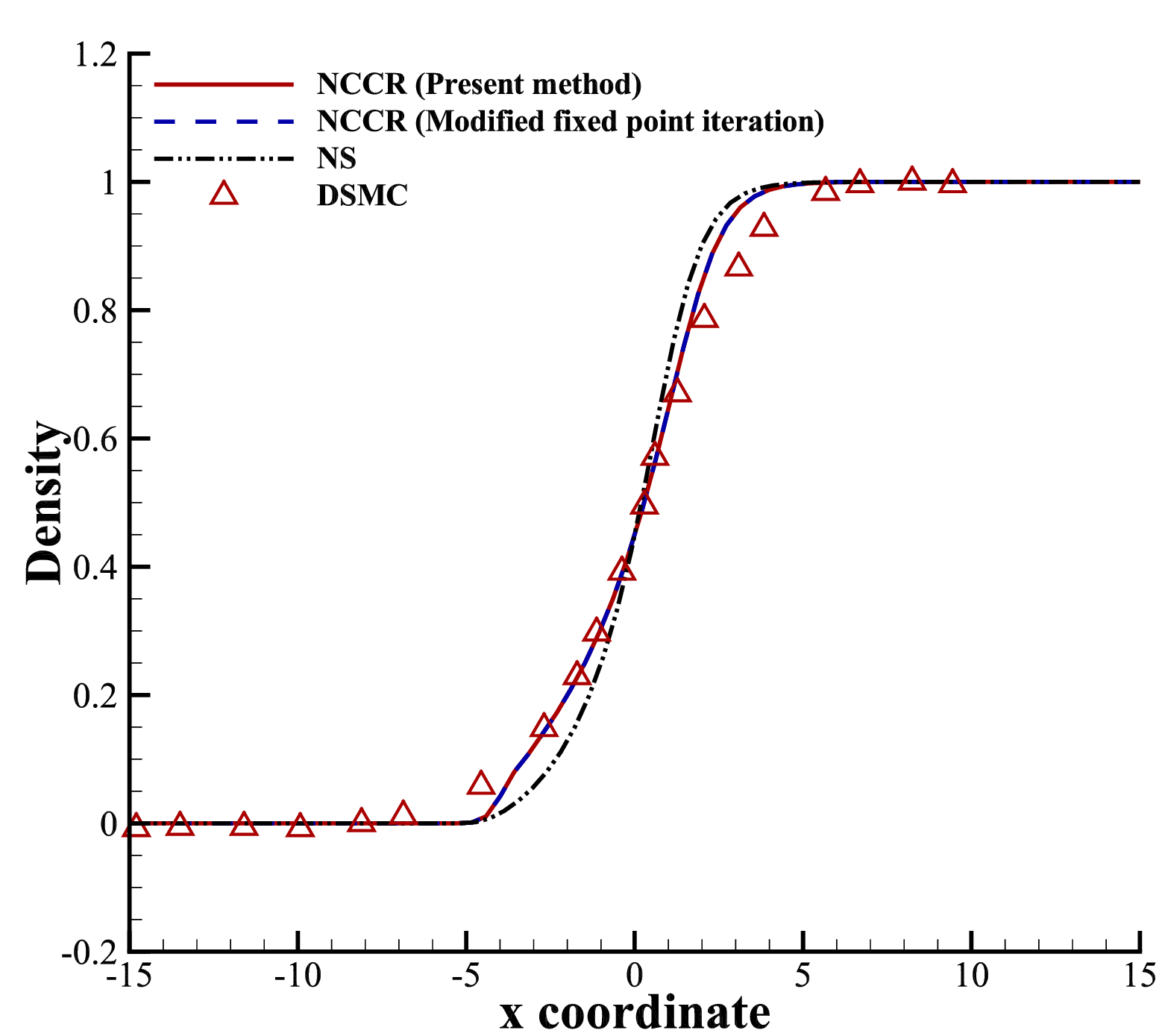}
		}
    \subfigure[]{
    		\includegraphics[width=0.45 \textwidth]{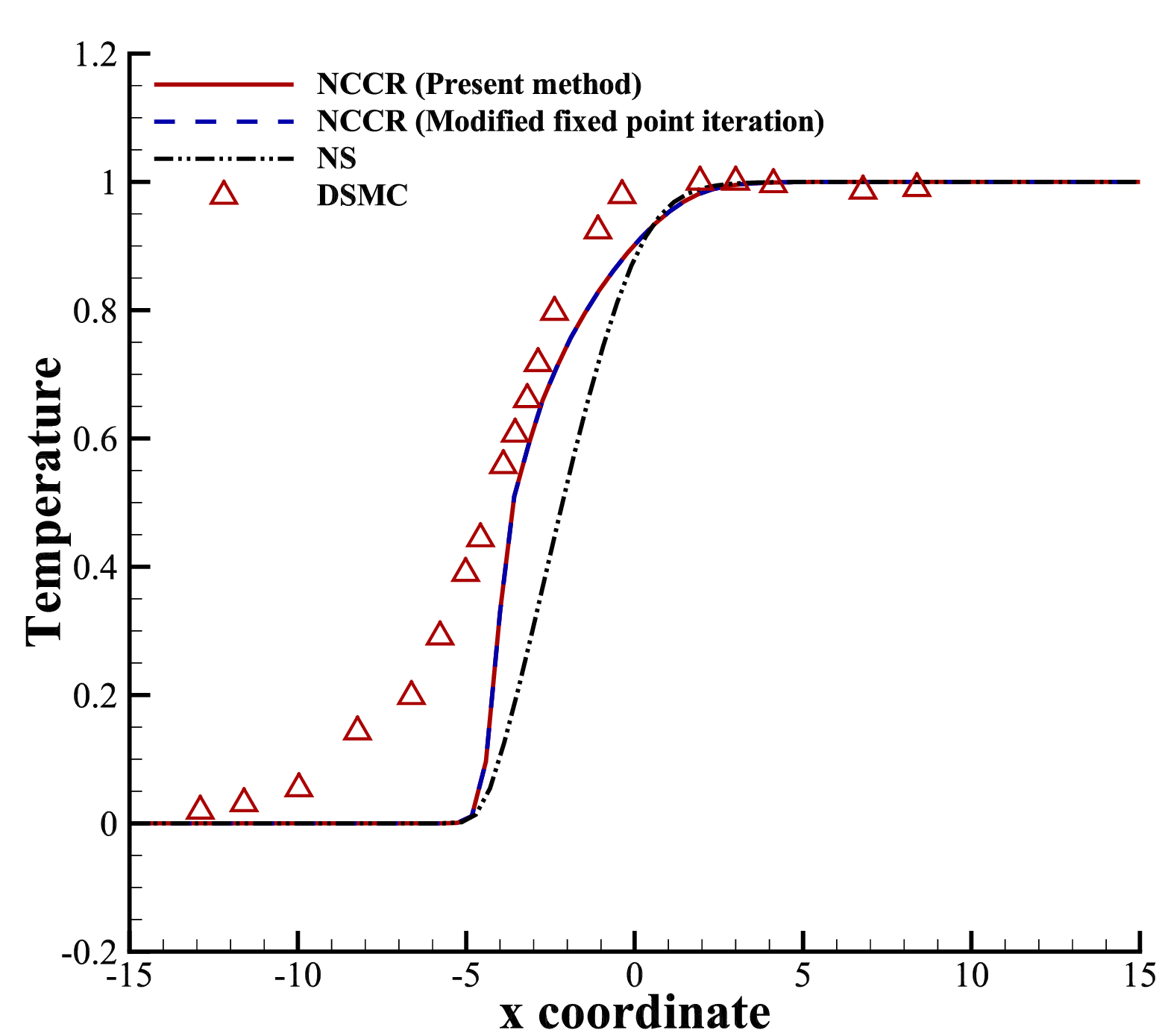}
    	}
	\caption{\label{fig8} Results of $\rm{Ma}=20.0$ argon gas shock structure: (a) Density, (b) temperature.}
\end{figure}

Other parameters are set referring to Ref.\cite{egks,dugks2015}, as follows. When Mach number is $3.0$, $\alpha=1.0$ and $\omega=0.5$. The reference data are results of Boltzmann equation\cite{boltzmannohwada} and UGKS\cite{ma3ugks}. Fig.\ref{fig5} shows the comparison results. The x coordinate is nondimensionalized through $0.5\sqrt{\pi}\lambda$. And the vertical coordinates are nondimensionalized as follows:
\begin{equation}\label{eq:ma3}
\begin{aligned}
\hat{\rho} &= \frac{\rho}{\rho_{\rm{up}}},\hat{T} &= \frac{T}{T_{\rm{up}}},\\
\hat{\Pi} &= \frac{\Pi}{p_{\rm{up}}},\hat{Q} &= \frac{Q}{p_{\rm{up}}\sqrt{2RT_{\rm{up}}}},
\end{aligned}
\end{equation}
where subscript ``up'' denotes inflow parameter. When Mach number is $8.0$, $\alpha=1.0$ and $\omega=0.68$. The reference data are results of DSMC\cite{ma8dsmc} and UGKS\cite{ma3ugks}. Fig.\ref{fig6} shows the comparison results. The x coordinate is nondimensionalized through $\beta\pi\lambda$. And the vertical coordinates are nondimensionalized as follows:
\begin{equation}\label{eq:ma8}
\begin{aligned}
\hat{\rho} &= \frac{\rho-\rho_{\rm{up}}}{\rho_{\rm{down}}-\rho_{\rm{up}}},\hat{T} &= \frac{T-T_{\rm{up}}}{T_{\rm{down}}-T_{\rm{up}}},\\
\hat{\Pi} &= \frac{\Pi}{\rho_{\rm{up}}(2RT_{\rm{up}})},\hat{Q} &= \frac{Q}{\rho_{\rm{up}}(2RT_{\rm{up}})^{1.5}},
\end{aligned}
\end{equation}
where subscript ``down'' denotes outflow parameter. When Mach number is $5.0$ or $20.0$, $\alpha=1.5325$ and $\omega=0.75$. The reference data are results of DSMC\cite{dsmc520}. Fig.\ref{fig7} and Fig.\ref{fig8} show the comparison results. The x coordinate is nondimensionalized through $\beta\pi\lambda$. And the vertical coordinates are nondimensionalized as Eq.\ref{eq:ma8}.

Fig.\ref{fig5} shows that results of NCCR equations are accurate when Mach number is $3.0$. With Mach number increasing, the results of NCCR equations deviate from reference results, especially temperature curves. The curves seem to be nipped at the pre-shock. But they are still more accurate than results of NS equations, particularly at the peak of heat flux curves. Besides, results of MFPI method and the proposed method are almost the same. This phenomenon can be explained that the situation in this case is similar to the left part in Fig.\ref{fig3}.

\subsection{Shock structures in diatomic nitrogen gas at different Mach numbers}\label{sec:ssn2}
\begin{figure}
	\centering
	\subfigure[]{
			\includegraphics[width=0.45 \textwidth]{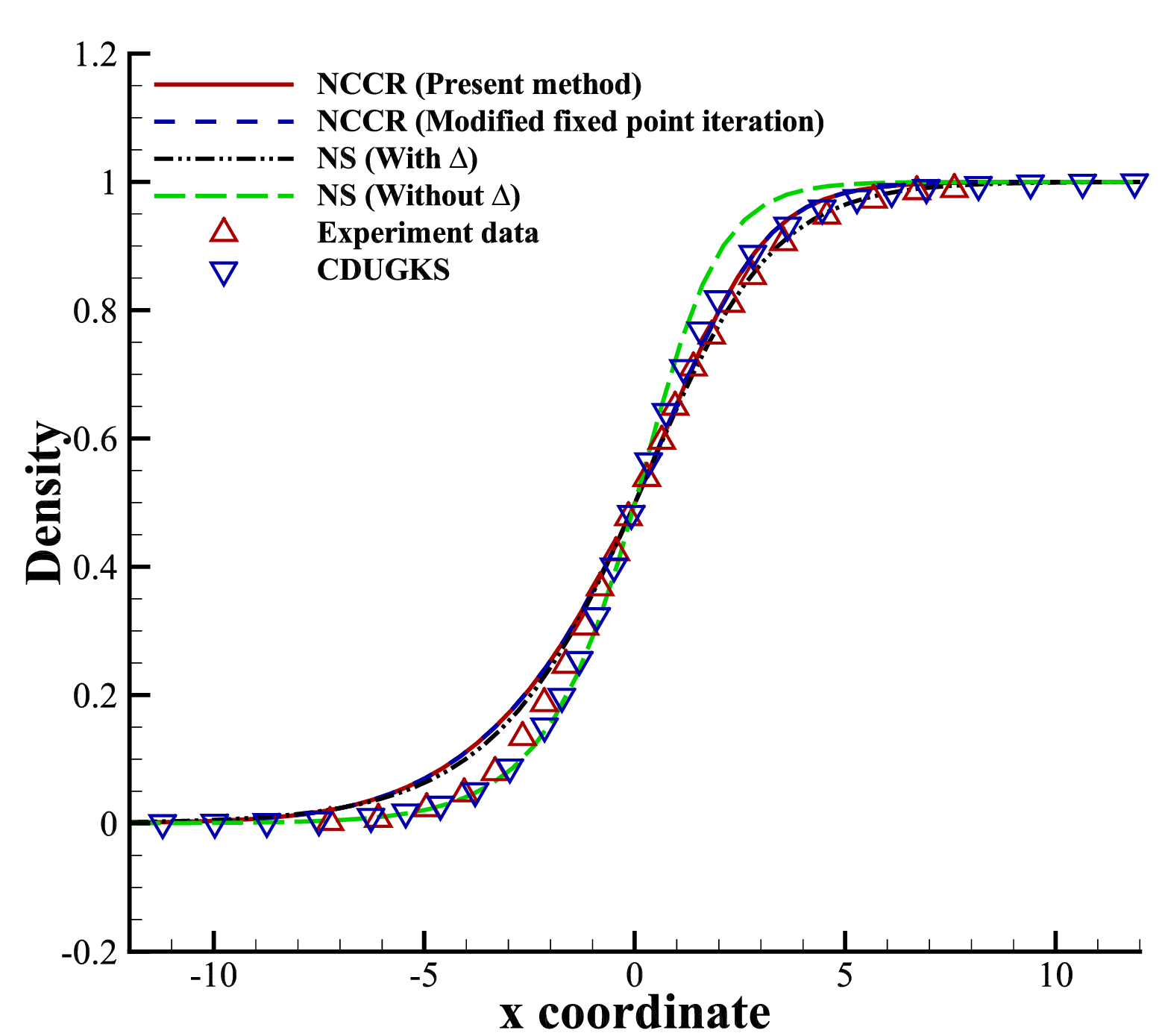}
		}
    \subfigure[]{
    		\includegraphics[width=0.45 \textwidth]{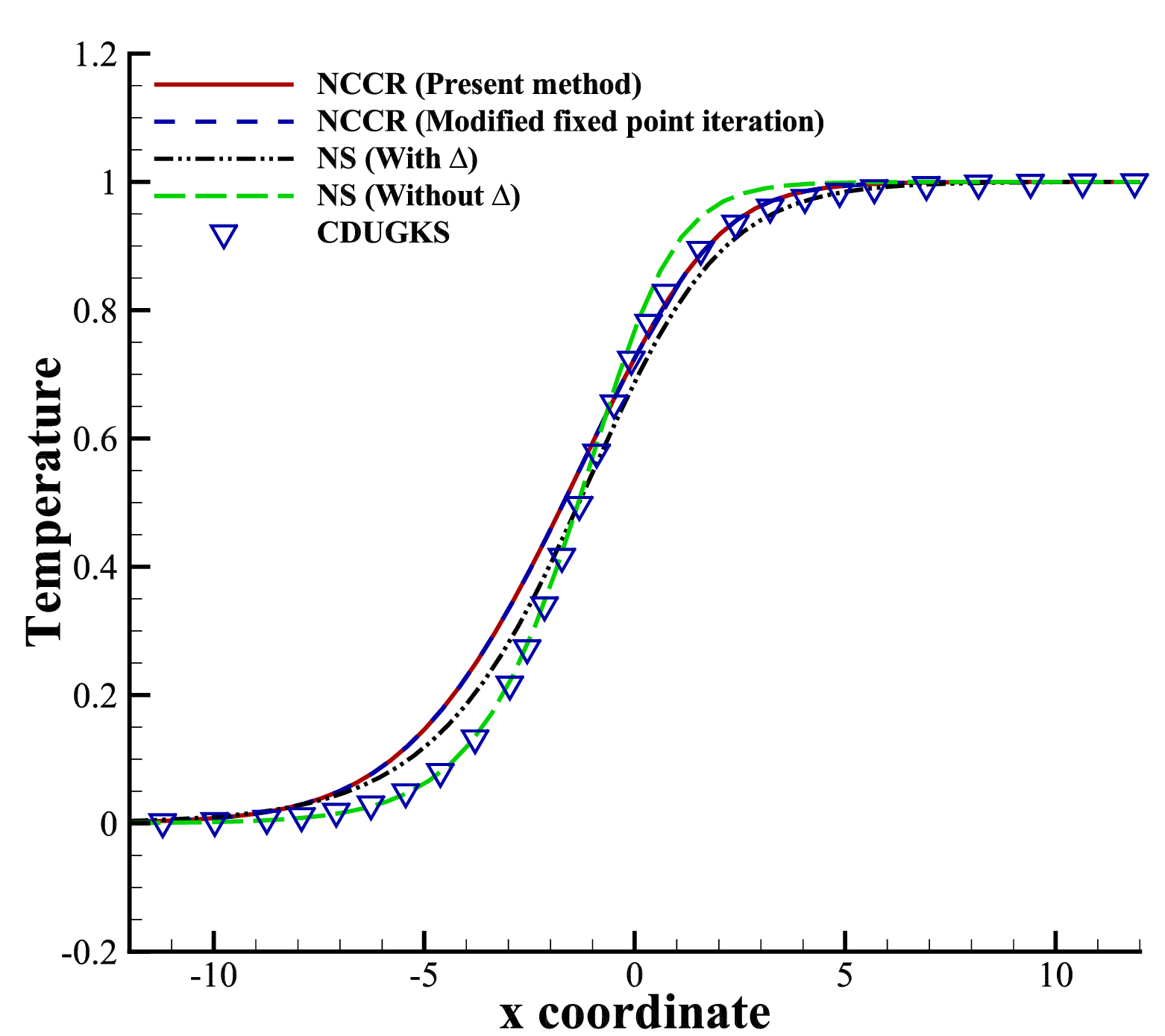}
    	}
	\caption{\label{fig9} Results of $\rm{Ma}=1.7$ nitrogen gas shock structure: (a) Density, (b) temperature.}
\end{figure}

\begin{figure}
	\centering
	\subfigure[]{
			\includegraphics[width=0.45 \textwidth]{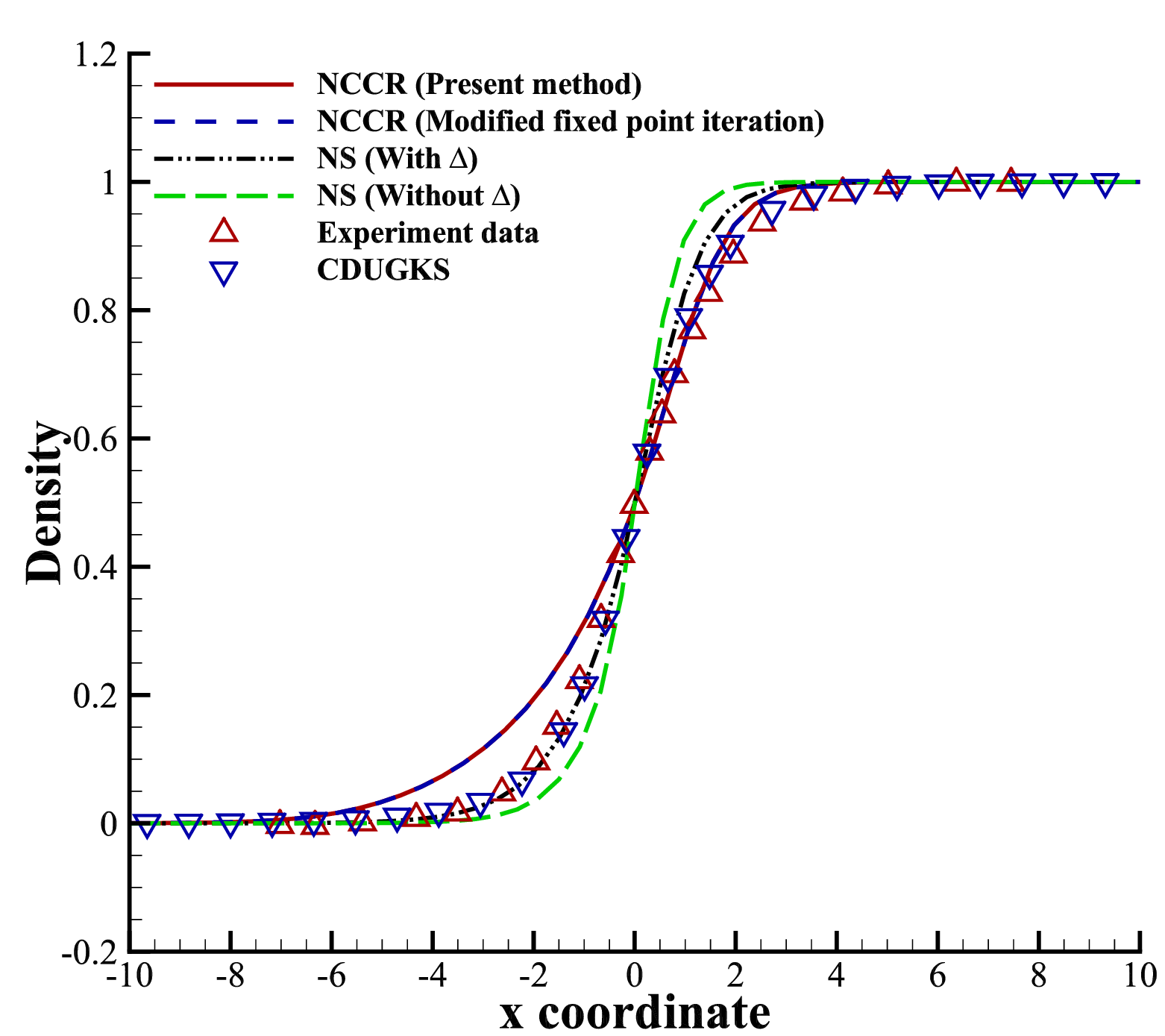}
		}
    \subfigure[]{
    		\includegraphics[width=0.45 \textwidth]{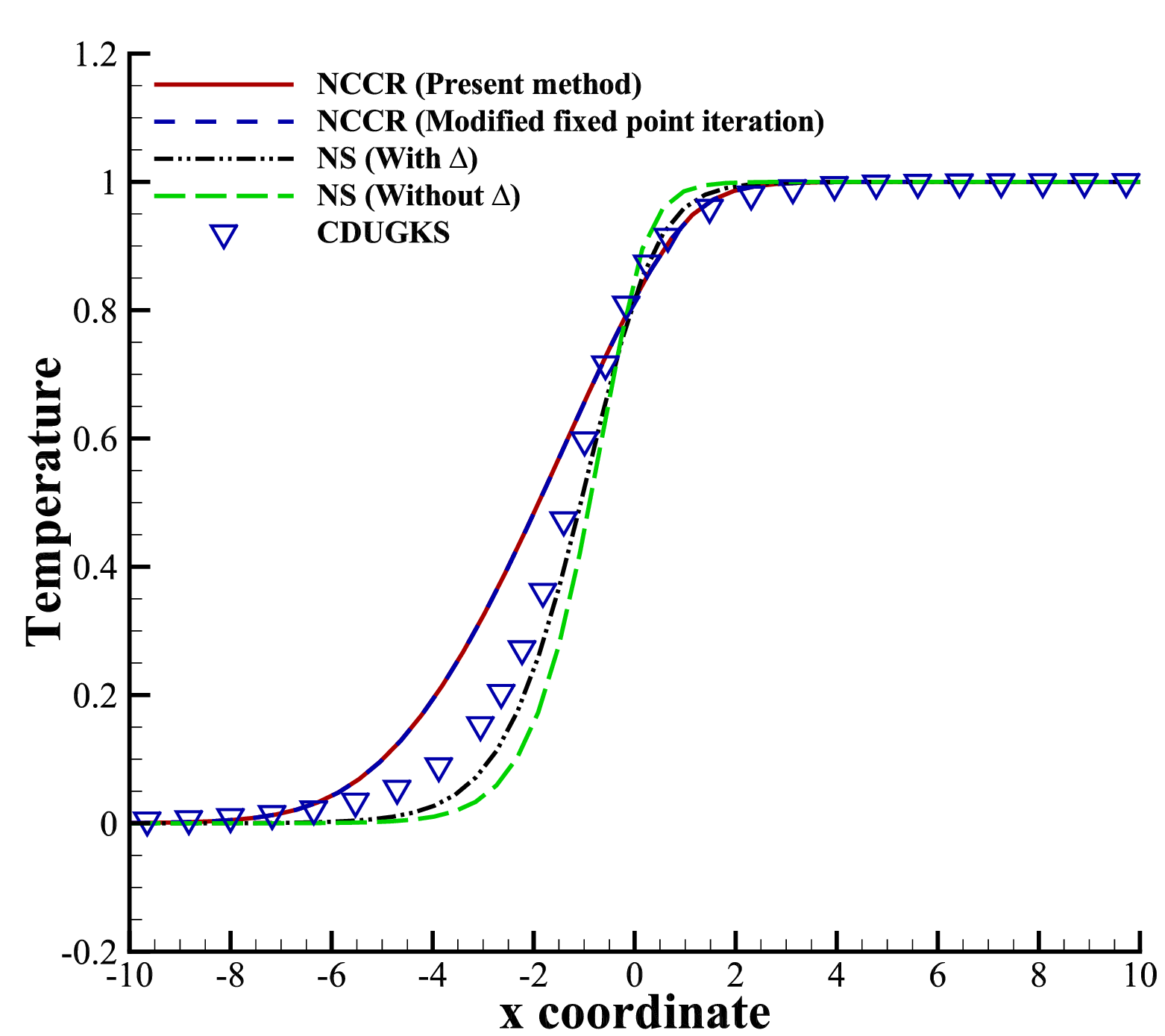}
    	}
	\caption{\label{fig10} Results of $\rm{Ma}=2.4$ nitrogen gas shock structure: (a) Density, (b) temperature.}
\end{figure}

\begin{figure}
	\centering
	\subfigure[]{
			\includegraphics[width=0.45 \textwidth]{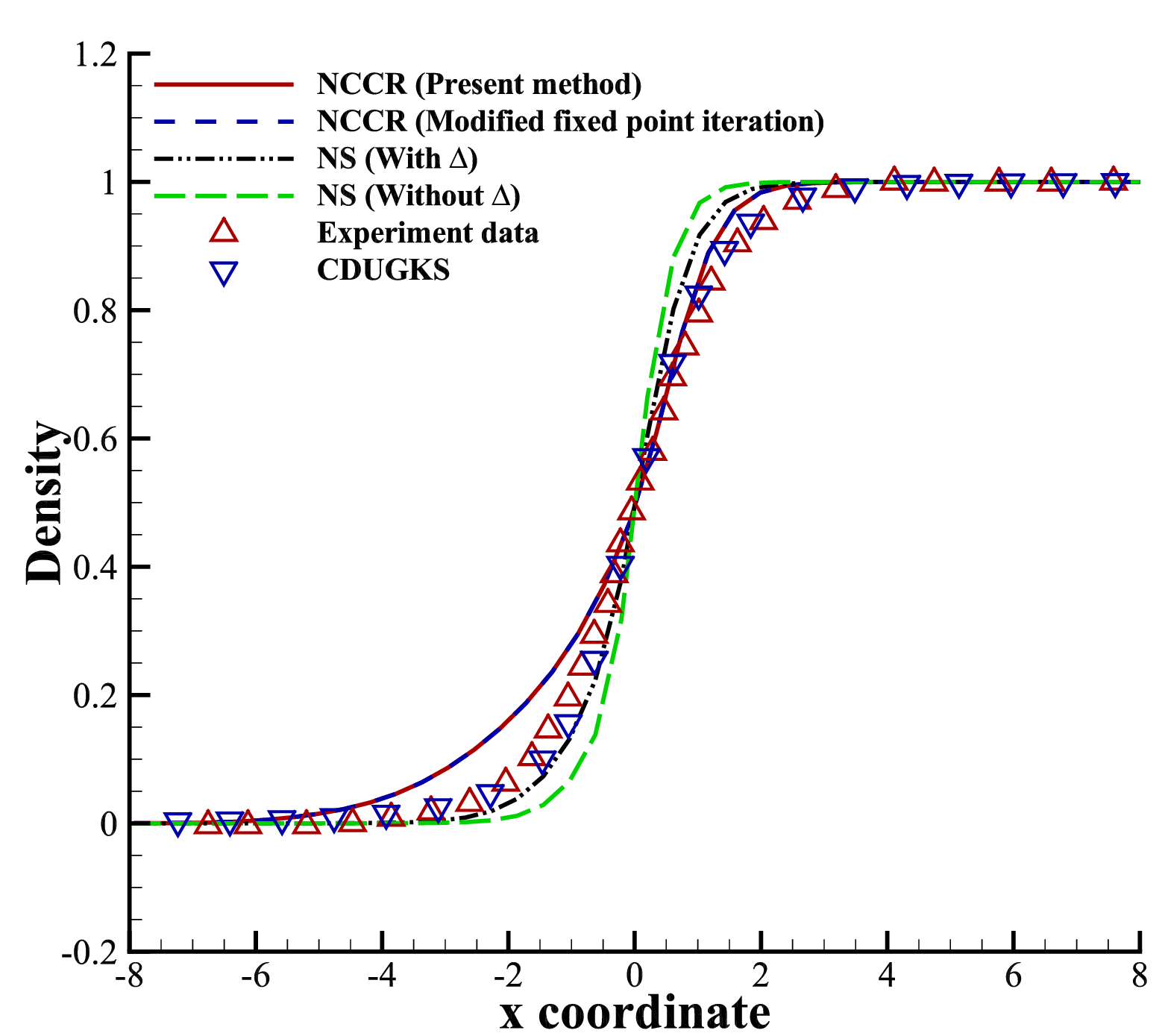}
		}
    \subfigure[]{
    		\includegraphics[width=0.45 \textwidth]{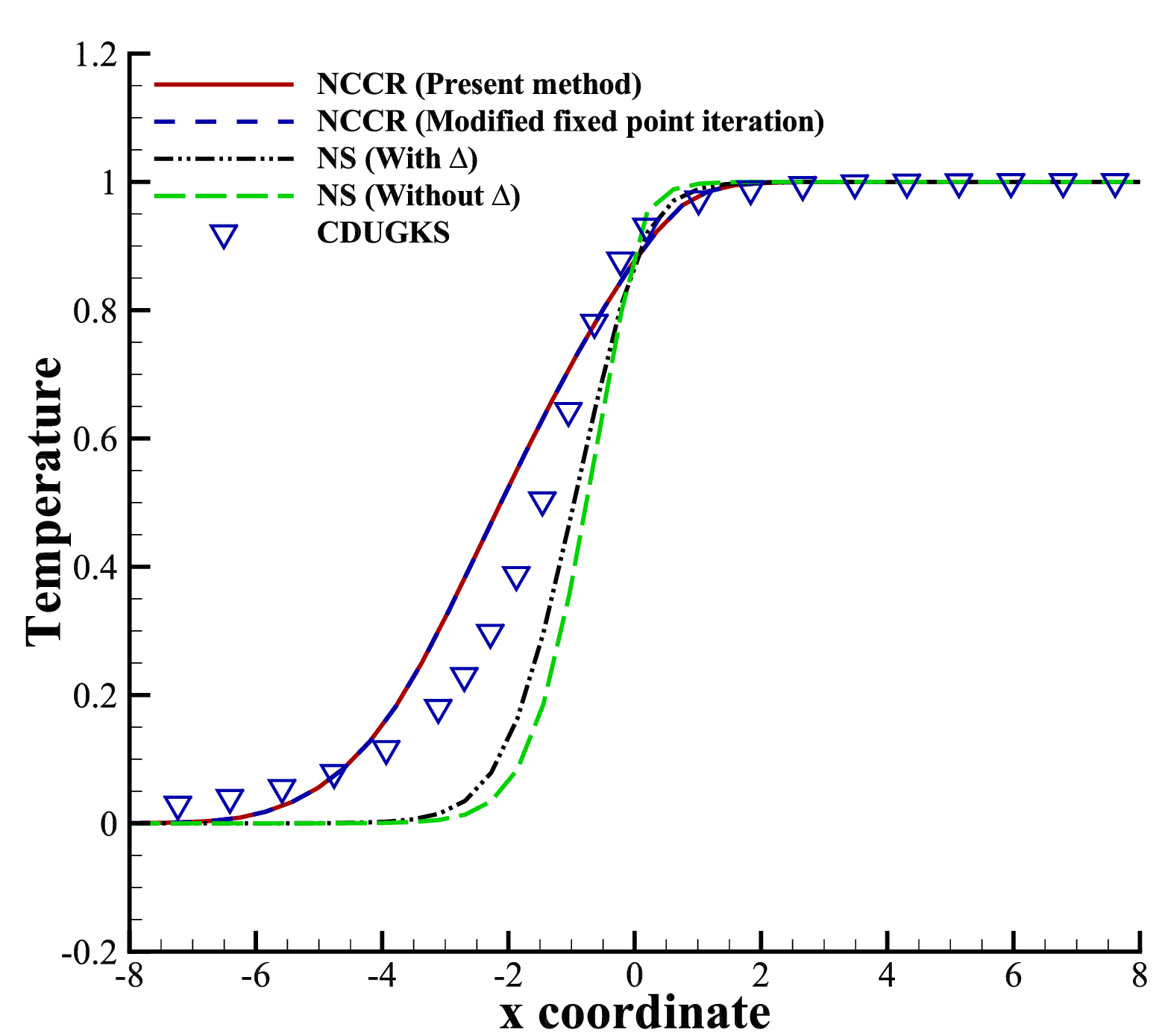}
    	}
	\caption{\label{fig11} Results of $\rm{Ma}=3.8$ nitrogen gas shock structure: (a) Density, (b) temperature.}
\end{figure}

\begin{figure}
	\centering
	\subfigure[]{
			\includegraphics[width=0.45 \textwidth]{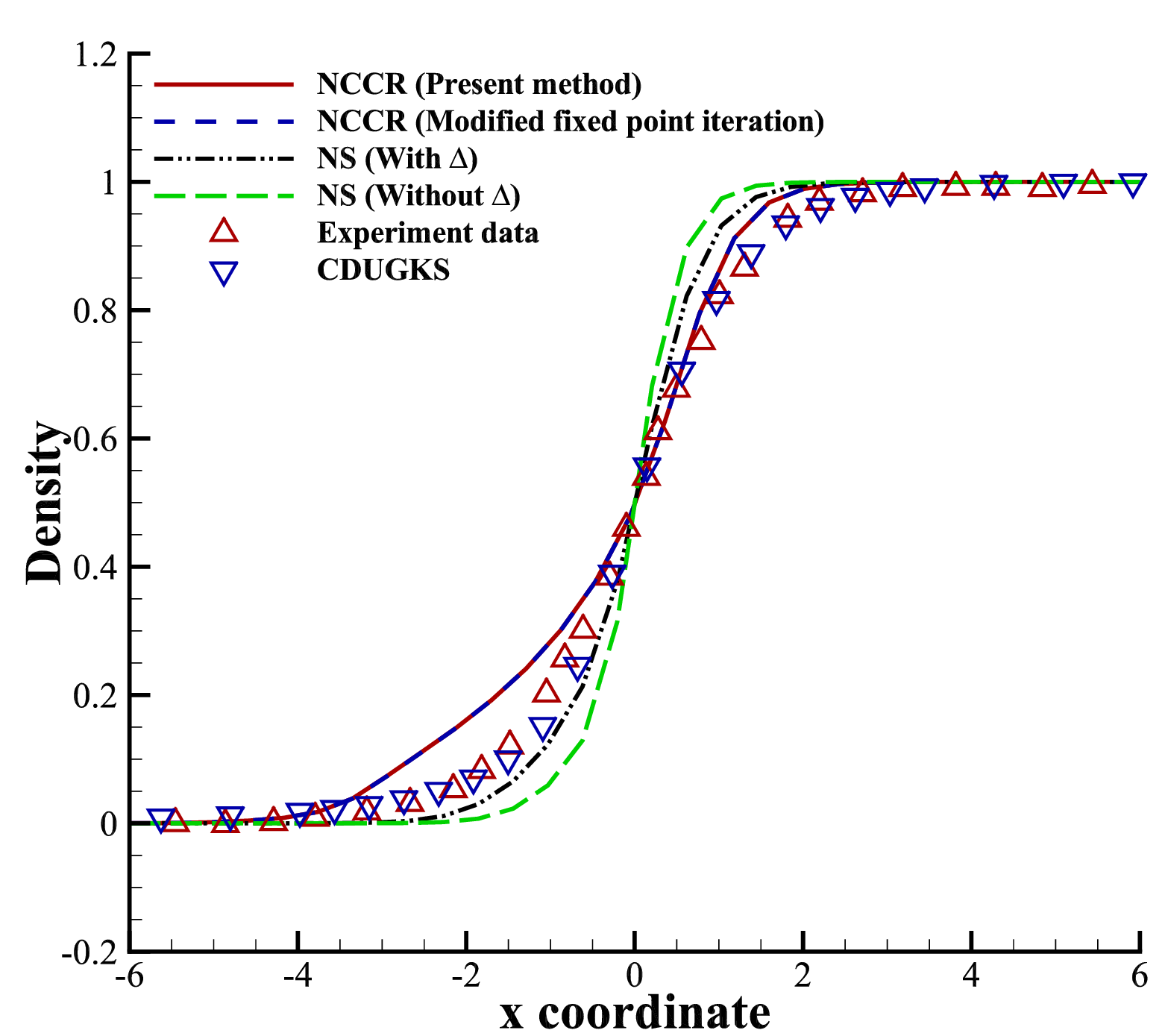}
		}
    \subfigure[]{
    		\includegraphics[width=0.45 \textwidth]{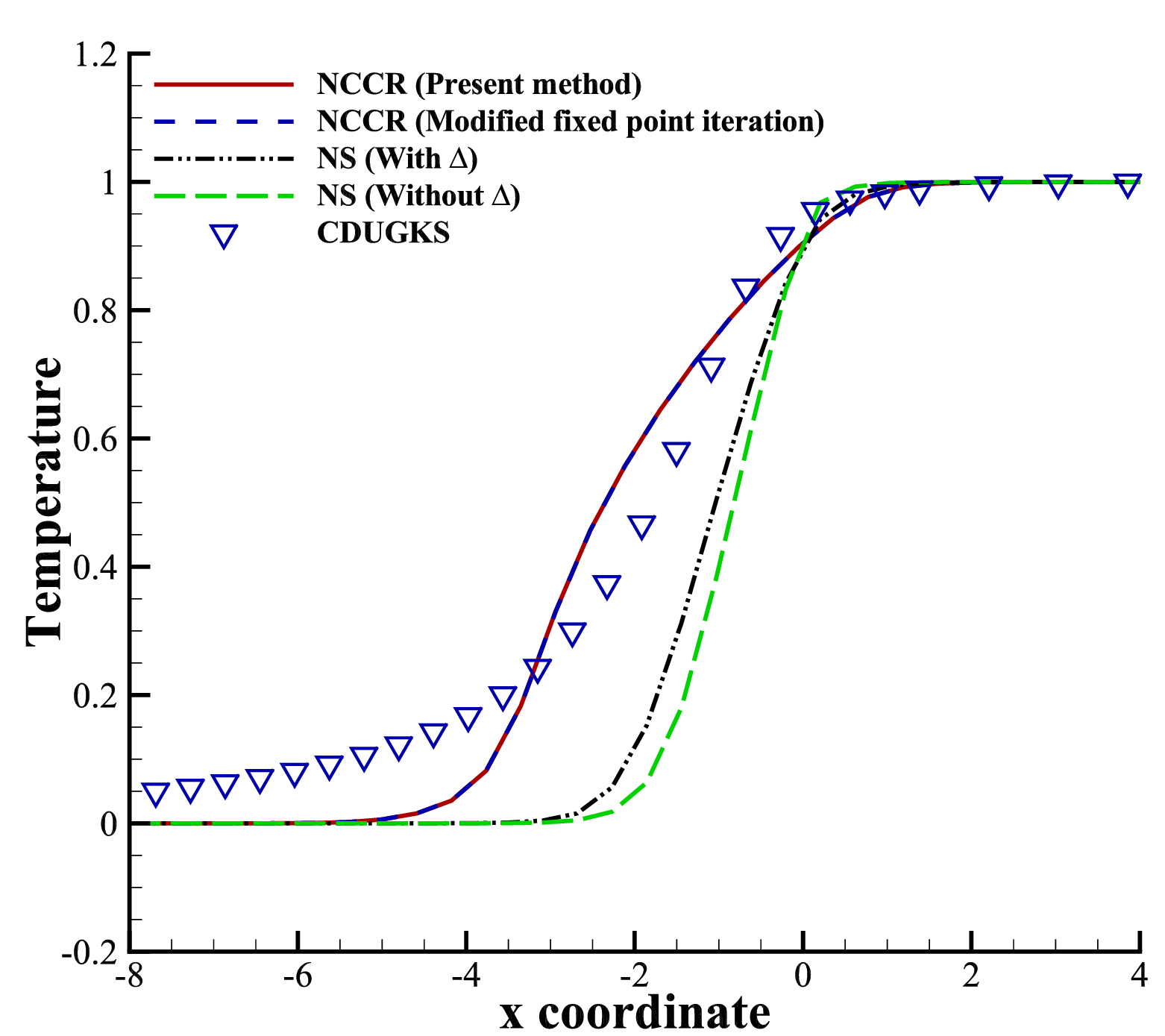}
    	}
	\caption{\label{fig12} Results of $\rm{Ma}=6.1$ nitrogen gas shock structure: (a) Density, (b) temperature.}
\end{figure}

\begin{figure}
	\centering
	\subfigure[]{
			\includegraphics[width=0.45 \textwidth]{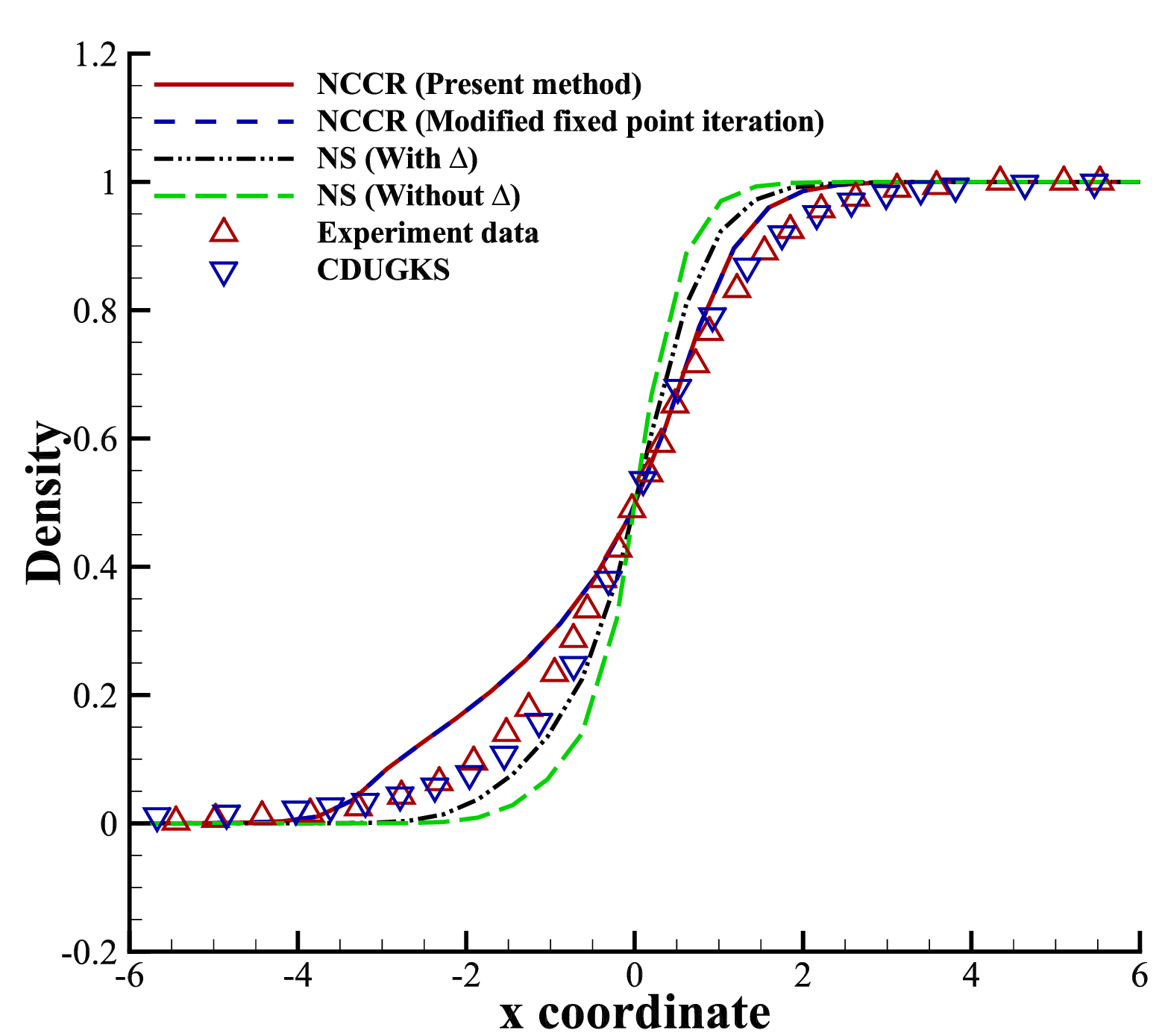}
		}
    \subfigure[]{
    		\includegraphics[width=0.45 \textwidth]{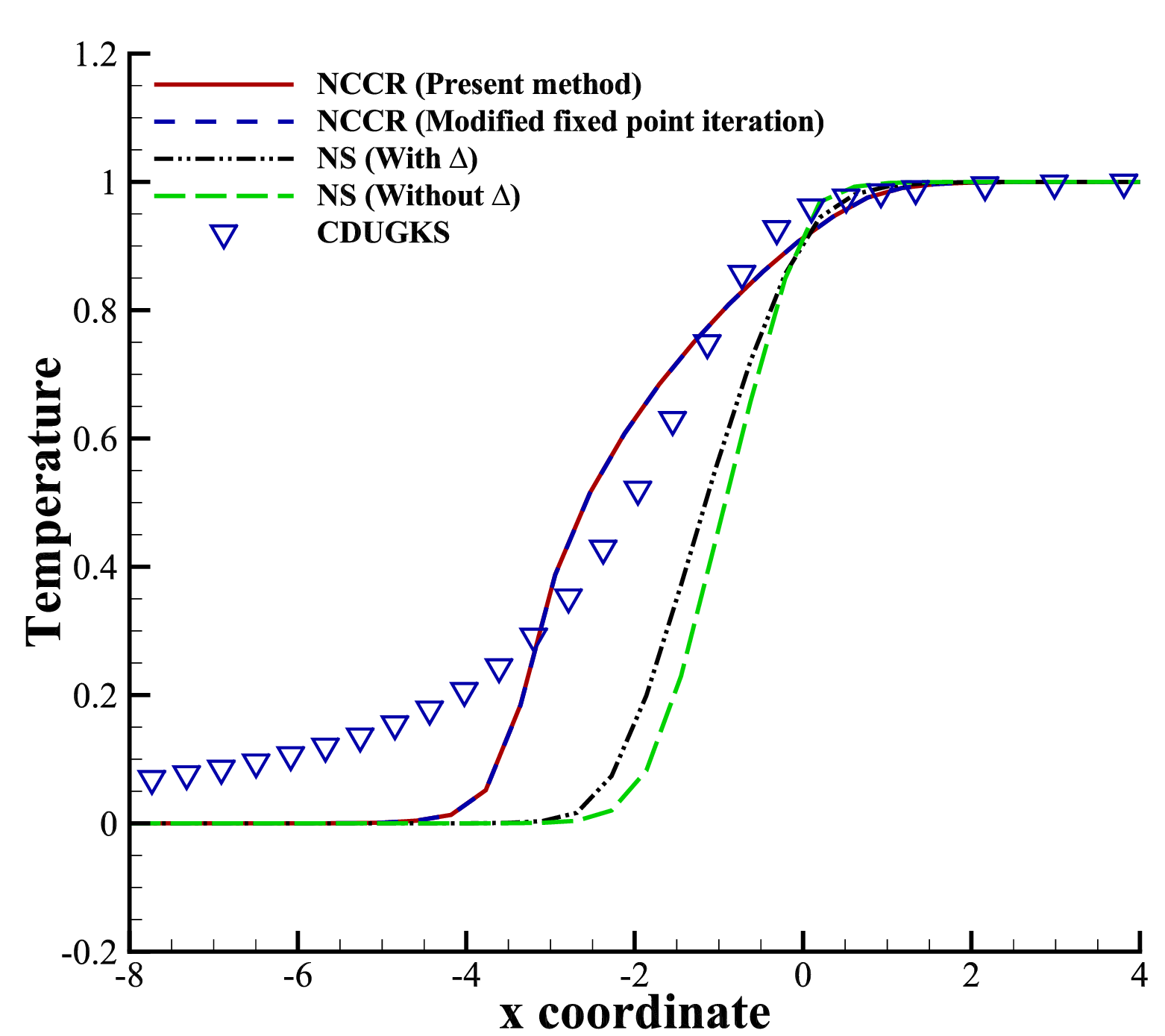}
    	}
	\caption{\label{fig13} Results of $\rm{Ma}=8.4$ nitrogen gas shock structure: (a) Density, (b) temperature.}
\end{figure}

\begin{figure}
	\centering
	\subfigure[]{
			\includegraphics[width=0.45 \textwidth]{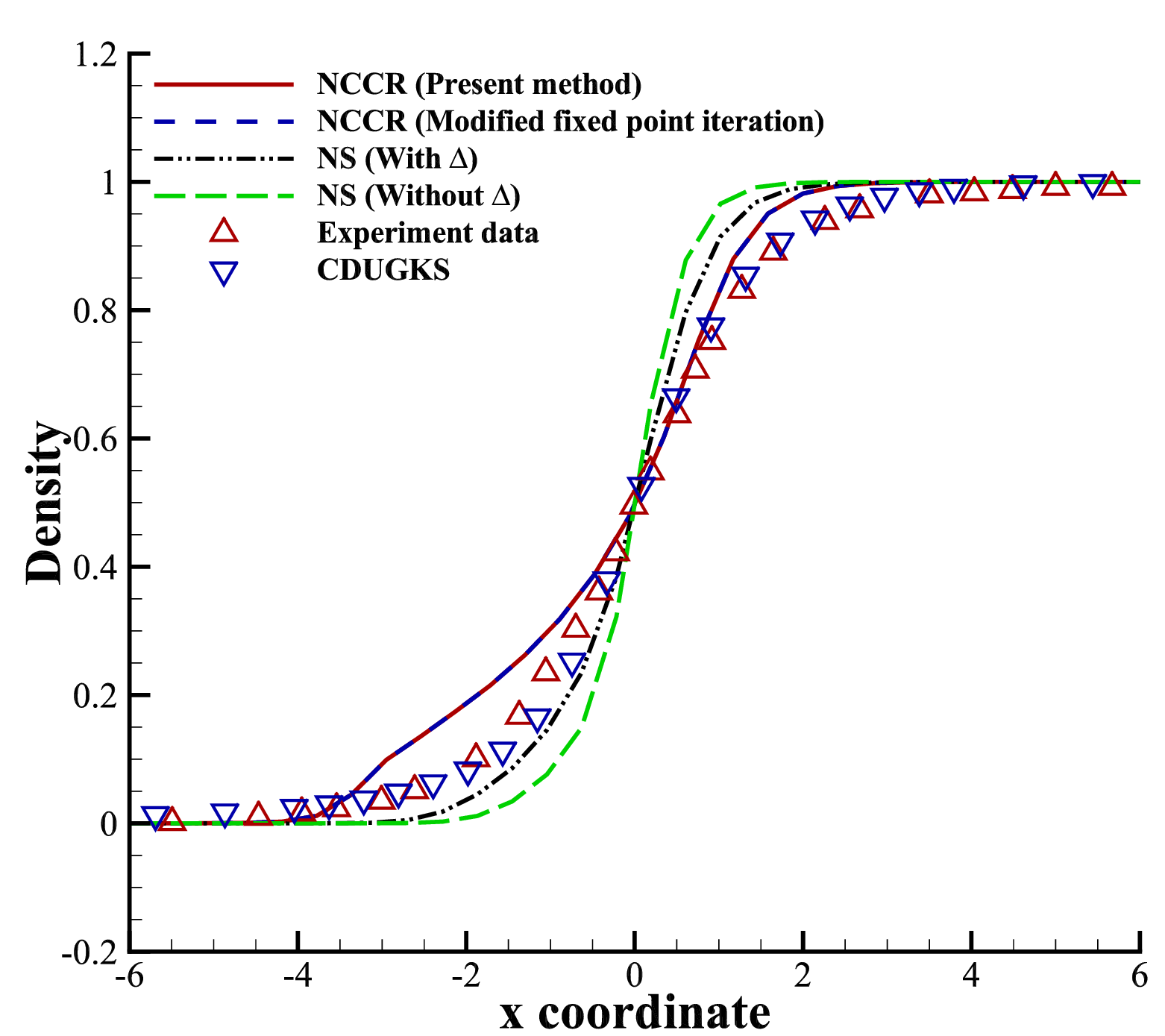}
		}
    \subfigure[]{
    		\includegraphics[width=0.45 \textwidth]{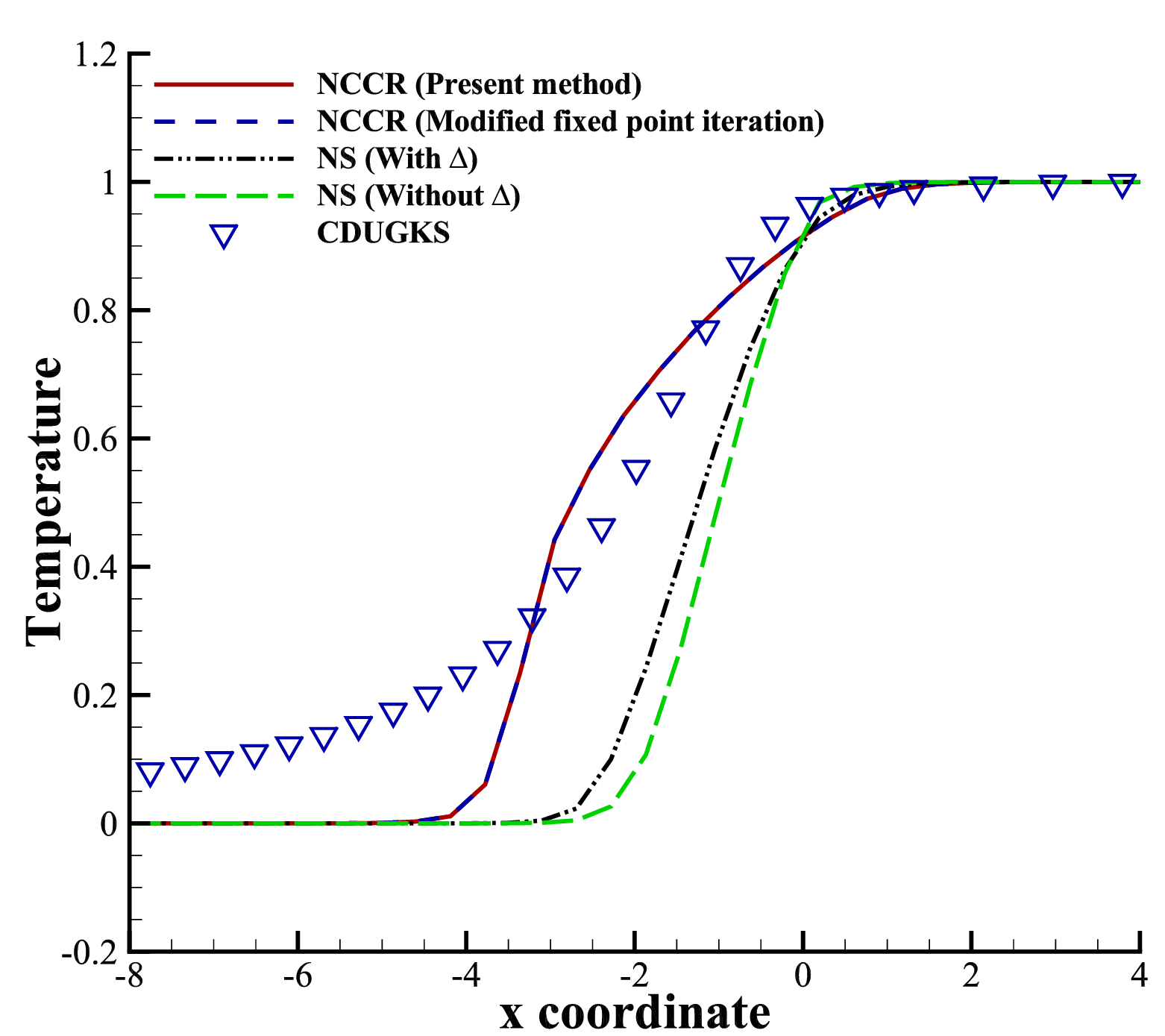}
    	}
    \subfigure[]{
            \label{fig14c}
    		\includegraphics[width=0.45 \textwidth]{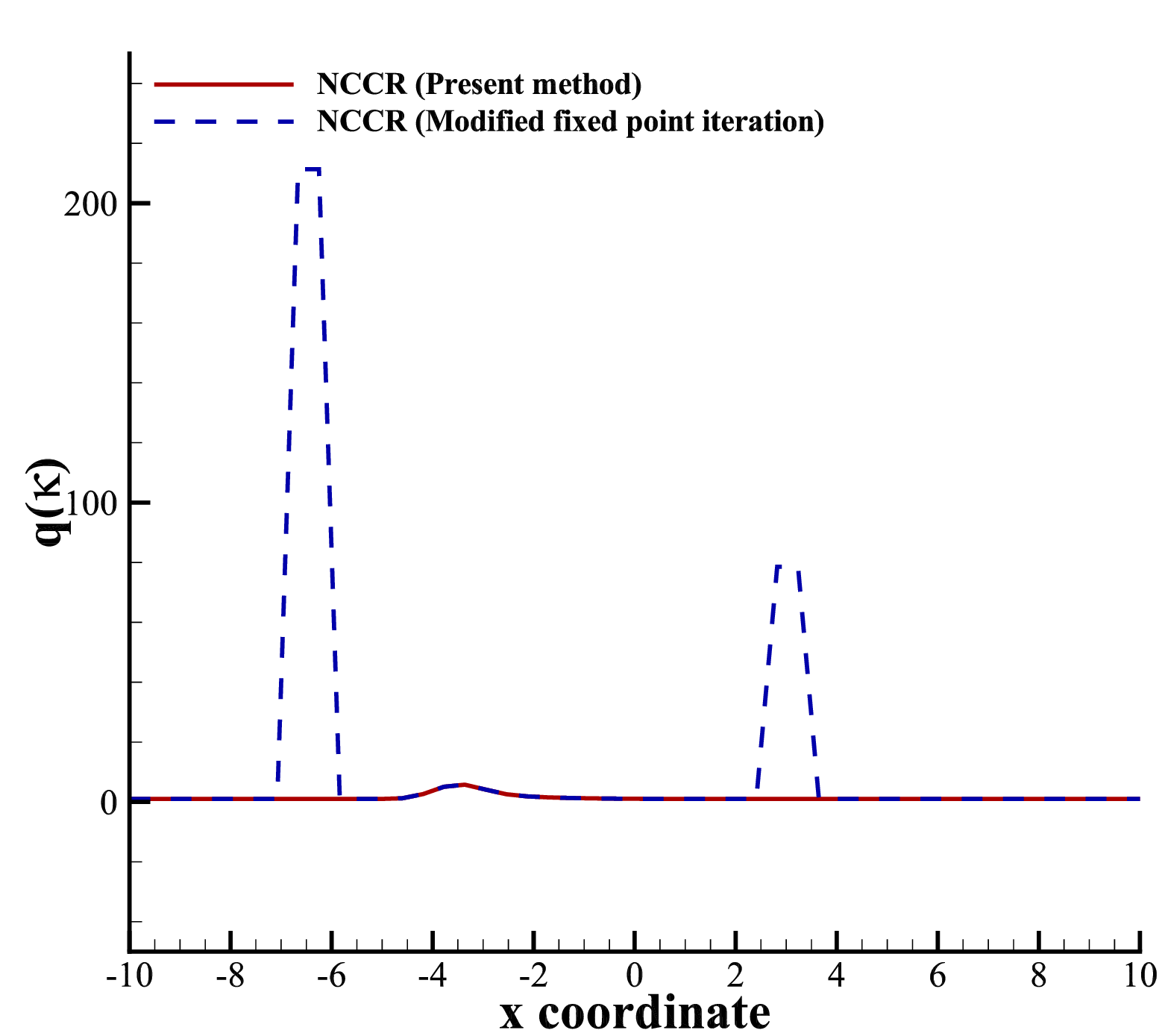}
    	}
	\caption{\label{fig14} Results of $\rm{Ma}=10.0$ nitrogen gas shock structure: (a) Density, (b) temperature, (c) $q_{\kappa}$.}
\end{figure}

Furthermore, in order to test the ability of proposed method to describe diatomic gas flow, shock structures in nitrogen gas at different Mach numbers are simulated (Mach numbers are $1.7$, $2.4$, $3.8$, $6.1$, $8.4$, $10.0$). Meanwhile, NS equations (with or without the excess normal stress $\Delta$ of bulk viscosity) and MFPI method for NCCR equations are tested as well. The inviscid flux used is KIF. Parameters are $\gamma=1.4$, $\rm{Pr}=0.72$ and bulk viscosity parameter $f_{\rm{b}}=0.8$. The dynamic viscosity is calculated through $\mu\sim T^{\omega}$. The size of mesh is set to be a half of molecular mean free path $\lambda$ as Eq.\ref{eq:lambda} and $\alpha=1.0$, $\omega=0.74$. Results are shown as Fig.\ref{fig9}-Fig.\ref{fig14}. The x coordinate is nondimensionalized through $1.212633\lambda$, which is the molecular mean free path of corresponding hard sphere (HS) model. And the vertical coordinates are nondimensionalized as Eq.\ref{eq:ma8}. Reference data are experiment data\cite{ssn2exp} and results of Conserved DUGKS (CDUGKS)\cite{ssn2cdugks}.

As results shown in Fig.\ref{fig9}-Fig.\ref{fig14}, results of NCCR equations and NS equations with $\Delta$ are all better than NS equations without $\Delta$. As to density profiles, results of NCCR equations are better at the post-shock and results of NS equations with $\Delta$ are better at the pre-shock. As to temperature profiles, more difference appears. With Mach number increasing, the results of NCCR equations deviate from reference results, especially temperature curves. Evidently, difficulty in accurately simulating diatomic gas shock structure is much higher than that of monatomic gas. In Sec.\ref{sec:ssar}, when simulating argon gas shock structures, results of NCCR equations are inline with reference data at $\rm{Ma}=3.0$. However, in this section, even at $\rm{Ma}=1.7$, obvious difference exists at the pre-shock. Similar with Sec.\ref{sec:ssar}, results of MFPI method and the proposed method are almost the same. Nevertheless, $q_{\kappa}$ profile is exhibited in Fig.\ref{fig14c}. There are two rather large $q_{\kappa}$ at the pre-shock and post-shock in the results of MFPI method. However, this does not affect the results in other profiles.

\subsection{The rarefied Couette flow}
In this section, the performance of methods for low speed flows is tested. The rarefied Couette flow is a typical low speed rarefied flows and is broadly studied\cite{couette,egks}. The working gas is set to be argon gas, with parameters $\gamma=5/3$, $\rm{Pr}=2/3$, and gas constant is set to $208.14JK/kg$. The Couette flow is driven by two infinite plates with parallel movement on opposite directions. The velocity of plates is set to $\pm0.5\sqrt{RT_{\rm{wall}}}$ and the wall temperature is set to $273K$. Three different Kn numbers are considered, $Kn=\lambda_{\rm{HS}}/H=0.2/\sqrt{\pi},2.0/\sqrt{\pi},20.0/\sqrt{\pi}$, and $H$ is the distance between the two plates.

\begin{figure}
	\centering
	\subfigure{
			\includegraphics[width=0.5 \textwidth]{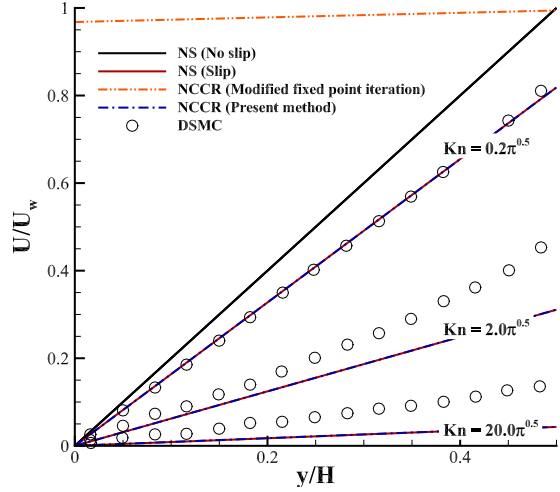}
		}
	\caption{\label{figcouette} Velocity profiles of Couette flow. (Only $Kn=0.2/\sqrt{\pi}$ profile for MFPI method is shown.)}
\end{figure}

Results are shown as Fig.\ref{figcouette}. It shows that as to macroscopic methods, accuracy of velocity profiles is mainly related to slip boundary condition. And there are little difference between results of NS equations and NCCR equations. However, in this test case, MFPI method shows bad stability and its convergence efficiency is quiet slow.

\subsection{Supersonic/hypersonic rarefied cylinder flows in diatomic nitrogen gas}\label{sec:n2bb}
Hypersonic rarefied cylinder flow is a typical test case. Firstly, diatomic nitrogen gas is set to be working gas. Parameters are set following Ref.\cite{n2bb}, as: inflow Kn number is $0.01$, inflow density is $7.4857\times10^{-5}kg/m^3$, inflow temperature is $217.45K$, gas constant is $296.72JK/kg$, and $\gamma=1.4$, $\rm{Pr}=0.72$. The dynamic viscosity is calculated through
\begin{equation}\label{eq:miu}
\mu=\mu_{\rm{ref}}\left({\frac{T}{T_{\rm{ref}}}}\right)^{\omega},
\end{equation}
where $\omega=0.75$, $T_{\rm{ref}}=273K$, $\mu_{\rm{ref}}=1.6734\times10^{-5}Ns/m^2$. These parameters are chosen referring to the state at $70km$ altitude. Three Mach numbers are taken for simulation, $3.0$, $6.0$ and $12.0$, whose corresponding wall temperatures are $300K$, $500K$, $1000K$ and Re numbers are $382.7$, $765.5$, $1530.8$, respectively. The reference length is the diameter of cylinder, $0.08m$. The first layer of mesh is set to $8.0\times10^{-6}m$. The inviscid flux used is KIF. Tested methods are the proposed method for solving NCCR equations and NS solvers (whether the slip boundary condition is implemented and whether the $\Delta$ term is considered). Results of DSMC are taken as reference\cite{n2bb}.

\begin{figure}
	\centering
	\subfigure[]{
			\includegraphics[width=0.45 \textwidth]{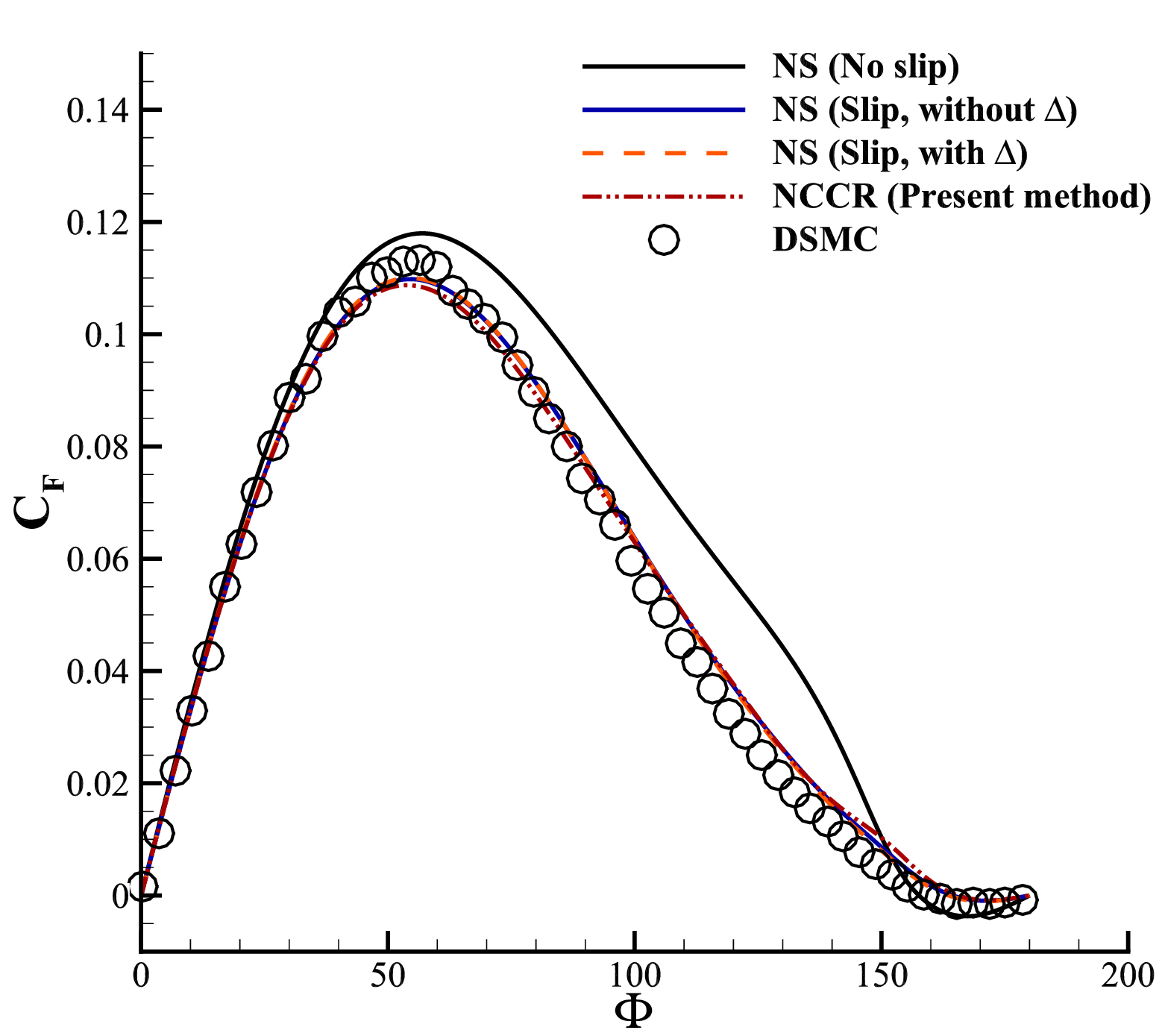}
		}
    \subfigure[]{
    		\includegraphics[width=0.45 \textwidth]{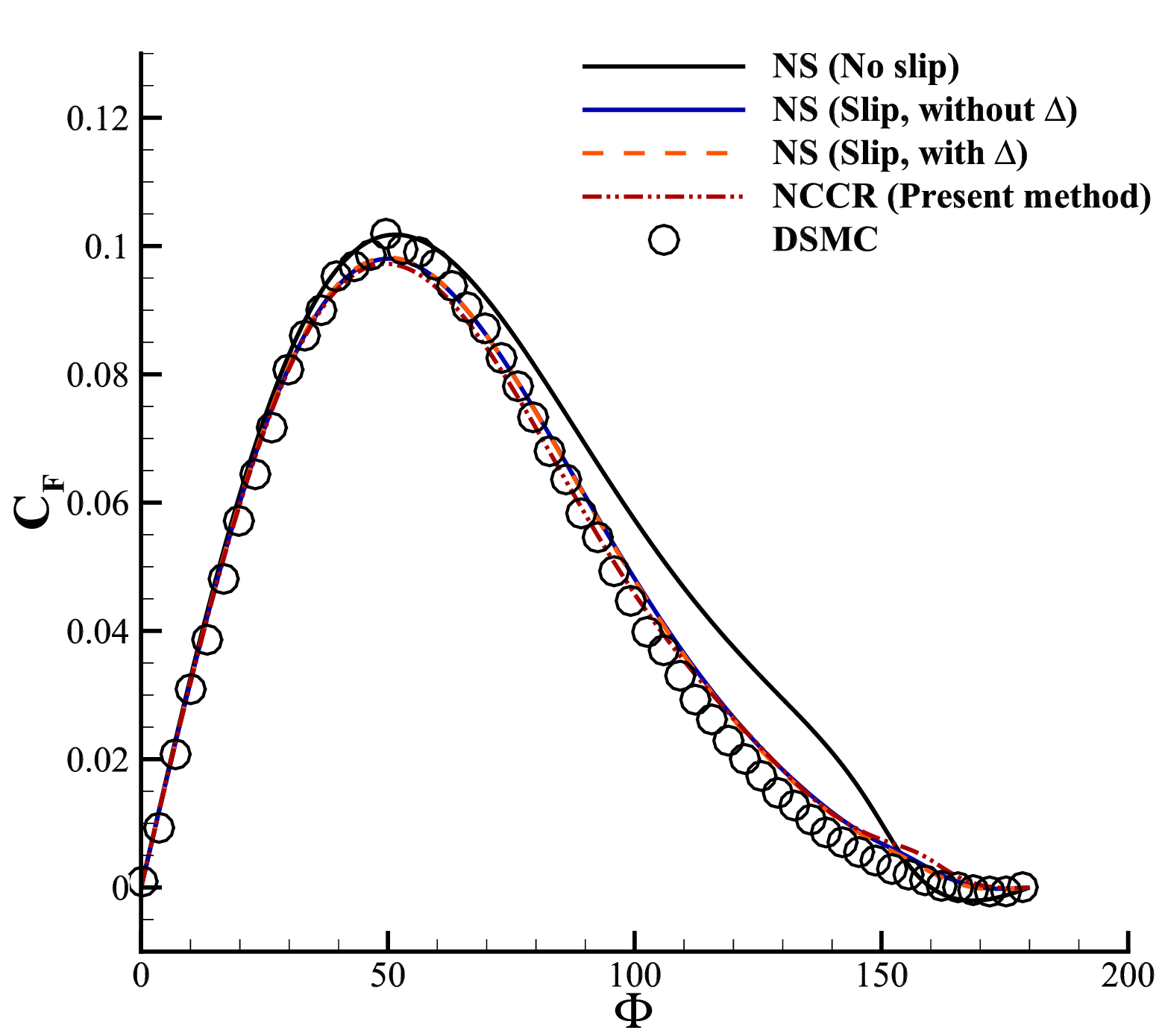}
    	}
    \subfigure[]{
    		\includegraphics[width=0.45 \textwidth]{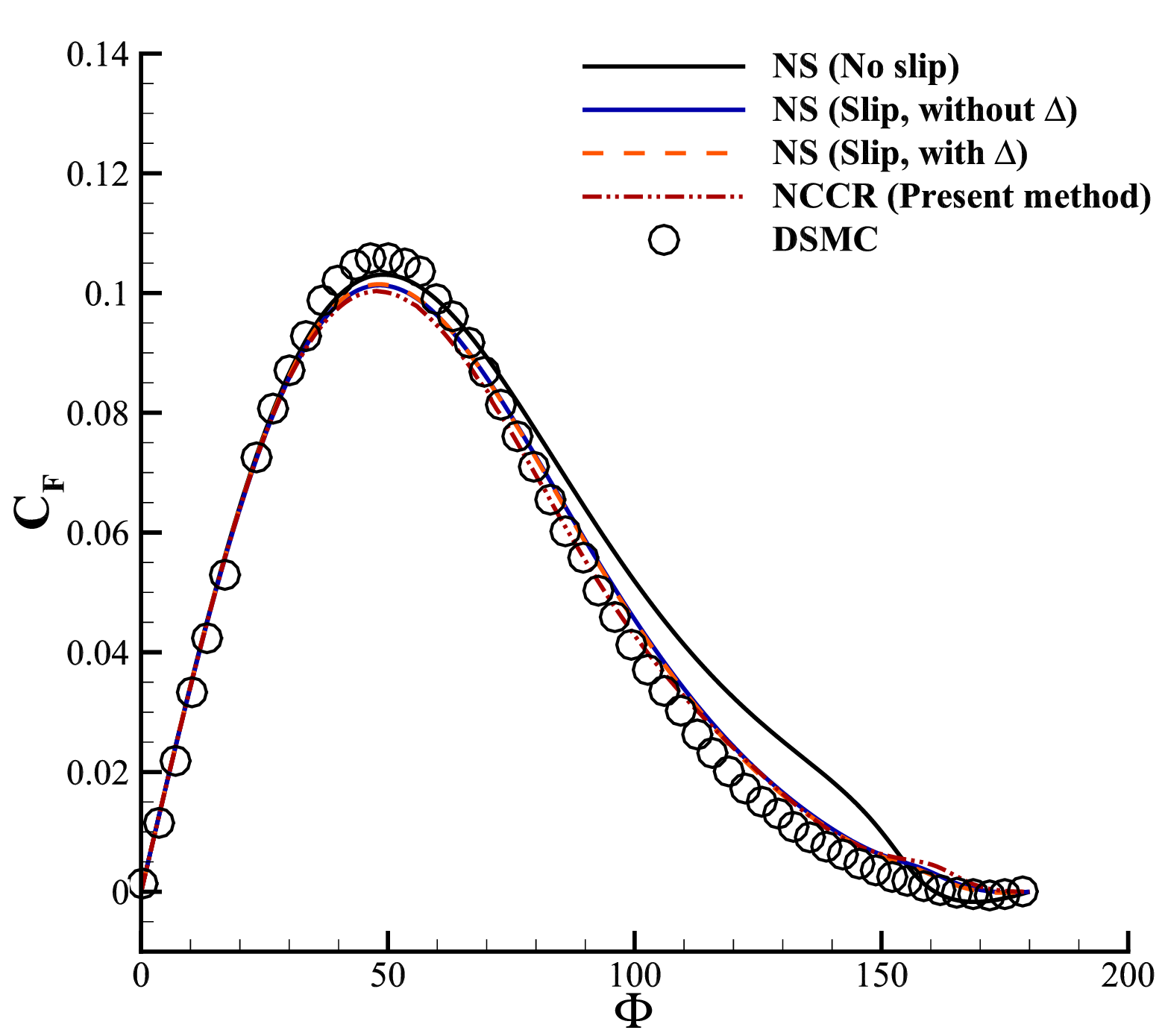}
    	}
	\caption{\label{fig15} Shear stress coefficient at wall for the supersonic/hypersonic rarefied nitrogen gas cylinder flow: (a) $Ma=3.0$, (b) $Ma=6.0$, (c) $Ma=12.0$.}
\end{figure}

\begin{figure}
	\centering
	\subfigure[]{
			\includegraphics[width=0.45 \textwidth]{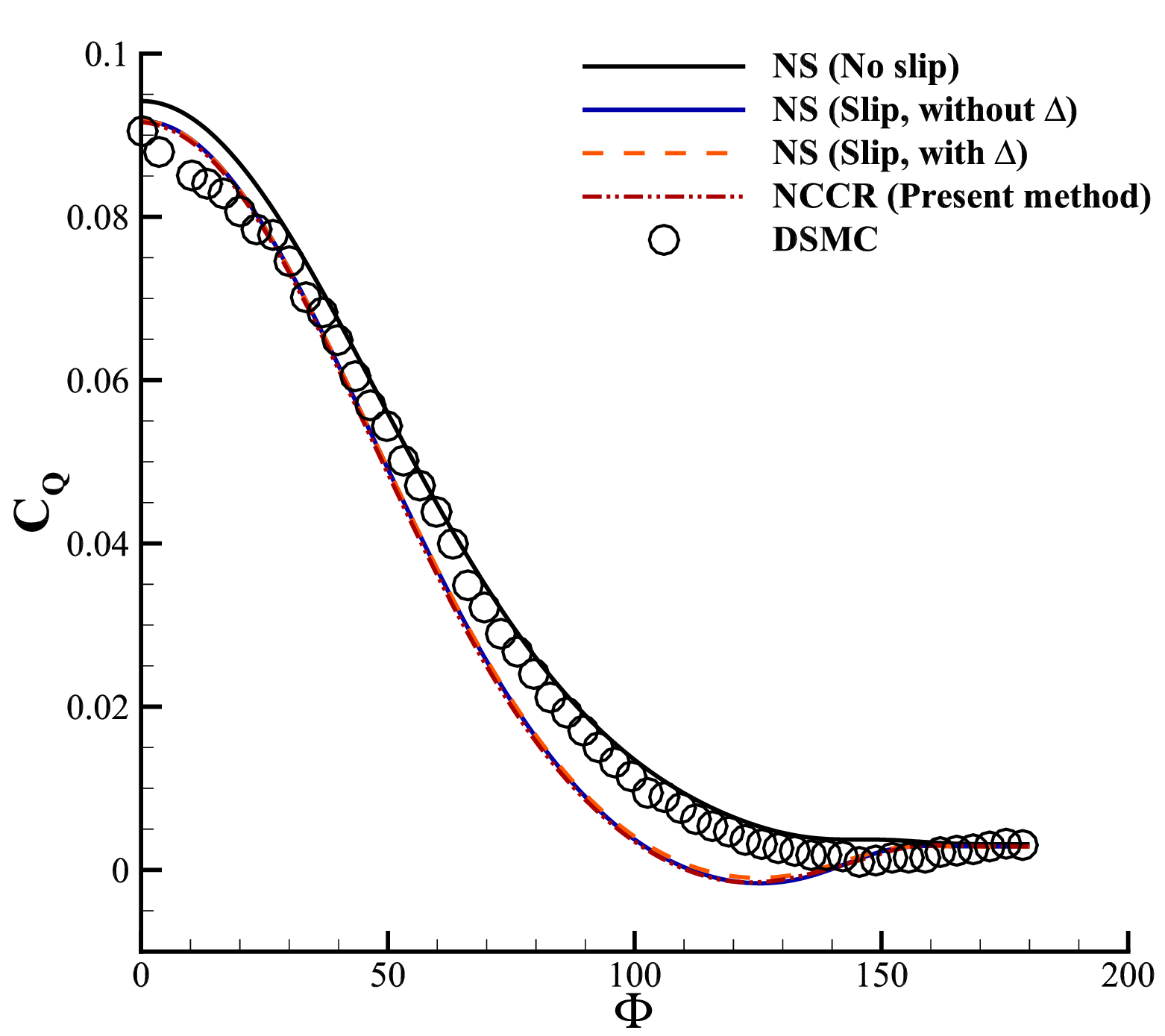}
		}
    \subfigure[]{
    		\includegraphics[width=0.45 \textwidth]{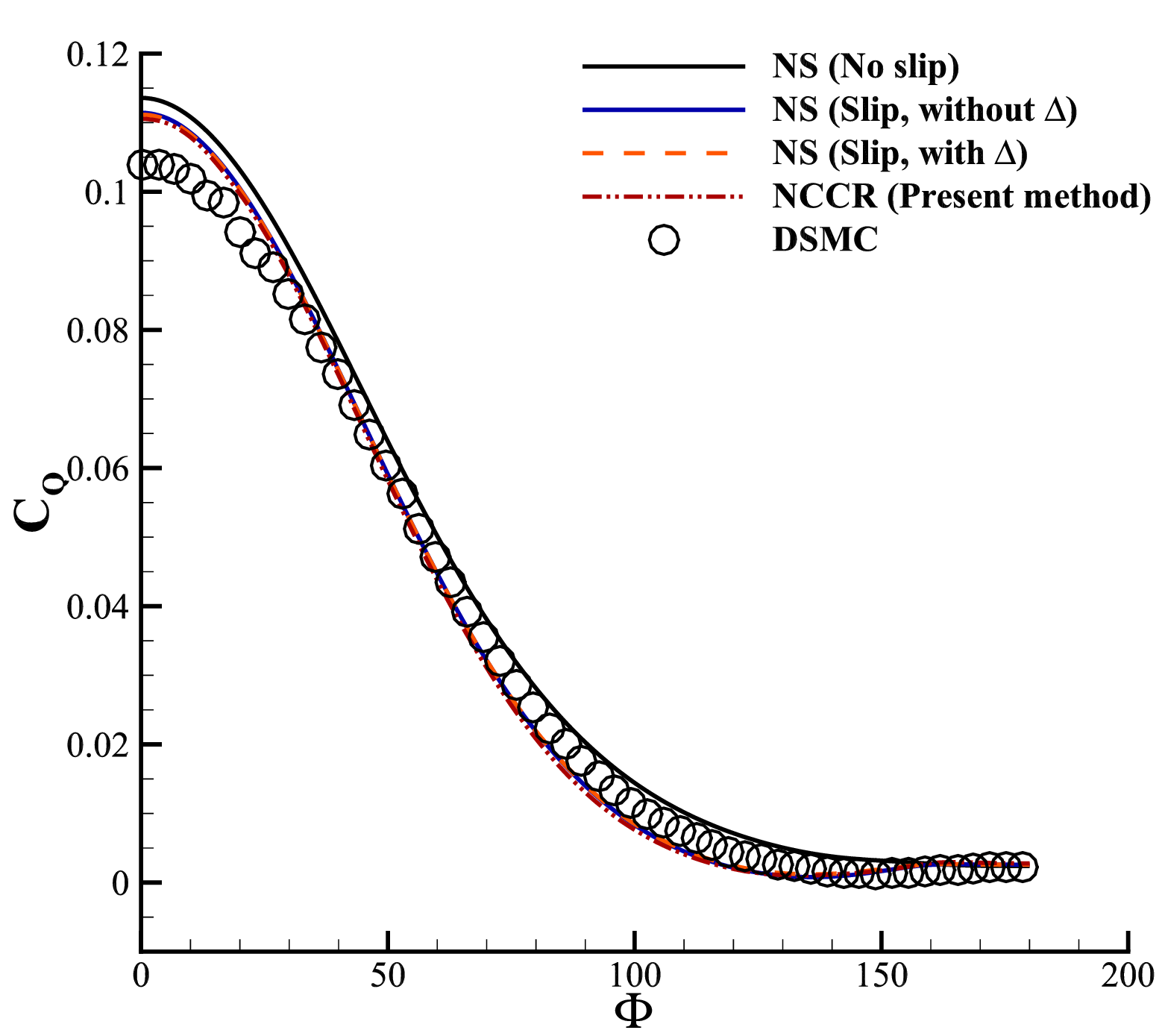}
    	}
    \subfigure[]{
    		\includegraphics[width=0.45 \textwidth]{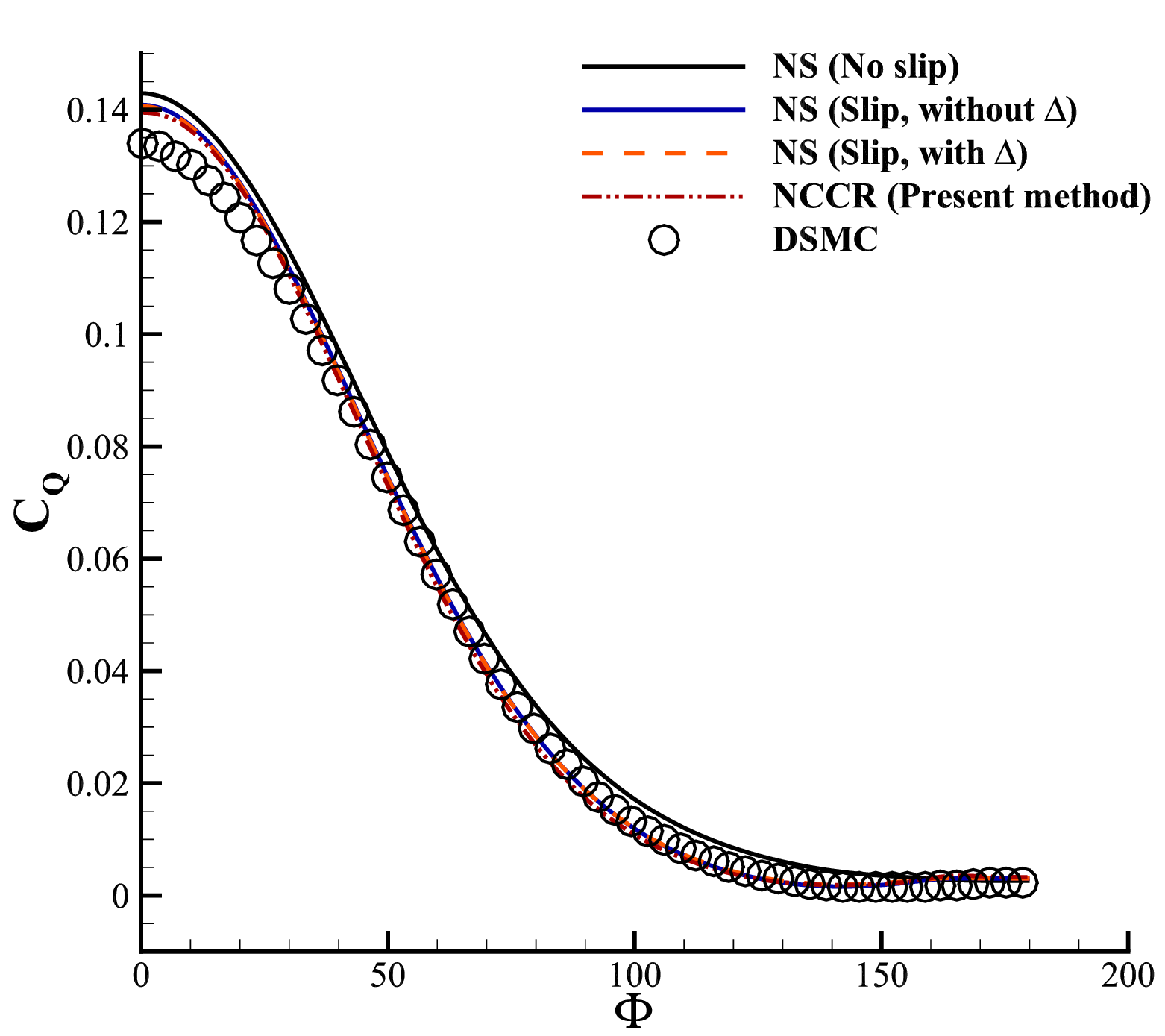}
    	}
	\caption{\label{fig16} Heat flux coefficient at wall for the hypersonic rarefied nitrogen gas cylinder flow: (a) $Ma=3.0$, (b) $Ma=6.0$, (c) $Ma=12.0$.}
\end{figure}

Shear stress coefficients at wall are shown in Fig.\ref{fig15} and heat flux coefficients are shown in Fig.\ref{fig16}. Force coefficients are nondimensionalized through $\frac{1}{2}\rho_{\infty}U_{\infty}^2$ and heat coefficients are nondimensionalized through $\frac{1}{2}\rho_{\infty}U_{\infty}^3$. It is found that the slip boundary condition plays a more important role than the first order constitutive relation, for the overall solution around a moving cylinder. And the differences between the results of NCCR equations and NS equations are marginal under the same slip boundary condition. If NCCR model is utilized in the slip boundary condition (second order), results may be better.

\begin{figure}
	\centering
	\subfigure[]{
			\includegraphics[width=0.45 \textwidth]{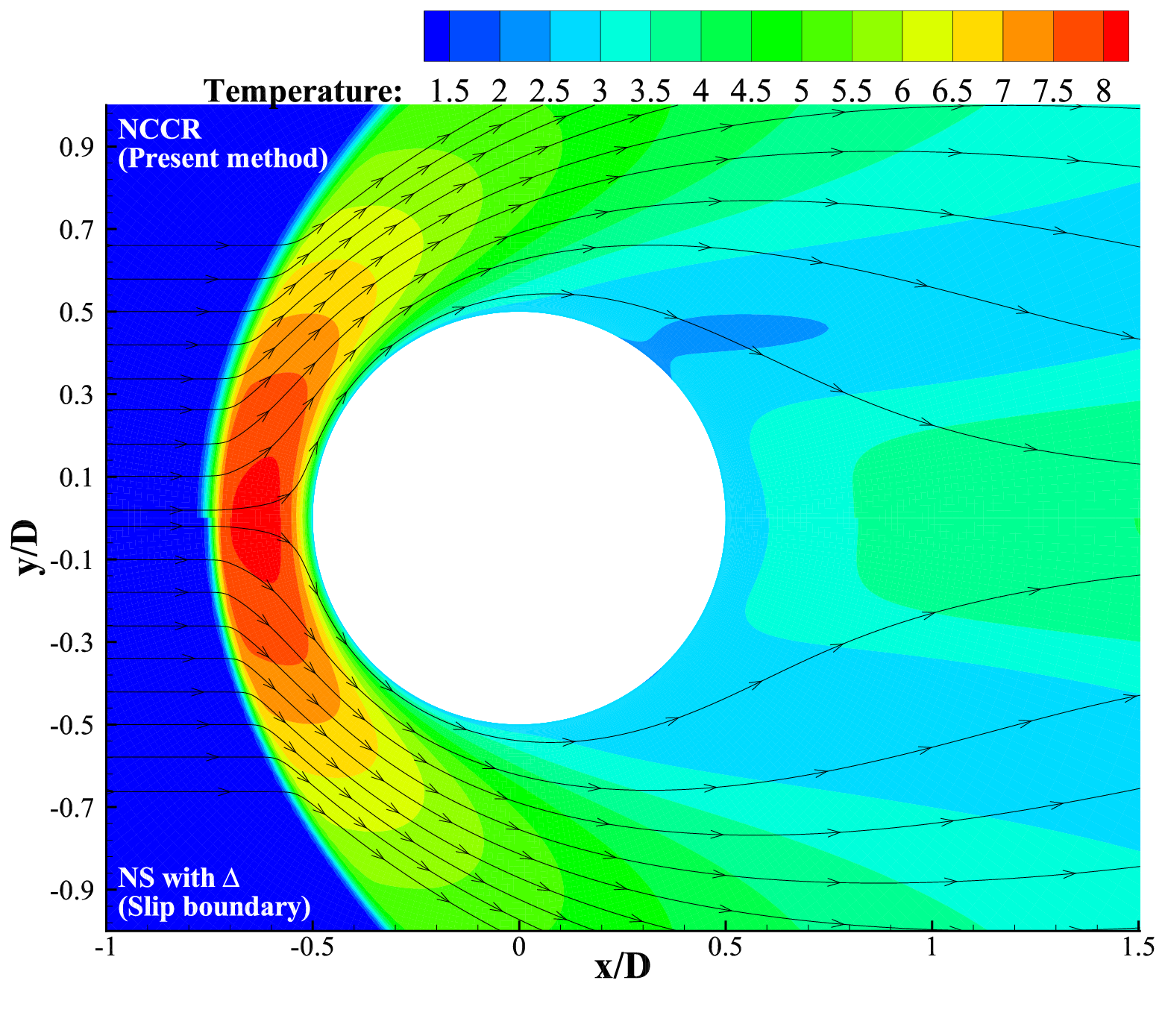}
		}
    \subfigure[]{
    		\includegraphics[width=0.45 \textwidth]{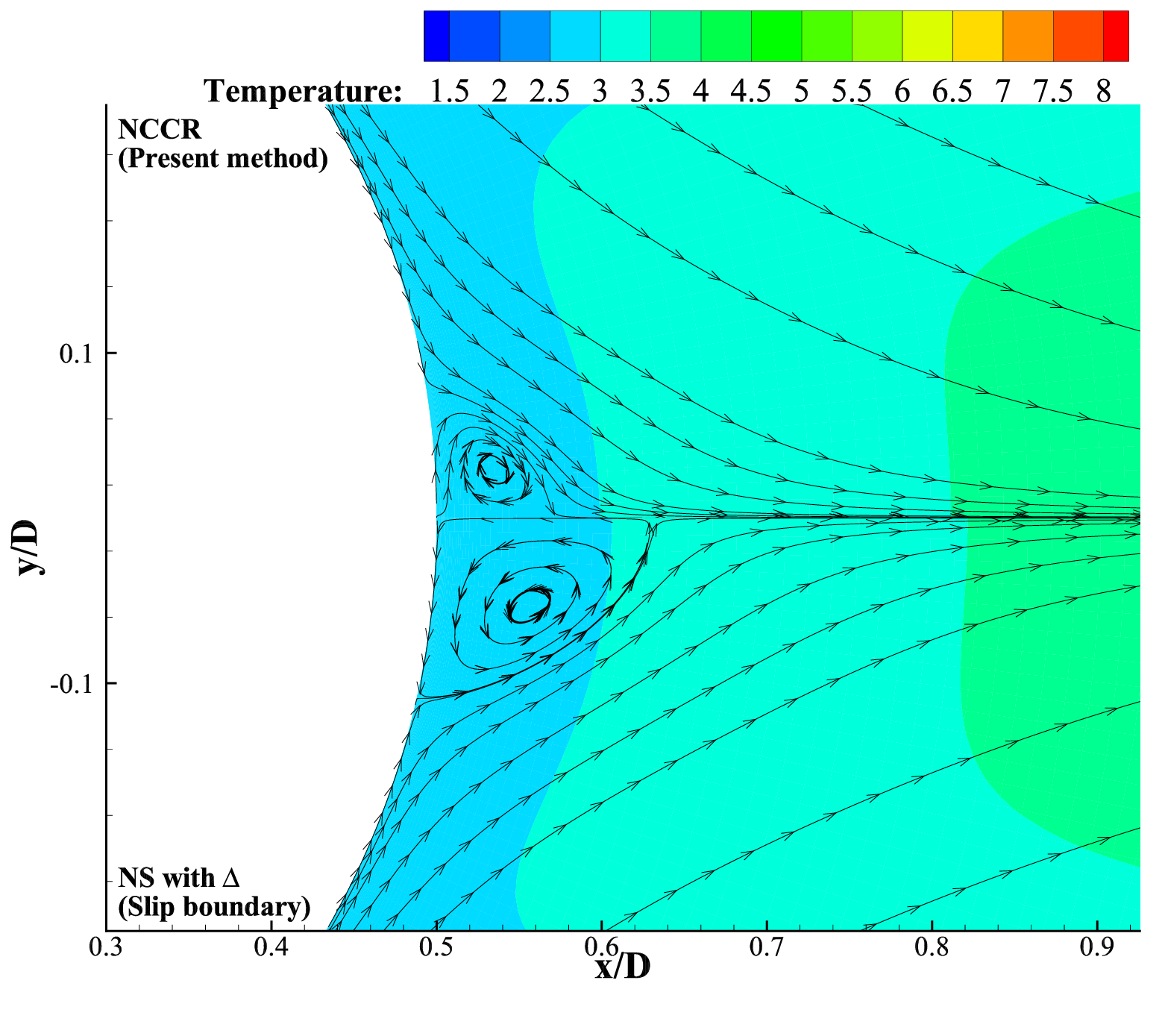}
    	}
	\caption{\label{fig17} Temperature contour and streamline at $Ma=6.0$ for the supersonic/hypersonic rarefied nitrogen gas cylinder flow: (a) Entirety, (b) wake of cylinder.}
\end{figure}

\begin{figure}
	\centering
	\subfigure[]{
			\includegraphics[width=0.60 \textwidth]{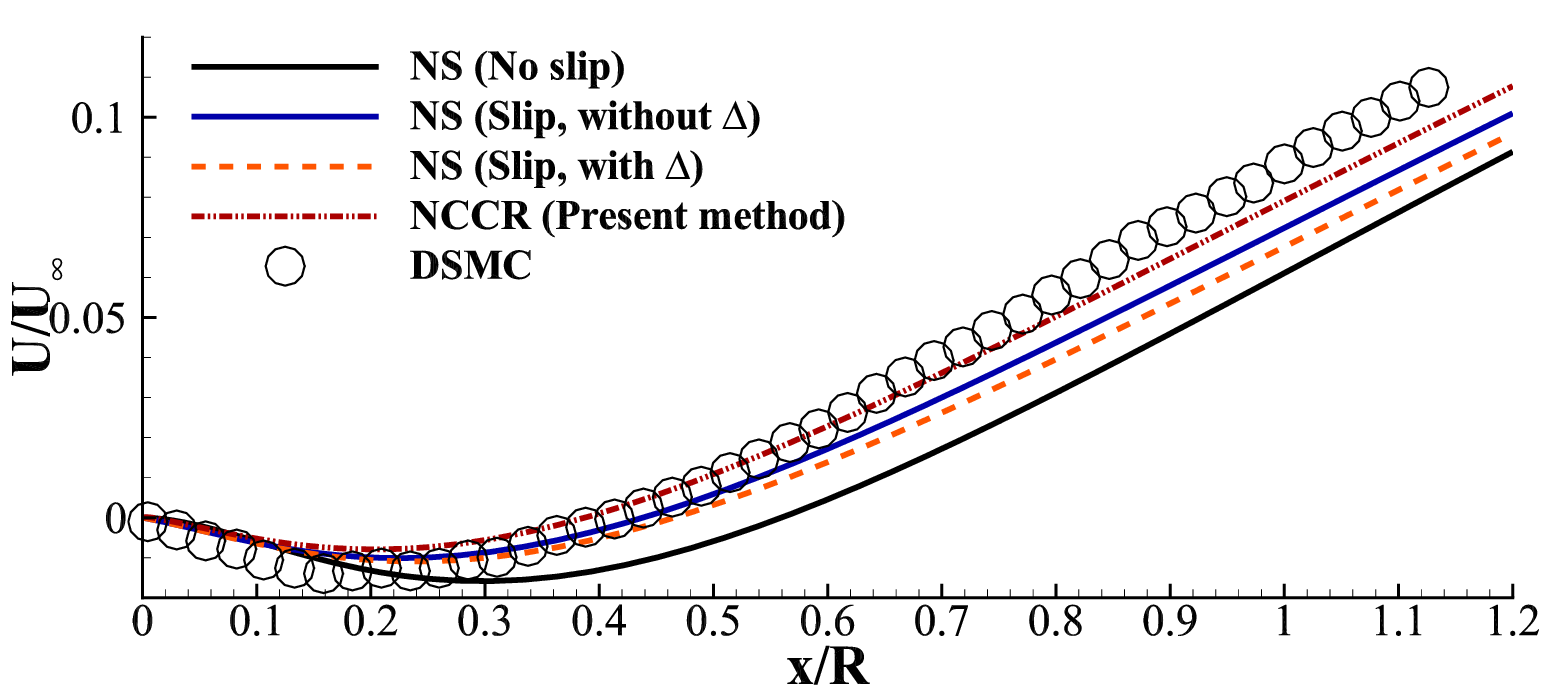}
		}
    \subfigure[]{
    		\includegraphics[width=0.60 \textwidth]{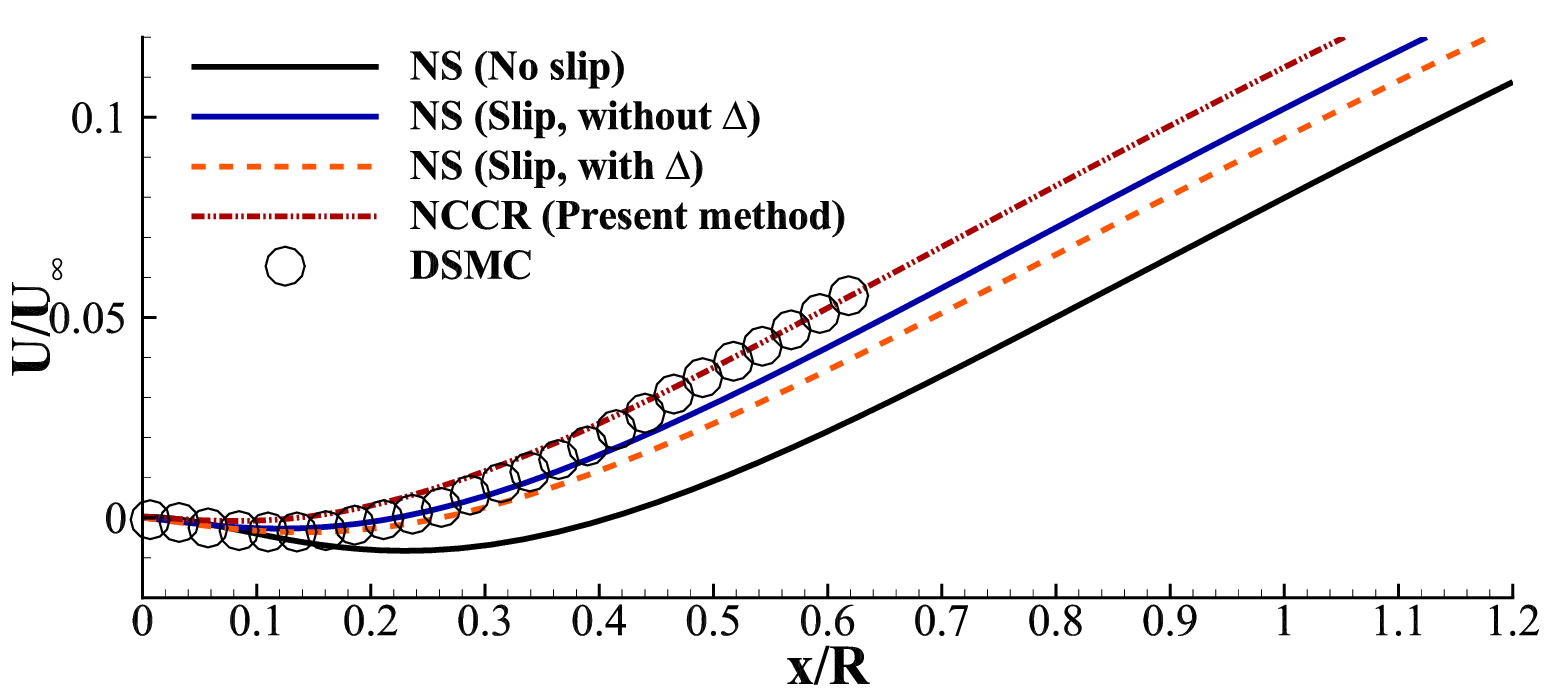}
    	}
    \subfigure[]{
    		\includegraphics[width=0.60 \textwidth]{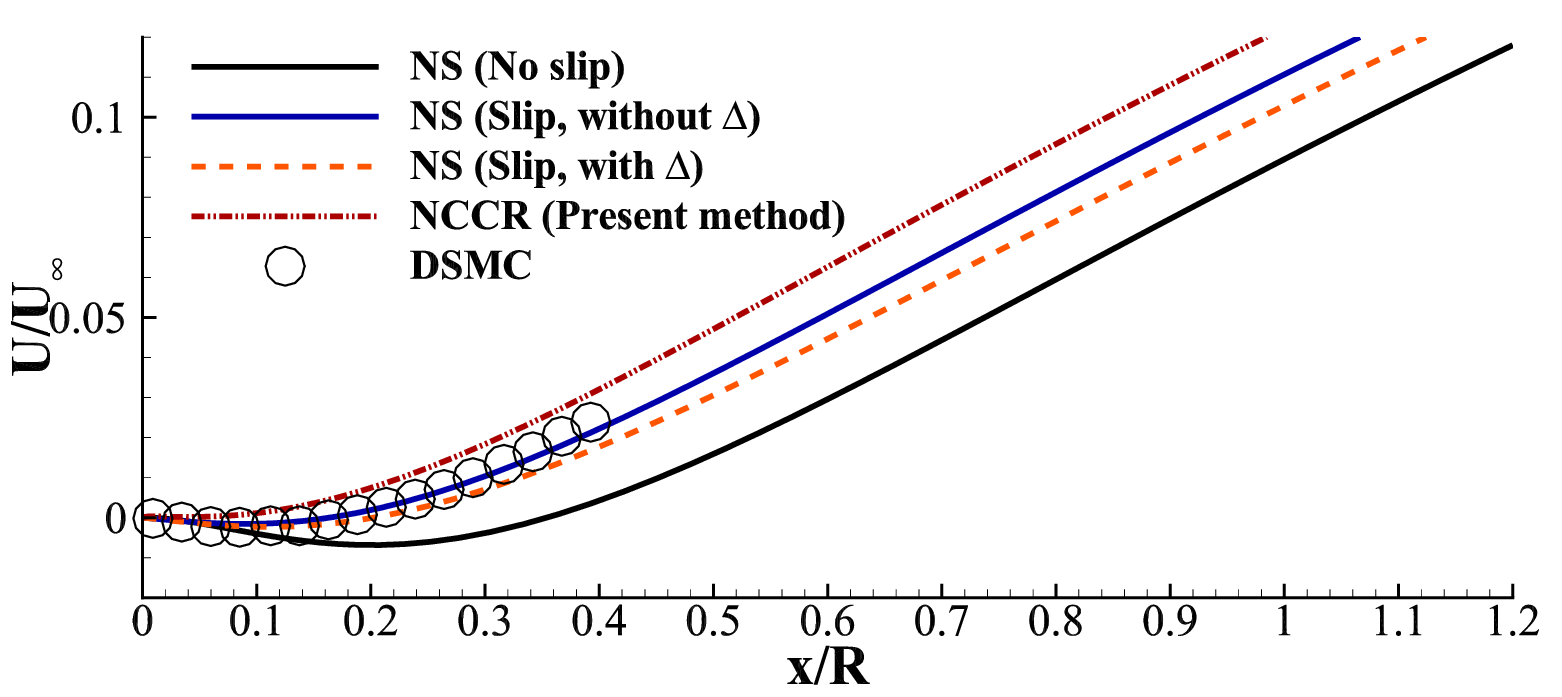}
    	}
	\caption{\label{fig18} x-velocity profiles along the symmetry axis in the wake of cylinder: (a) $Ma=3.0$, (b) $Ma=6.0$, (c) $Ma=12.0$.}
\end{figure}

Now that wall parameters are similar between NS equations and NCCR equations, main difference is in the details of flow field. As Fig.\ref{fig17} shows, different performances are mainly in the shock structure and the wake of cylinder. In Sec.\ref{sec:ssar} and Sec.\ref{sec:ssn2}, difference in the shock structure is analyzed adequately. The quantitative comparison for x-velocity profile in the wake is shown in Fig.\ref{fig18}. Results of NCCR equations are better for $Ma=3.0$ and $Ma=6.0$, and result of NS equations without $\Delta$ is better for $Ma=12.0$, where NCCR equations give higher velocity than reference.

\subsection{The hypersonic rarefied cylinder flow in monatomic argon gas}\label{sec:arbb}
\begin{figure}
	\centering
	\subfigure[]{
			\includegraphics[width=0.45 \textwidth]{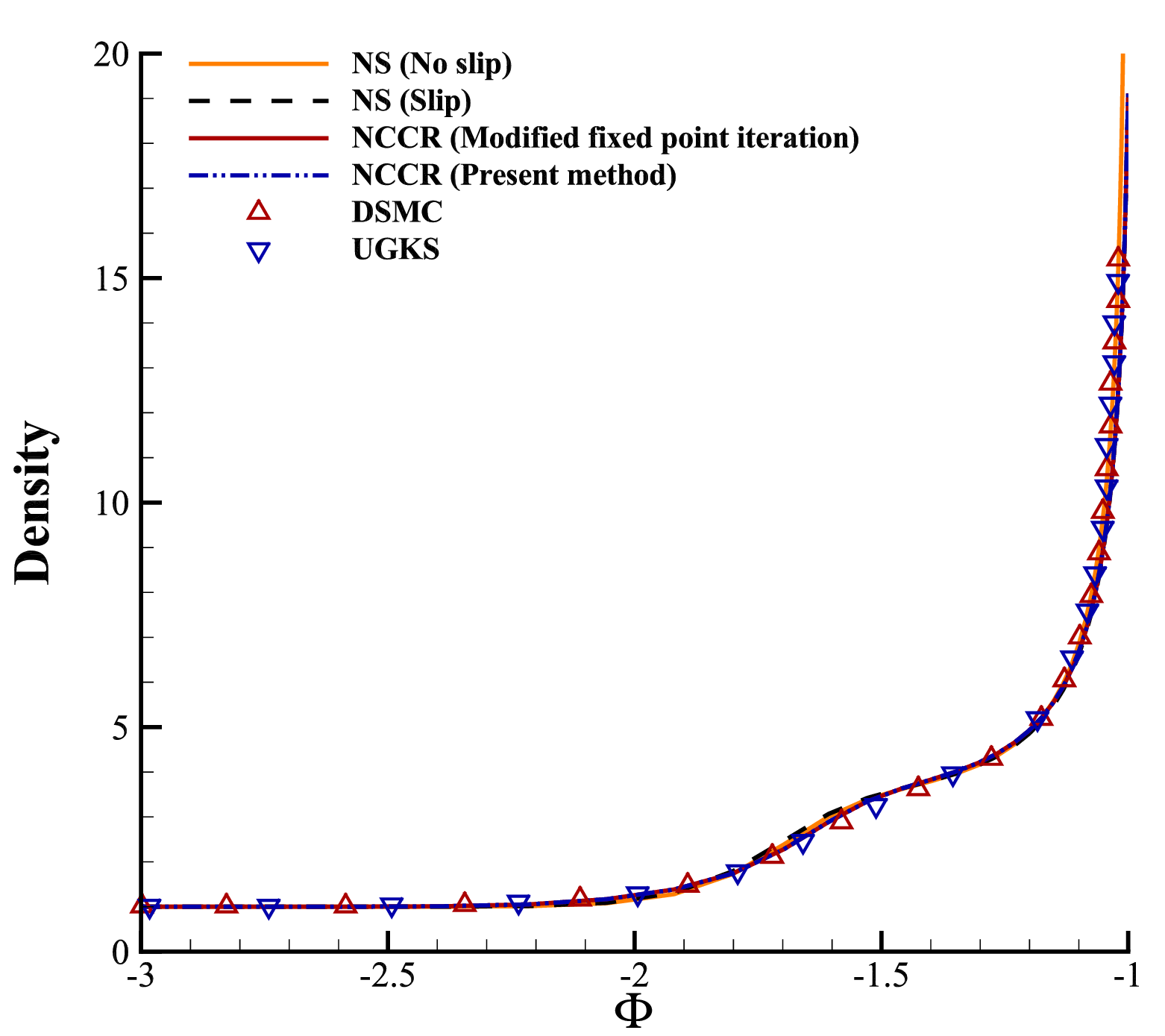}
		}
    \subfigure[]{
    		\includegraphics[width=0.45 \textwidth]{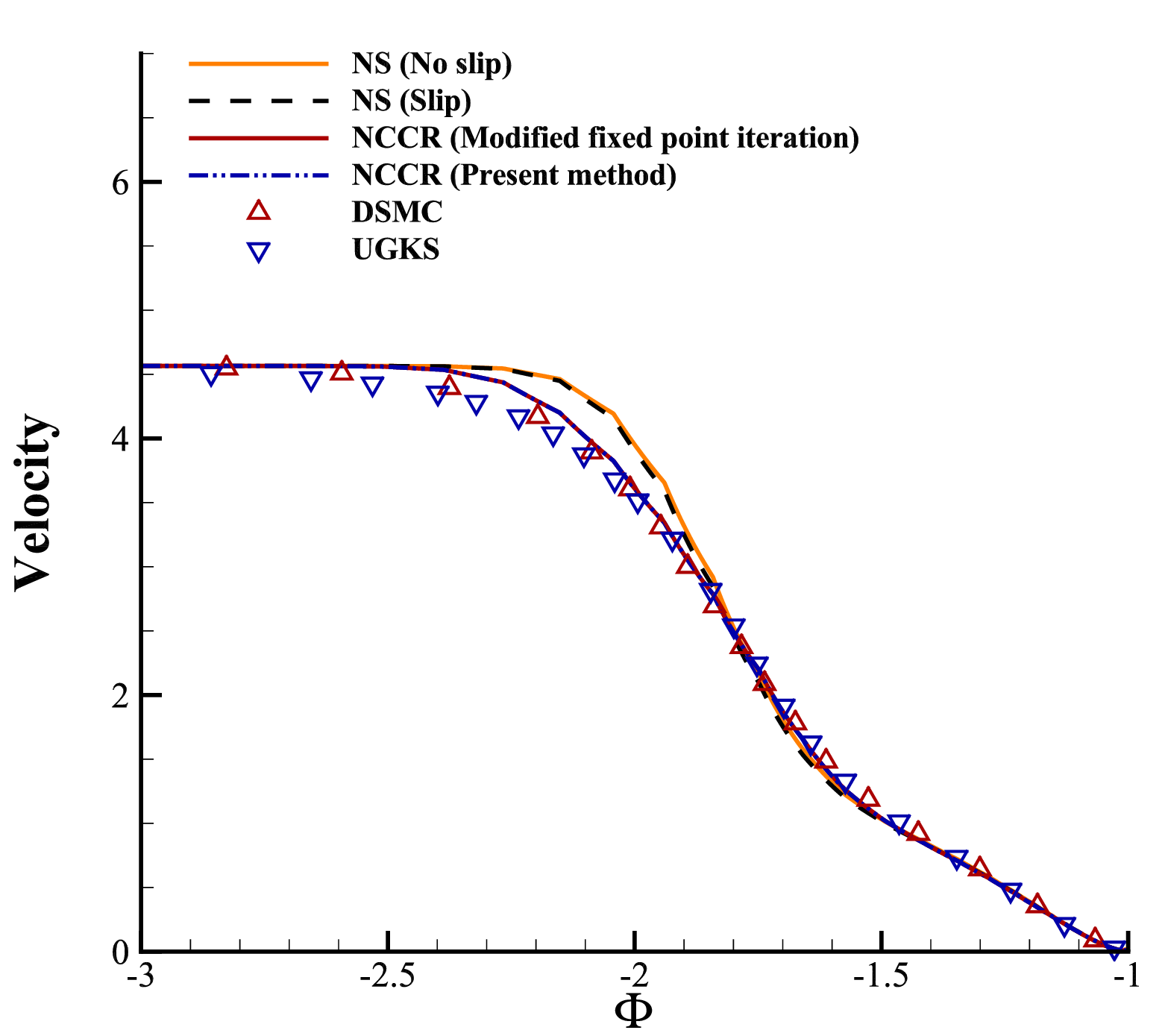}
    	}
    \subfigure[]{
    		\includegraphics[width=0.45 \textwidth]{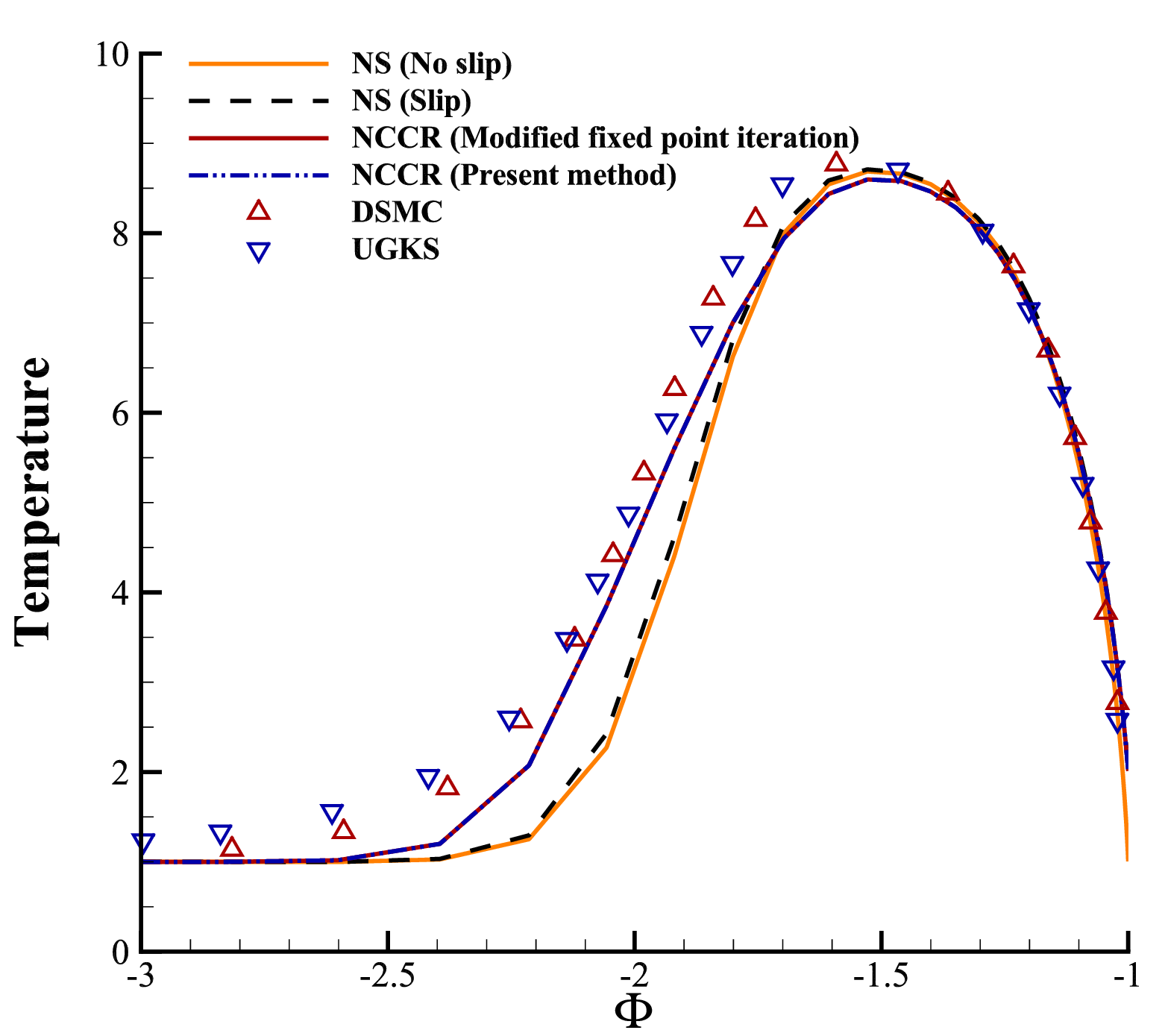}
    	}
	\caption{\label{fig19} Results at the stagnation point line for the hypersonic rarefied argon gas cylinder flow: (a) Density, (b) velocity, (c) temperature.}
\end{figure}

\begin{figure}
	\centering
	\subfigure[]{
			\includegraphics[width=0.45 \textwidth]{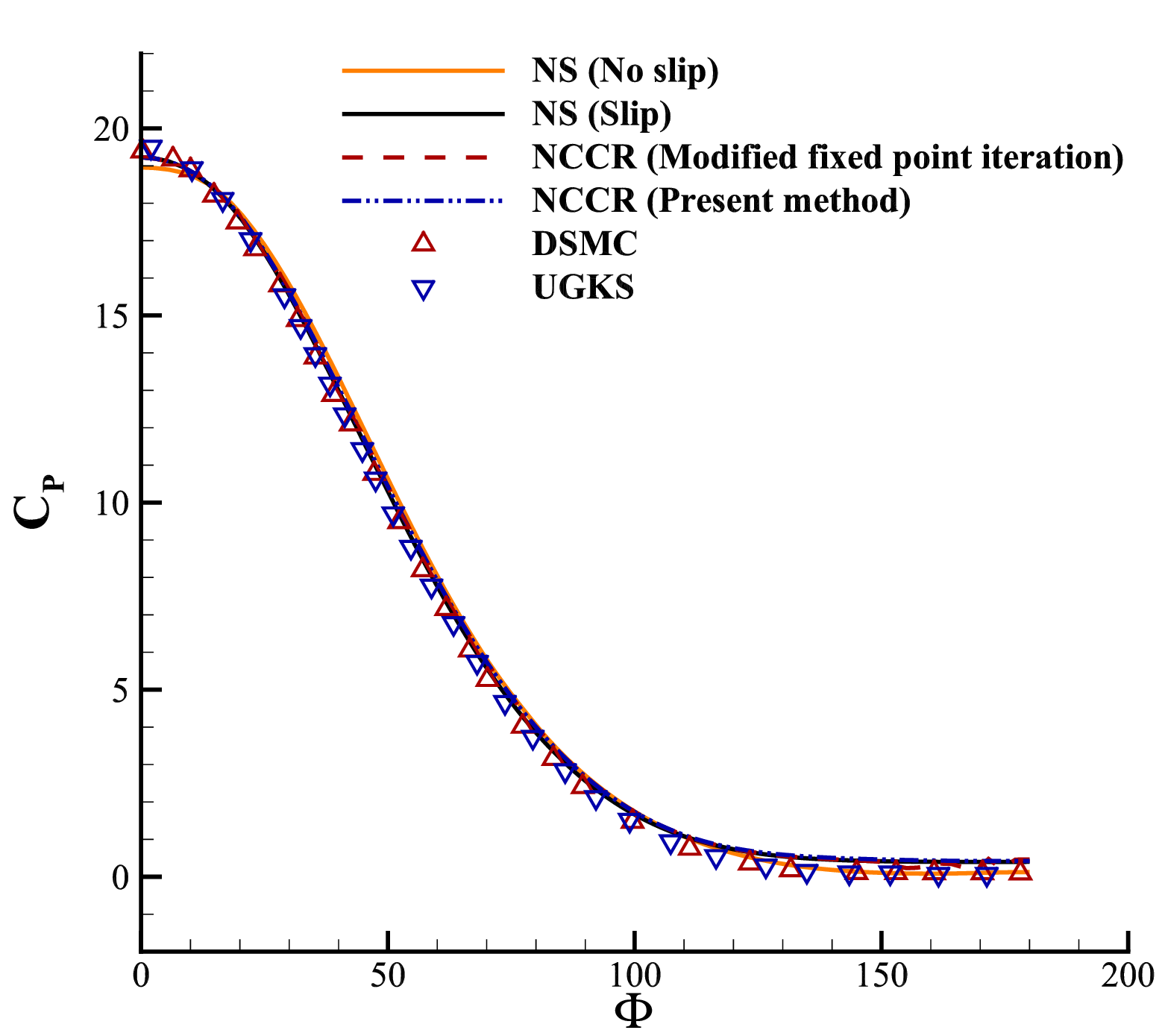}
		}
    \subfigure[]{
    		\includegraphics[width=0.45 \textwidth]{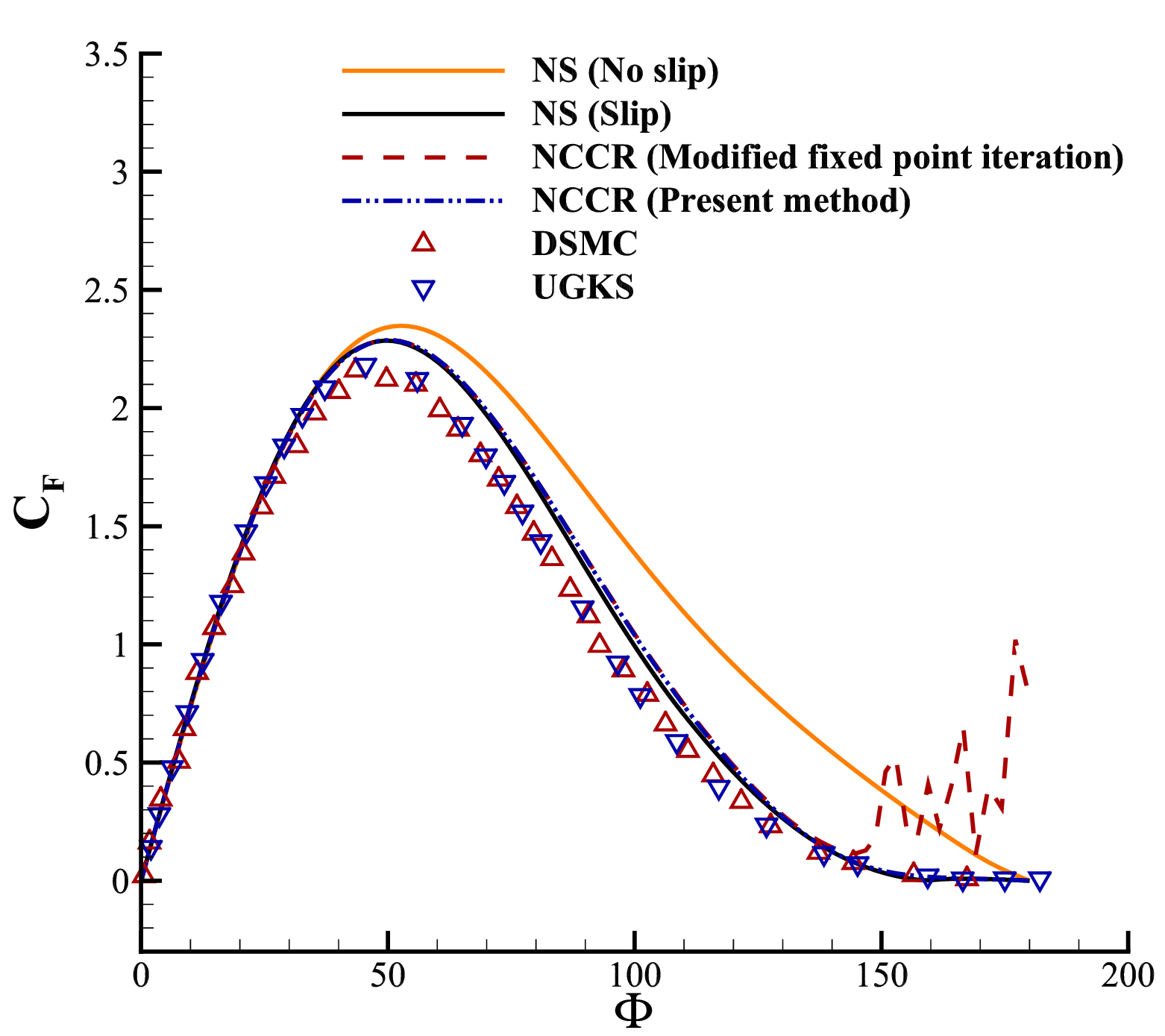}
    	}
    \subfigure[]{
    		\includegraphics[width=0.45 \textwidth]{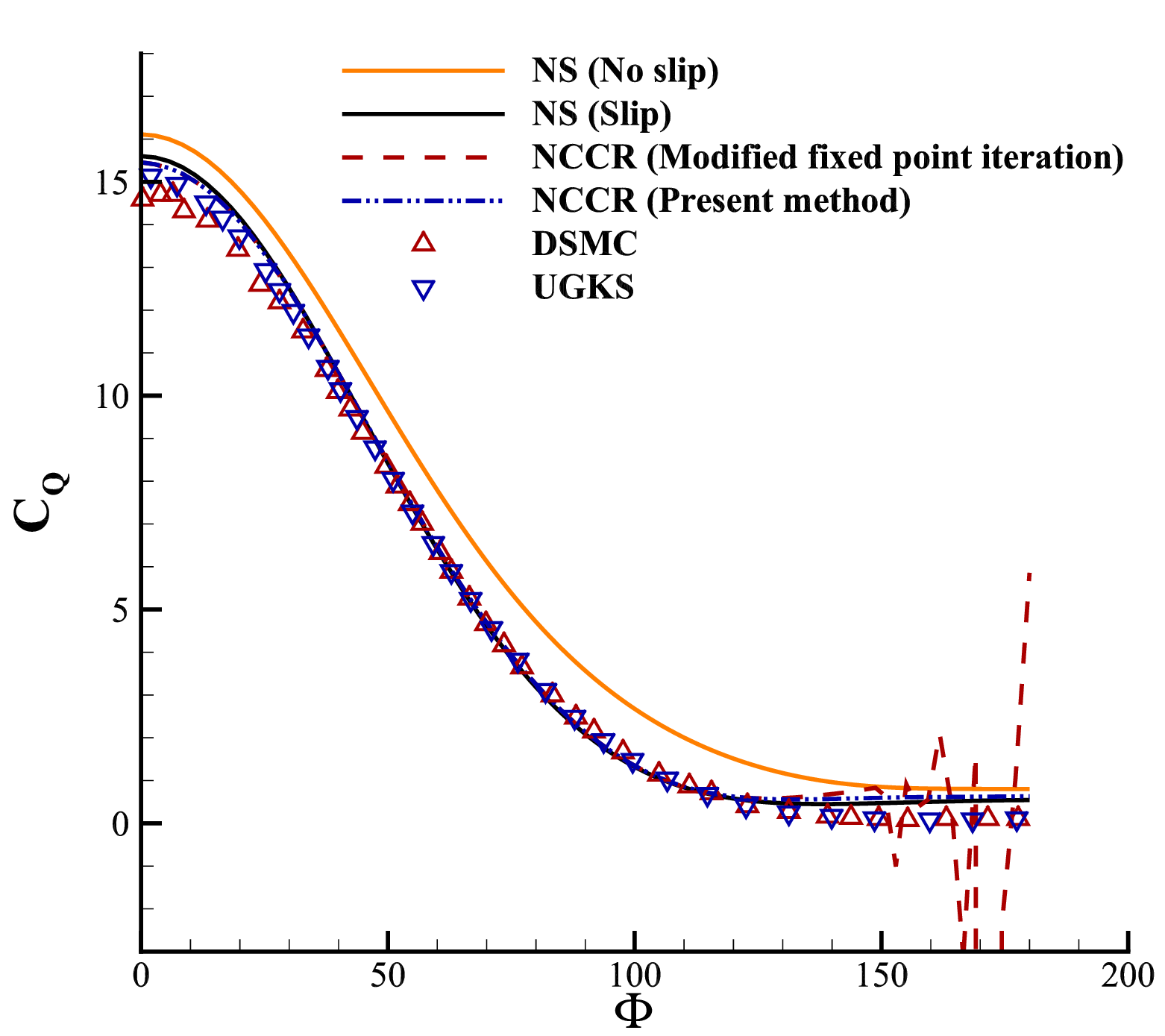}
    	}
	\caption{\label{fig20} Wall parameters for the hypersonic rarefied argon gas cylinder flow: (a) Pressure coefficient, (b) shear stress coefficient, (c) heat flux coefficient.}
\end{figure}

In this section, the hypersonic cylinder flow of monatomic argon gas is simulated. In this case, inflow Mach number is set to $5.0$, $Kn$ number is set to $0.1$, and the corresponding Reynolds number (Re) is $62.44$. The reference length is the radius, $1.0$. The inflow temperature and wall temperature are both set to be $273K$. $\rm{Pr}=0.6667$, $\gamma=1.6667$, and gas constant is set to $208.14JK/kg$. The dynamic viscosity is calculated through Eq.\ref{eq:miu} where $\omega=0.81$, $T_{\rm{ref}}=273K$ and $\mu_{\rm{ref}}=2.116\times10^{-5}Ns/m^2$. The first layer of mesh is set to $0.001$. Results are shown in Fig.\ref{fig19} and Fig.\ref{fig20}. The reference data are results of UGKS and DSMC from Ref.\cite{ugks2}. Tested methods are proposed method for NCCR equations, MFPI method for NCCR equations and the method for NS equations. All above methods are coupled with slip boundary condition and the NS solver without slip boundary condition is shown for comparison as well. The inviscid flux used is KIF. Density and Temperature are nondimensionalized through inflow values $\rho_{\infty}$ and $T_{\infty}$. Velocity is nondimensionalized through $C_{\infty}=\sqrt{2RT_{\infty}}$. Force coefficients are nondimensionalized through $\rho_{\infty}C_{\infty}^2$ and heat coefficients are nondimensionalized through $\rho_{\infty}C_{\infty}^3$.

Fig.\ref{fig19} shows density, velocity and temperature profiles along the stagnation point line. As to the density profile, results of all methods are similar. Results of NCCR are better than NS equations at velocity and temperature profiles, which is consistent with the phenomenon in shock structure test cases (Sec.\ref{sec:ssar} and Sec.\ref{sec:ssn2}). Fig.\ref{fig20} shows wall parameters, pressure coefficient, shear stress coefficient and heat flux coefficient. As to wall parameter curves, methods with slip boundary condition are all more accurate than without it. Results of NCCR equations and NS equations are similar at pressure and shear stress coefficient curves. At the heat flux curve, peak value of NCCR equations is a little lower than NS equations, which is more consistent with the result of DSMC. In conclusion, in this test case, performance of NCCR equations is better than NS equations at the stagnation point line. But there is not much difference about the wall parameters. As the statement in Sec.\ref{sec:arbb}, if second order constitutive relation is utilized in the slip boundary condition, results may be better.

Different from Sec.\ref{sec:arbb}, in this test case, the Kn number is much larger. At the wedge, the rarefaction degree is quiet high, and the $q(\kappa)$ there increases endlessly until the calculation blows up. This means the NCCR model is not stable itself. As Ref.\cite{egks} comments, "In Eu's non-equilibrium thermodynamics, $q(\kappa)$ physically means that there is an extra increment of the entropy relaxation rate towards the equilibrium compared with the linear relaxation when the distribution function is far from the equilibrium. But the extra entropy relaxation rate is actually bounded\cite{villani1,villani2}", which explains why the NCCR model loses its accuracy and stability when $q(\kappa)$ is too huge or the flow is too rarefied. As a result, a correction technique is proposed. At $q(\kappa)_{\rm{NCCR}} > 2.0$, the stress, heat flux and $\Delta$ are used as follows:

\begin{equation}
\begin{aligned}
{\Pi _{\rm{ij}}} &= \chi{\Pi _{\rm{ij,NCCR}}}+(1-\chi){\Pi _{\rm{ij,NS}}},\\
{Q_{\rm{i}}} &= \chi{Q_{\rm{i,NCCR}}}+(1-\chi){Q_{\rm{i,NS}}},\\
\Delta &= \chi\Delta_{\rm{NCCR}}+(1-\chi)\Delta_{\rm{NS}},\\
\chi &= (q(\kappa)_{\rm{NCCR}}-1)^{-2}.
\end{aligned}
\end{equation}

\begin{figure}
	\centering
	\subfigure{
			\includegraphics[width=0.5 \textwidth]{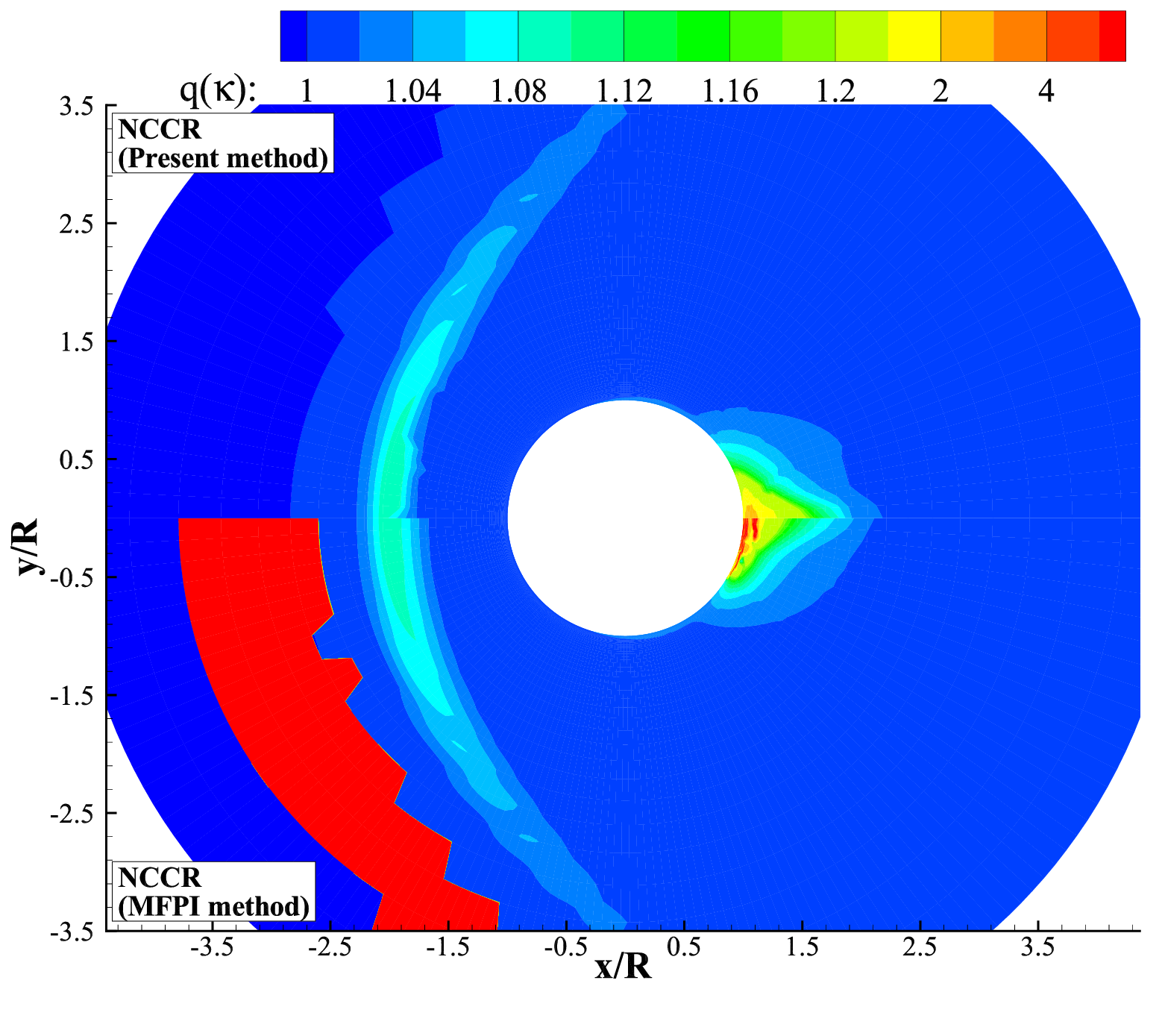}
		}
	\caption{\label{figarbb} $q(\kappa)$ contour for the hypersonic rarefied argon gas cylinder flow.}
\end{figure}

As Fig.\ref{figarbb} shows, using this technique, the $q(\kappa)$ contour at the wedge of cylinder is clear. This technique is used in all test cases below. Another phenomenon is that in the result of MFPI method, the value of $q(\kappa)$ is huge at the pre-shock. This is the same with Fig.\ref{fig14c} of Sec.\ref{sec:ssn2}.

The efficiency is tested in this case. During the same iterations for the whole domain, the computation time of proposed method is $58.5$ times of method for NS equations, and $2.8$ times of MFPI for NCCR equations. Because MFPI is taken as the pretreatment of Jiang's coupling method, the efficiency of proposed method is as the same magnitude order of Jiang' method.

\subsection{Hypersonic rarefied flat flow in diatomic nitrogen gas}
\begin{figure}
	\centering
	\subfigure{
			\includegraphics[width=0.5 \textwidth]{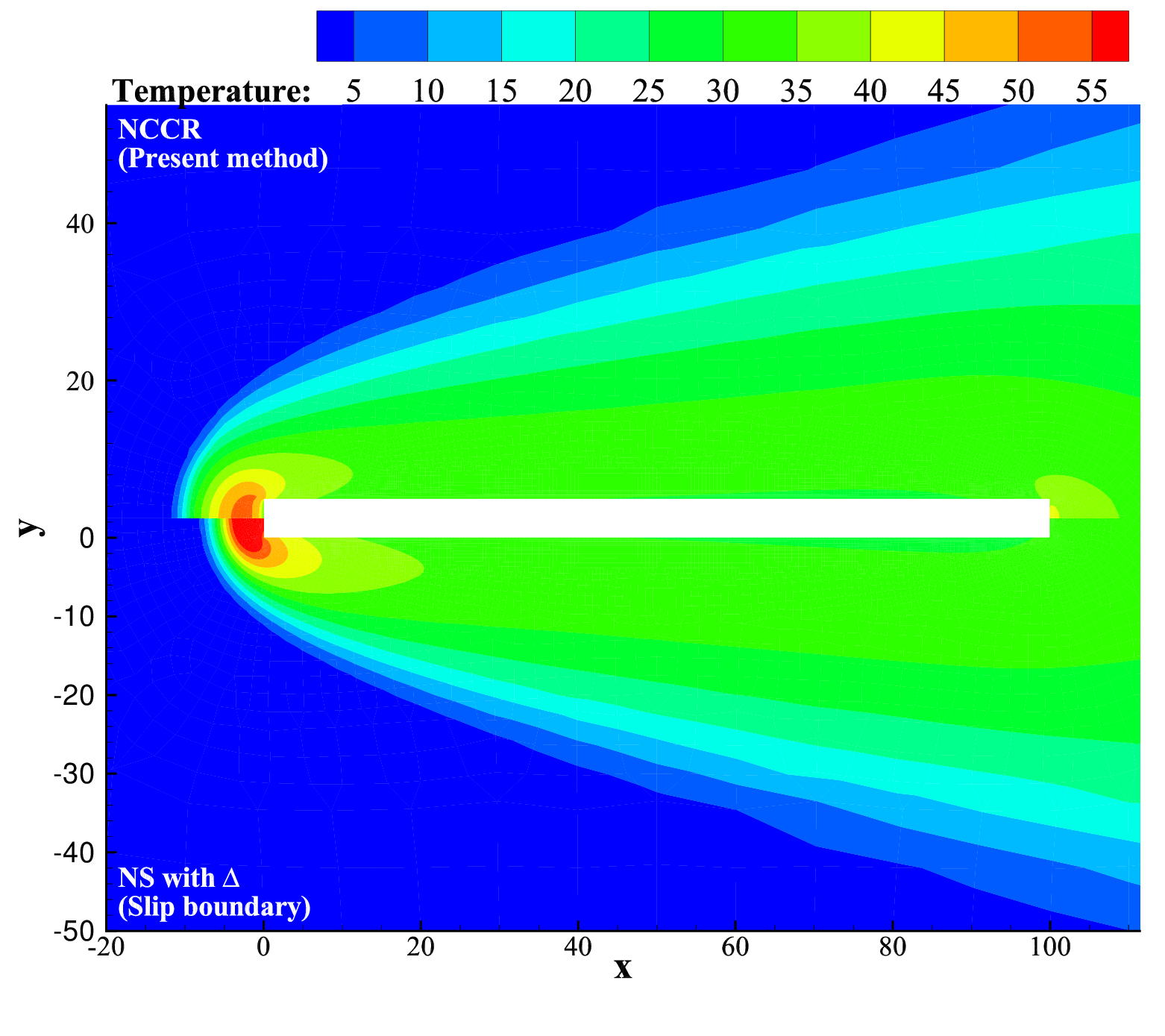}
		}
	\caption{\label{fig21} Temperature contour for the hypersonic rarefied argon gas flat flow.}
\end{figure}

In this section, another typical rarefied flow is studied, the hypersonic rarefied flat flow of nitrogen gas, which is more difficult to simulate. Experiment case in Ref.\cite{flatexp} is taken for simulation, which is widely used for method tests or software validations\cite{flat1,flat2}. Except for experiment results, DSMC results are used as reference as well\cite{flat1,flat2}. The length of the flat is $100mm$, and the thickness is $5mm$. The inflow Mach number is $20.2$, Kn number is $0.016$. The inflow temperature is $13.32K$ and the wall temperature is set to $300K$. The inflow density is $1.7278\times10^{-5}kg/m^3$. The gas constant is $296.72JK/kg$, and $\gamma=1.4$, $\rm{Pr}=0.72$. The dynamic viscosity is calculated through Eq.\ref{eq:miu}, where $\omega=0.74$, $T_{\rm{ref}}=273K$, $\mu_{\rm{ref}}=1.6579\times10^{-5}Ns/m^2$. The inviscid flux used is AUSM+ -up scheme. The first layer of mesh is set to $0.005mm$. And the temperature contour is shown as Fig.\ref{fig21}.

\begin{figure}
	\centering
	\subfigure[]{
			\includegraphics[width=0.45 \textwidth]{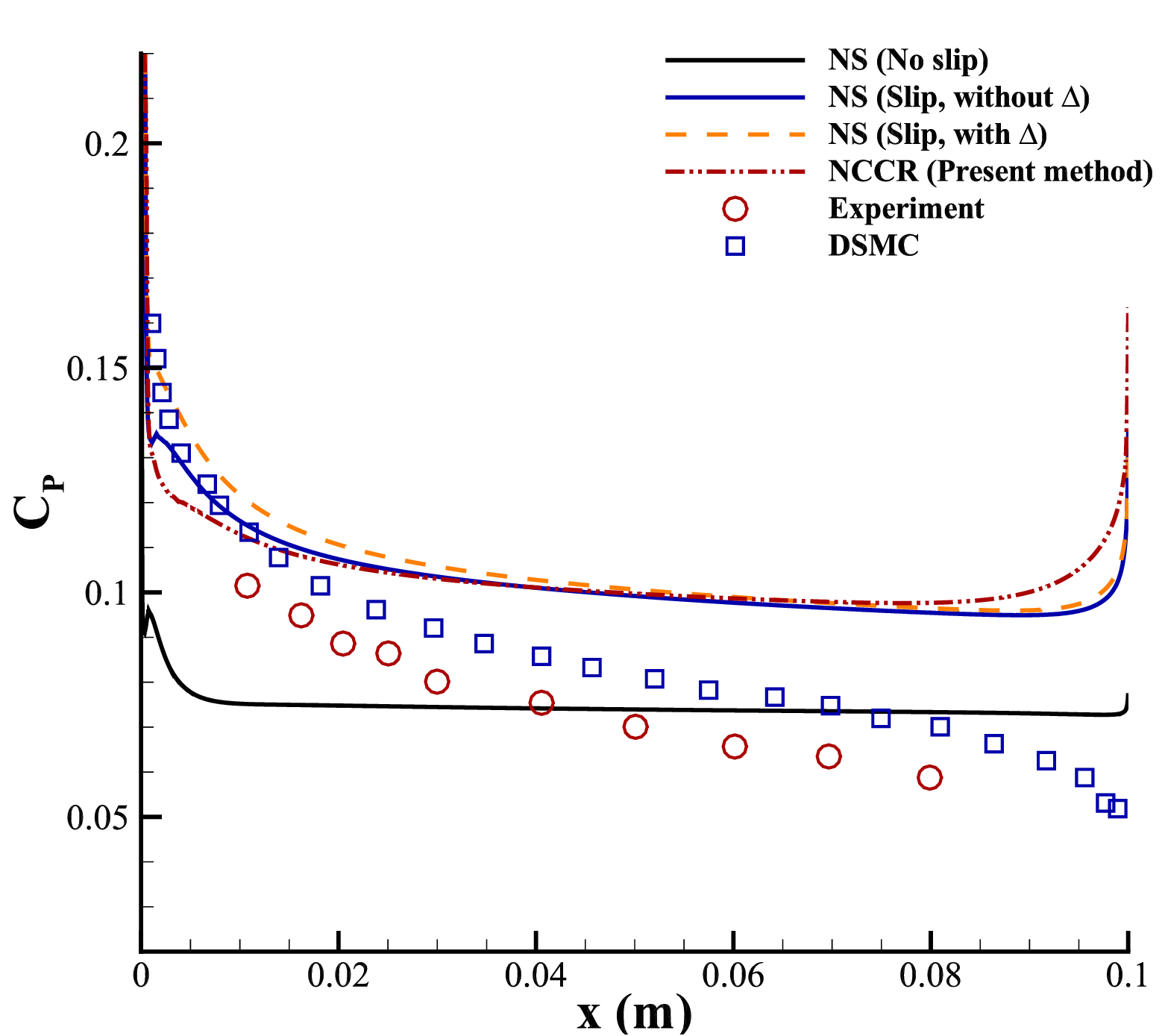}
		}
    \subfigure[]{
    		\includegraphics[width=0.45 \textwidth]{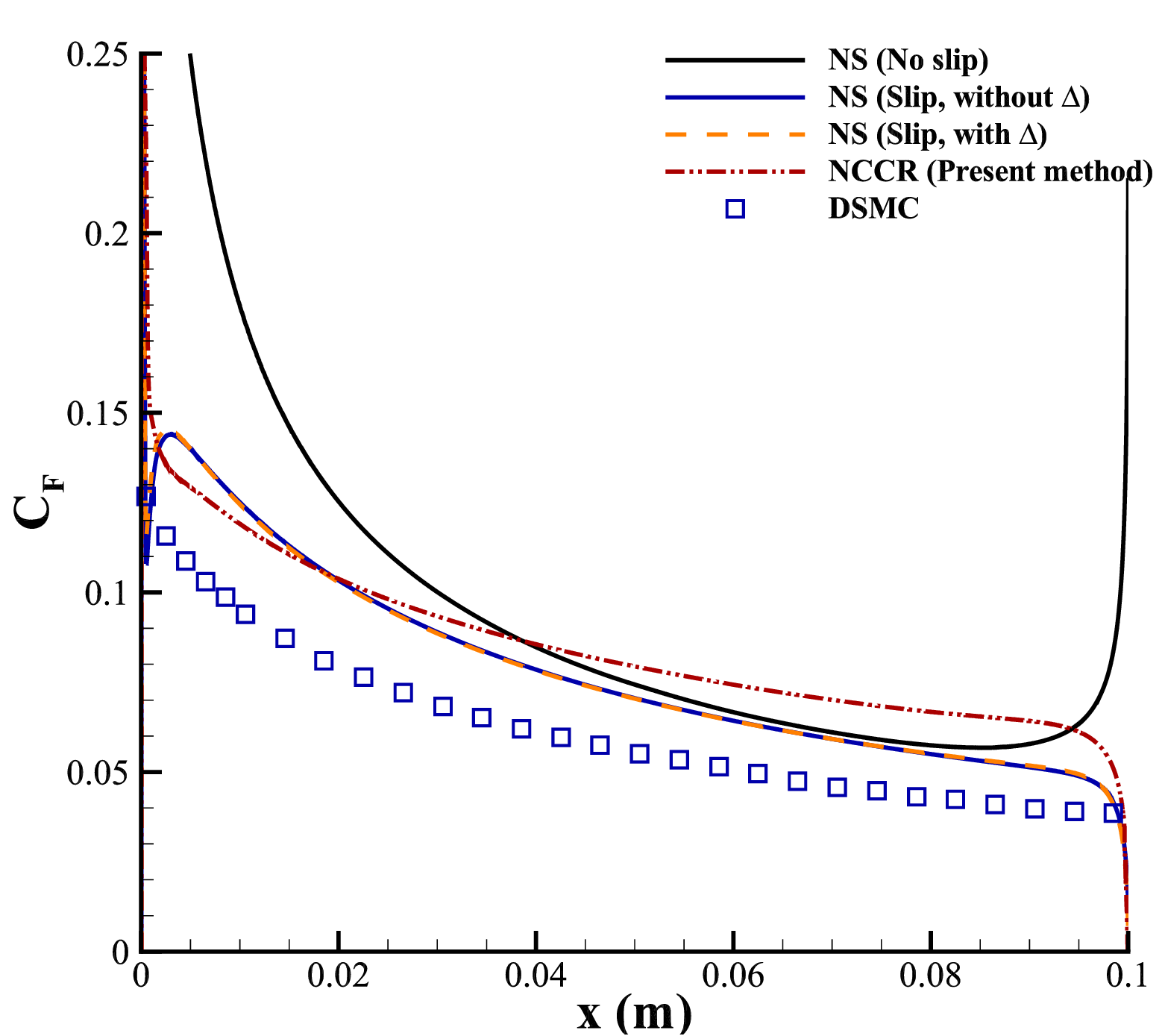}
    	}
    \subfigure[]{
    		\includegraphics[width=0.45 \textwidth]{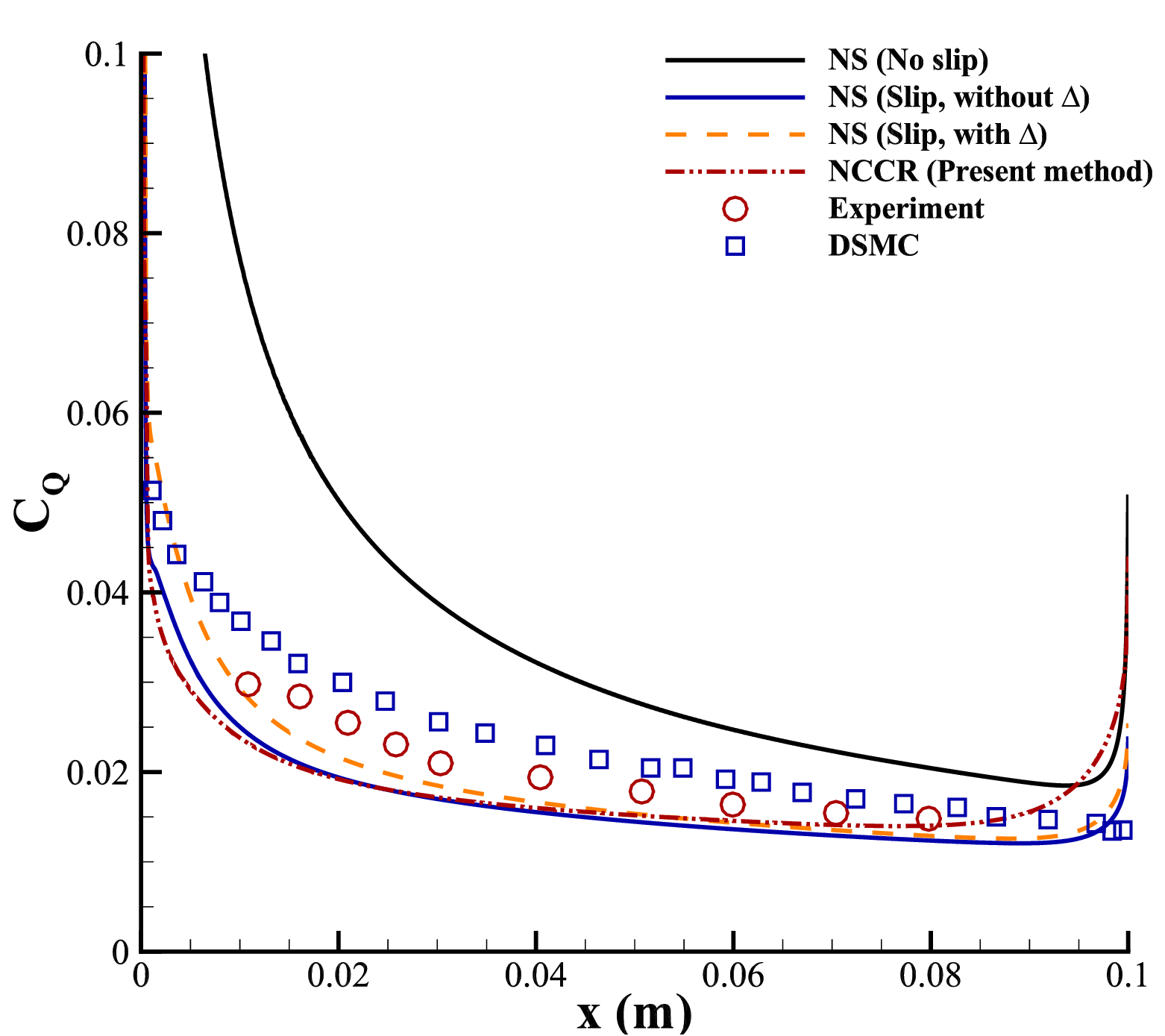}
    	}
	\caption{\label{fig22} Wall parameters for the hypersonic rarefied nitrogen gas flat flow: (a) Pressure coefficient, (b) shear stress (c) heat flux coefficient.}
\end{figure}

Wall parameter results are shown as Fig.\ref{fig22}. Force coefficients are nondimensionalized through $\frac{1}{2}\rho_{\infty}U_{\infty}^2$ and heat coefficients are nondimensionalized through $\frac{1}{2}\rho_{\infty}U_{\infty}^3$. The wall parameter curves are mainly influenced by the slip boundary condition. Methods with slip boundary condition give better heat flux results, but worse pressure results.

\subsection{Supersonic rarefied sphere flow in diatomic nitrogen gas}
\begin{figure}
	\centering
	\subfigure[]{
			\includegraphics[width=0.45 \textwidth]{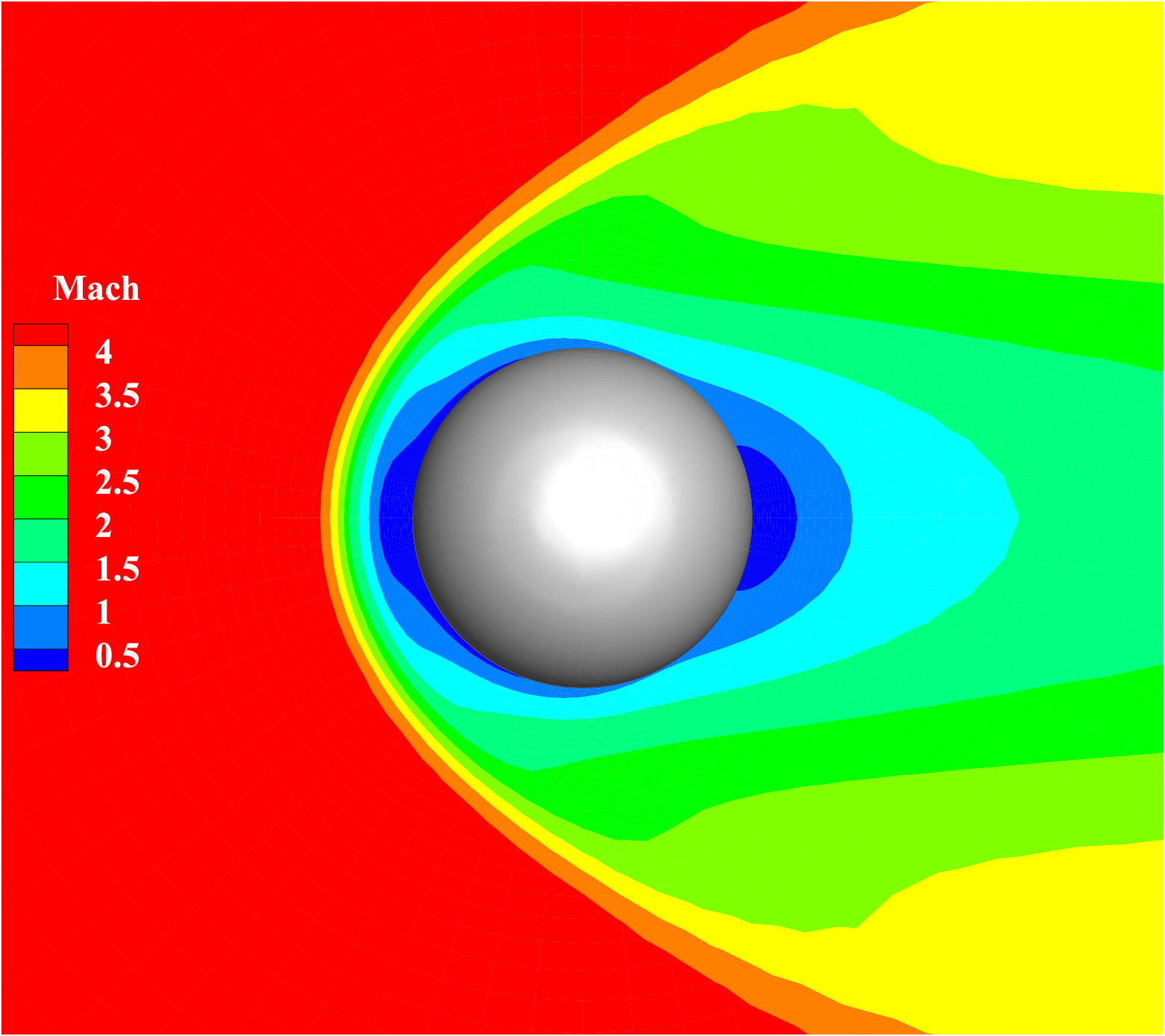}
		}
    \subfigure[]{
    		\includegraphics[width=0.45 \textwidth]{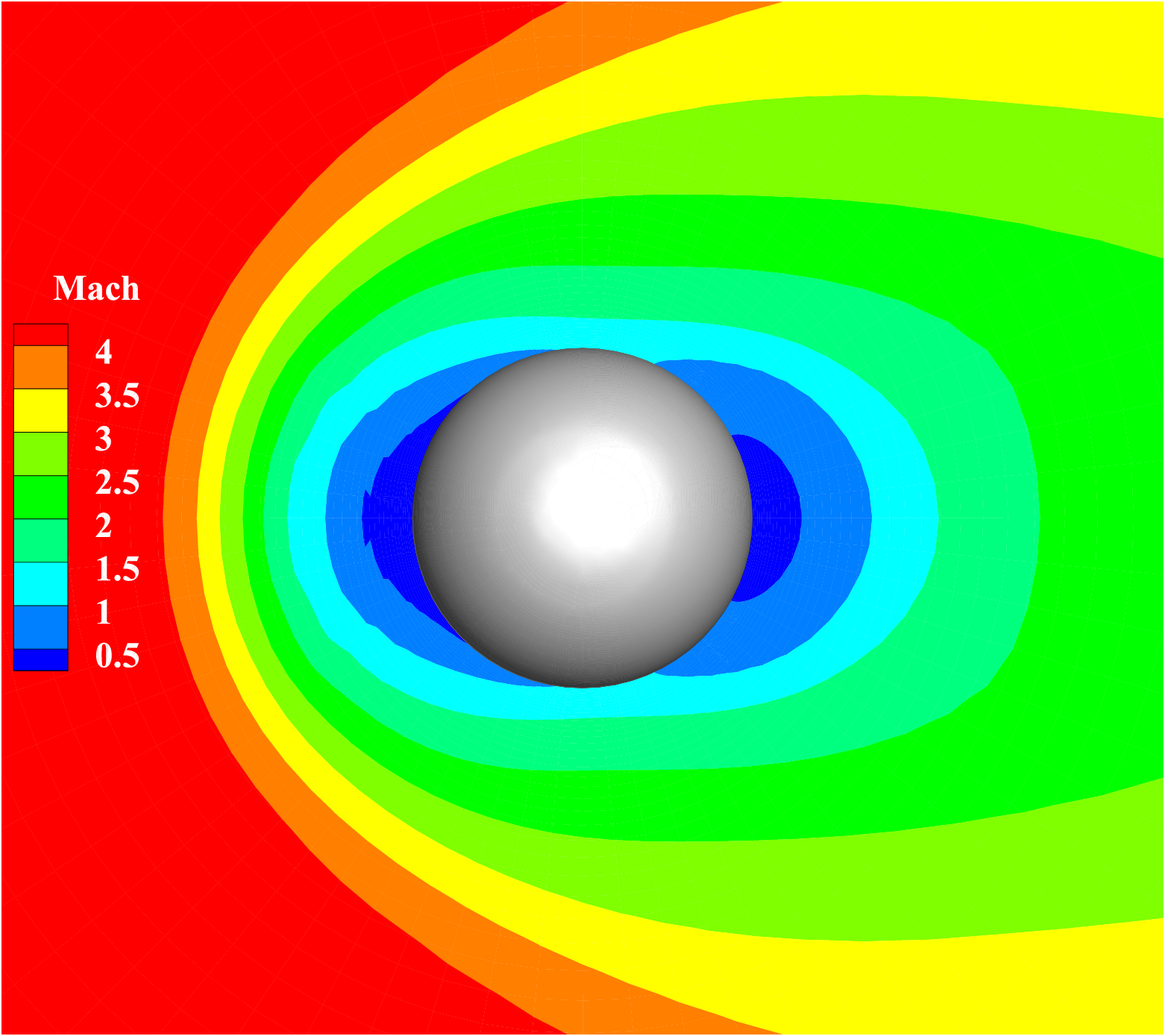}
    	}
	\caption{\label{fig23} Mach contours of proposed NCCR method for the supersonic rarefied nitrogen gas sphere flow: (a) $Kn=0.031$, (b) $Kn=0.121$.}
\end{figure}

In this section, a three-dimensional flow is simulated as Fig.\ref{fig23}. The inflow Mach number is $4.25$, Kn numbers are $0.031$ and $0.121$ and corresponding Re numbers are $210$ and $53$, respectively, whose reference length is the diameter of sphere, 2.0. The inflow temperature is $65K$ and the wall temperature is set to $302K$. The gas constant is $296.72JK/kg$, and $\gamma=1.4$, $\rm{Pr}=0.72$. The dynamic viscosity is calculated through Eq.\ref{eq:miu}, where $\omega=0.74$, $T_{\rm{ref}}=273K$, $\mu_{\rm{ref}}=1.6579\times10^{-5}Ns/m^2$. The inviscid flux used is KIF scheme. The first layer of mesh is set to $0.01$.

\begin{figure}
	\centering
	\subfigure[]{
			\includegraphics[width=0.45 \textwidth]{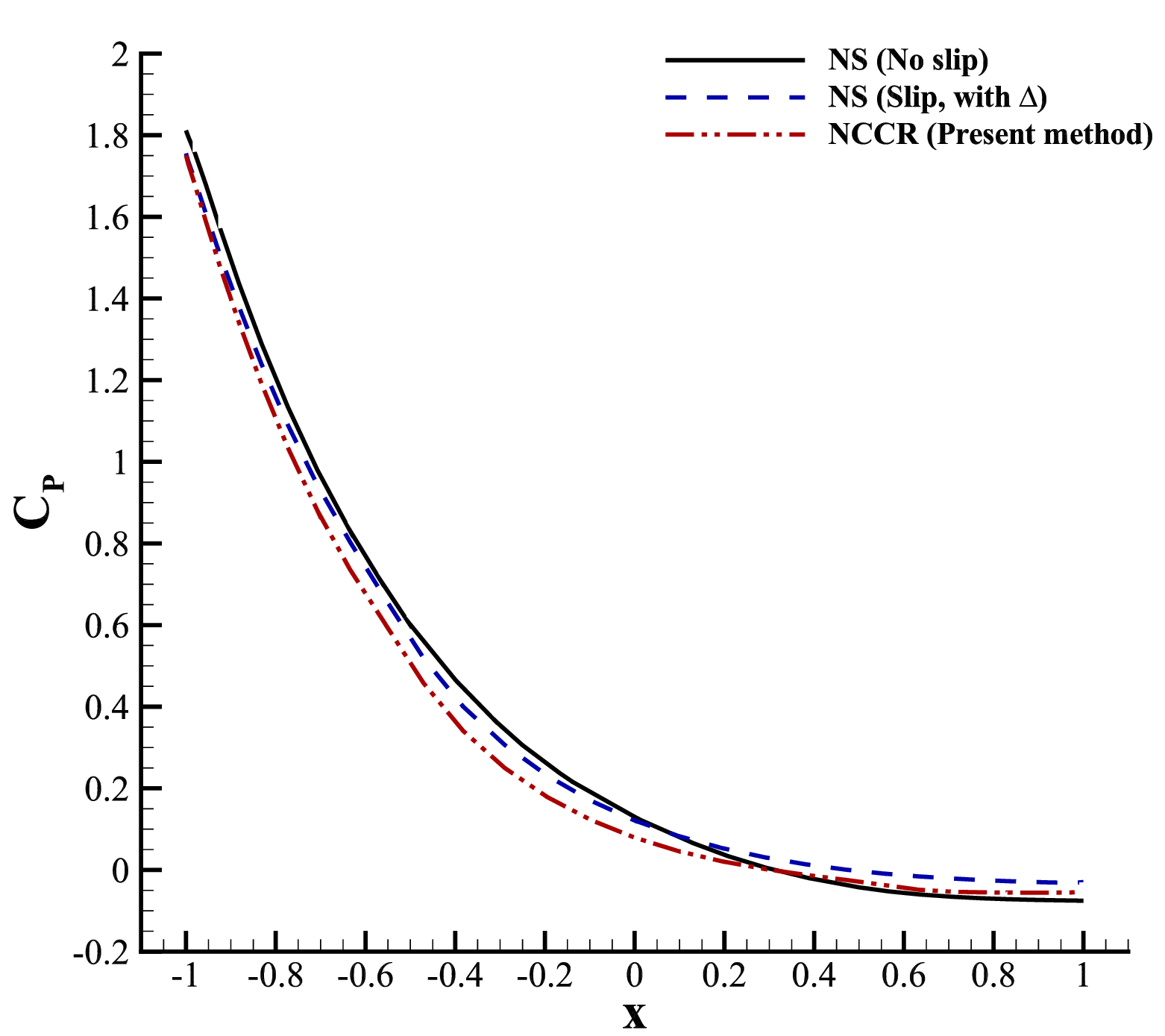}
		}
    \subfigure[]{
    		\includegraphics[width=0.45 \textwidth]{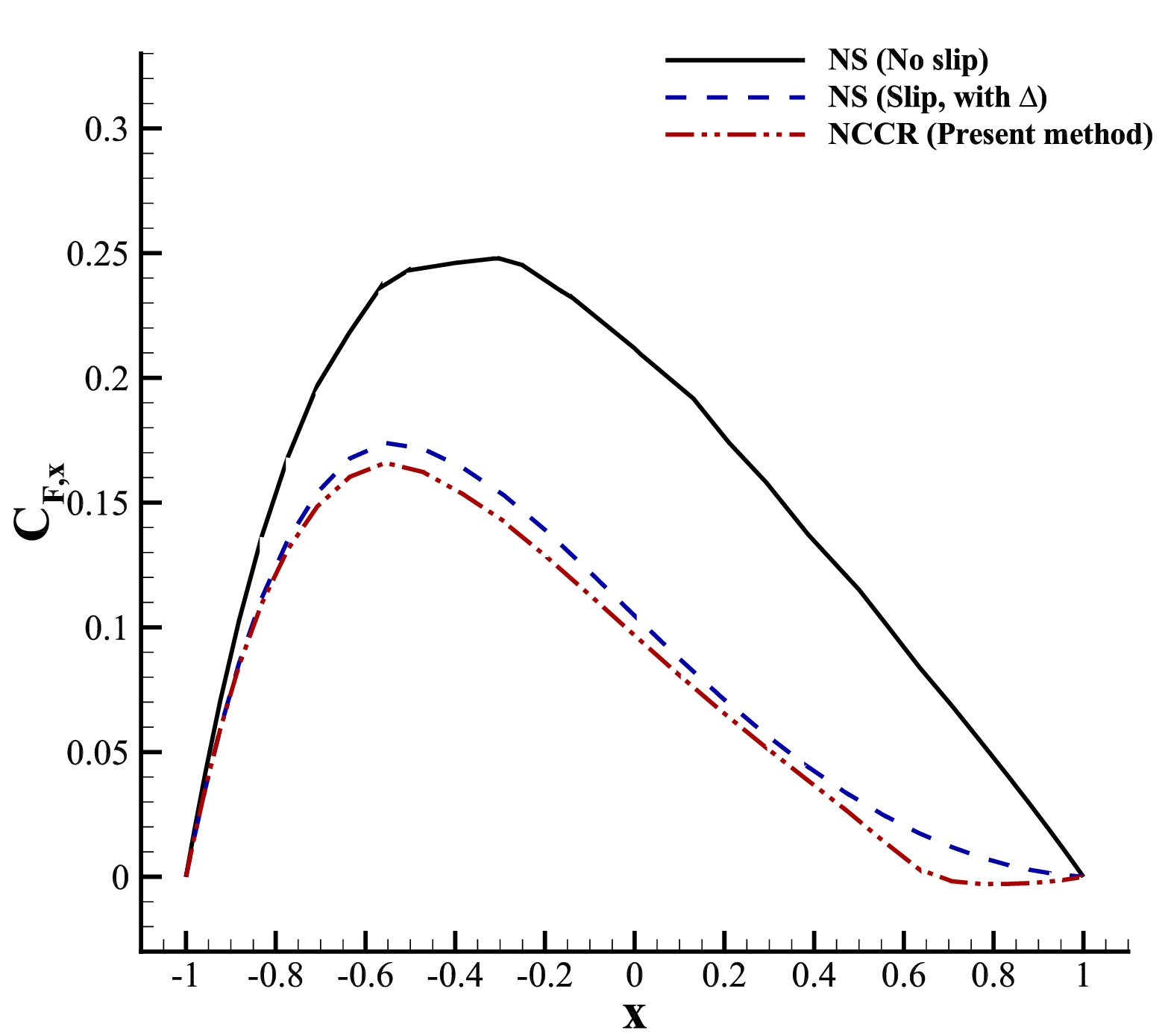}
    	}
    \subfigure[]{
			\includegraphics[width=0.45 \textwidth]{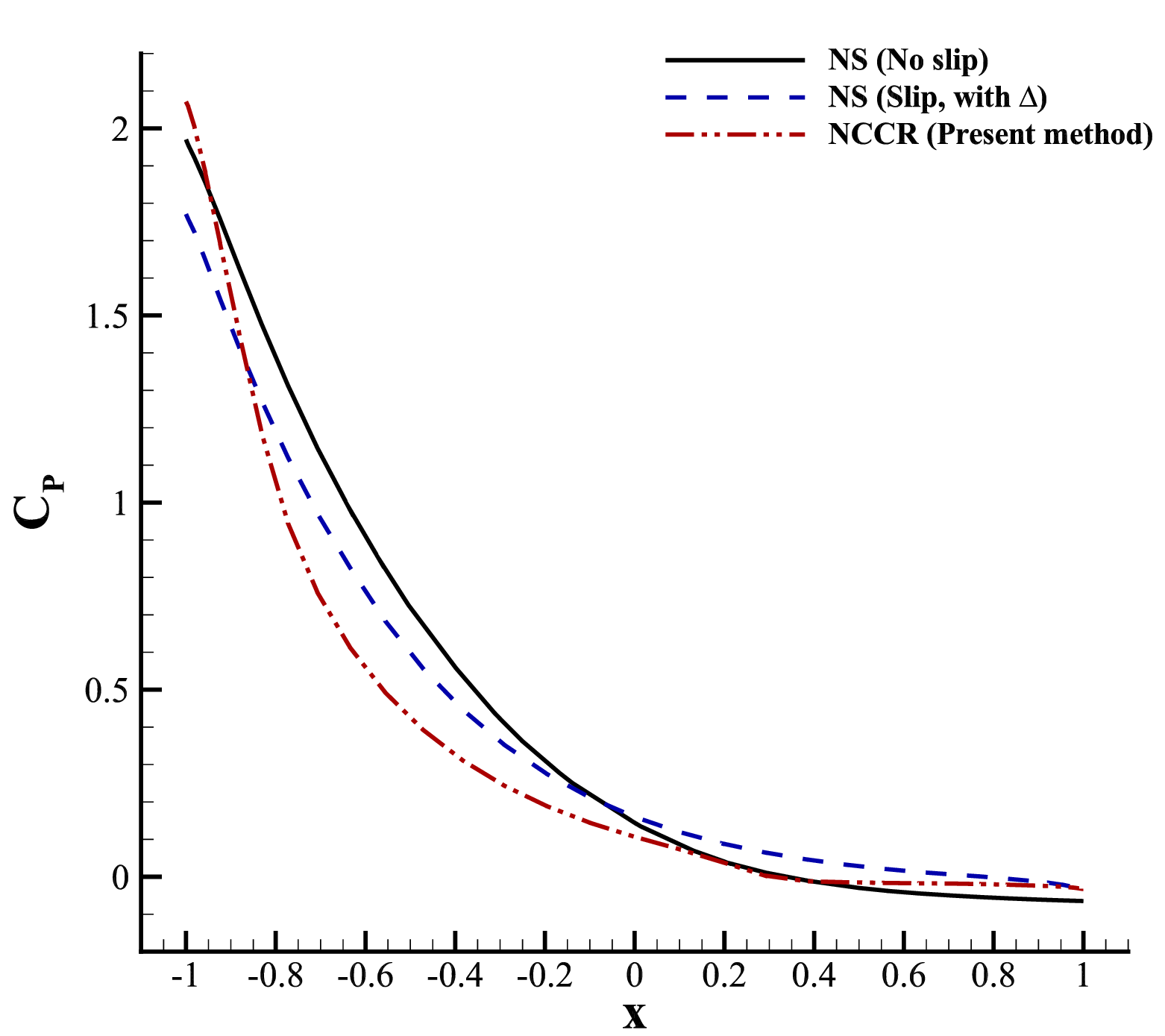}
		}
    \subfigure[]{
    		\includegraphics[width=0.45 \textwidth]{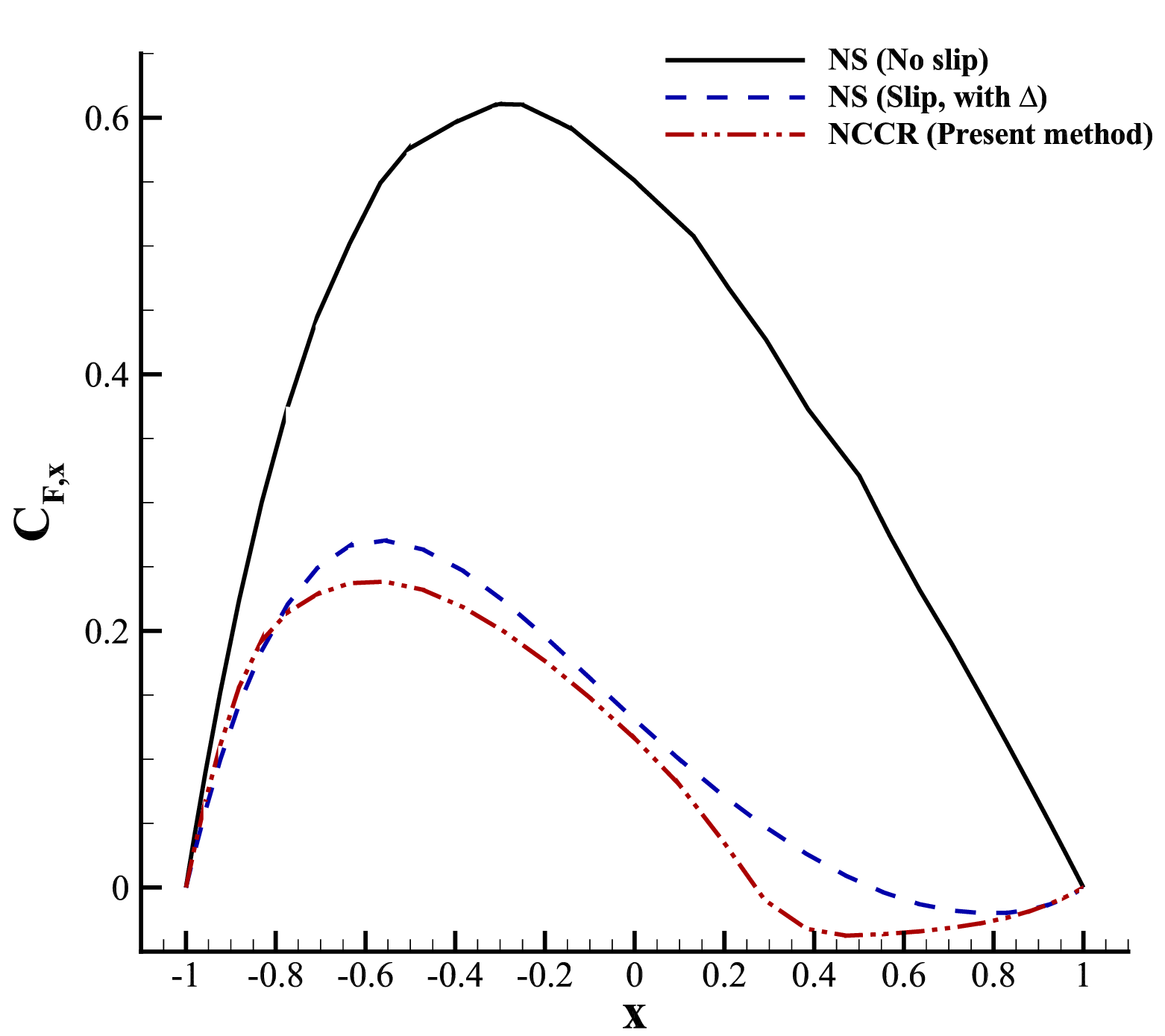}
    	}
	\caption{\label{fig24} Wall parameter results for the supersonic rarefied nitrogen gas sphere flow: (a) Pressure coefficient at $Kn=0.031$, (b) Shear stress coefficient in the $x$ direction at $Kn=0.031$, (c) Pressure coefficient at $Kn=0.121$, (d) Shear stress coefficient in the $x$ direction at $Kn=0.121$.}
\end{figure}

Drag coefficients are shown in Tab.\ref{tab:sphere} for comparison, whose reference area is $\pi R^2=\pi$. The reference data are experiment results\cite{sphereexp} (whose working gas is air), UGKS results\cite{sphereugks} and UGKWP results\cite{ugkwp5}. The results of unified methods like UGKS and UGKWP are accurate. The results of proposed method for NCCR equations are accurate as well, which are better than the results of NS equation methods with slip boundary condition. These results show that the NCCR model has advantages when predicting force coefficients. Details about wall parameters are shown in Fig.\ref{fig24}.

\begin{table}[h]\label{tab:sphere}
\centering
\caption{Comparison of the drag coefficients}
\begin{tabular}{*{5}{c}}
\hline
    \multirow{2}*{\textbf{Method}} & \multicolumn{2}{c}{$Kn=0.031$} & \multicolumn{2}{c}{$Kn=0.121$} \\
    \cmidrule(lr){2-3} \cmidrule(lr){4-5}
                                   & Drag coefficient & Relative error & Drag coefficient & Relative error \\
\hline
	Experiment                     & $1.35$           & -              & $1.69$           & -              \\ \hline
	UGKS                           & $1.355$          & $0.37\%$       & $1.694$          & $0.24\%$       \\ \hline
	UGKWP                          & $1.346$          & $0.30\%$       & $1.636$          & $3.20\%$       \\ \hline
    NS without slip boundary       & $1.684$          & $24.74\%$      & $3.855$          & $128.11\%$     \\ \hline
    NS-$\Delta$ with slip boundary & $1.378$          & $2.07\%$       & $1.883$          & $11.42\%$      \\ \hline
    NCCR (proposed method)         & $1.352$          & $0.15\%$       & $1.766$          & $4.50\%$       \\
\hline
\end{tabular}
\end{table}

\section{Conclusions}\label{sec:conclusion}
In this work, for solving the complicated NCCR equations, a parameter $q(\kappa)$ is used to design an objective function $F_{\rm{ob}}(\tilde{q})$. Then, the solutions of NCCR equtions are converted into a problem of finding zero point of $F_{\rm{ob}}(\tilde{q})$. The property of $F_{\rm{ob}}(\tilde{q})-\tilde{q}$ curve is studied systematically, and the multiple solutions of NCCR equations are discovered. The method for identifying the physical solution is proposed and used in the scheme for solving the NCCR equations. Furthermore, an iterative method is proposed, and the unique solution of NCCR model is obtained numerically and analyzed physically. Many typical rarefied flow problems are simulated to validate the numerical performance of the proposed method and the physical accuracy of NCCR model. Results of NCCR equations are more reliable than those of NS equations in the non-equilibrium regime, such as the shock structures, velocity profiles in the wake and the force coefficients around a moving sphere. However, the wall parameters, like pressure coefficient, shear stress coefficient and heat flux coefficient, are mainly influenced by the slip boundary condition, where the differences between the results of NCCR equations and NS equations are marginal under the same slip boundary condition. It seems that the slip boundary condition plays a more important role than the first order constitutive relation, for the overall solution around a flying body. At a large Kn number, in the wake of cylinder, the NCCR model itself is associated with unstable solution. A remedy is proposed to deal with this problem, by imposing a bound on extra entropy relaxation rate. Meanwhile, the accuracy and stability of MFPI method are tested as well. Accurate solutions can be obtained at the head of blunt body. However, at the pre-shock, post-shock, wake and low-speed shear flow, the stability of MFPI method is not satisfactory.

\section*{Appendix A: Modified fixed point iteration method}\label{Sec:mfpi}
It is a hard task to analyze the accuracy and convergency of MFPI method and indeed it is not accurate in some cases. However, this method is efficient and its accuracy and robustness are usually sufficient. As a result, it is tested in this study and introduced as follows. Firstly, multiplying corresponding value, Eq.\ref{eq:nccr1} is written as:
\begin{equation}\label{eq:nccrmfpi}
\begin{aligned}
\frac{1}{2p^2}{\Pi _{\rm{ij}}}:{\Pi _{\rm{ij}}} &= \frac{1}{{q(\kappa )}}\frac{\mu }{p}\frac{1}{2p^2}{\Pi _{\rm{ij}}}:\left\{ { - 2\frac{{\partial {U_{\rm{ < i}}}}}{{\partial {x_{\rm{k}}}}}{\Pi _{\rm{j > k}}} + 2(p - \Delta )\frac{{\partial {U_{\rm{ < i}}}}}{{\partial {x_{\rm{j > }}}}}} \right\},\\
\frac{\rm{Pr}}{Tp^2C_p}{Q_{\rm{i}}}\cdot{Q_{\rm{i}}} &= \frac{1}{{q(\kappa )}}\frac{\mu }{p}\frac{1}{{\Pr }}\frac{\rm{Pr}}{Tp^2C_p}{Q_{\rm{i}}}\cdot\left\{ {{C_p}(p - \Delta )\frac{{\partial T}}{{\partial {x_{\rm{i}}}}} - {C_p}{\Pi _{\rm{ik}}}\frac{{\partial T}}{{\partial {x_{\rm{k}}}}} - \frac{{\partial {U_{\rm{i}}}}}{{\partial {x_{\rm{k}}}}}{Q_{\rm{k}}}} \right\},\\
\frac{5-3\gamma}{2f_bp^2}\Delta\Delta  &= \frac{1}{{q(\kappa )}}\frac{5-3\gamma}{2f_bp^2}\Delta\left\{ {{\mu _{\rm{b}}}\frac{{\partial {U_{\rm{i}}}}}{{\partial {x_{\rm{i}}}}} - 3\frac{{{\mu _{\rm{b}}}}}{p}(\Delta {I_{\rm{ij}}} + {\Pi _{\rm{ij}}})\frac{{\partial {U_{\rm{i}}}}}{{\partial {x_{\rm{j}}}}}} \right\}.
\end{aligned}
\end{equation}
The sum of left part is abbreviated to $\hat{R}^2$ ($\hat{R}\geq 0$) and the sum of right part is abbreviated to $\hat{F}/q(\kappa)$, so,
\begin{equation}\label{eq:rf}
\hat{R}^2=\frac{\hat{F}}{q(\kappa)}\Rightarrow q(\kappa)=\frac{\hat{F}}{\hat{R}^2}.
\end{equation}
From Eq.\ref{eq:qkapa} and Eq.\ref{eq:kapa}, we know that:
\begin{equation}\label{eq:rfq}
q(\kappa)=\frac{\rm{sinh}\left({\frac{\pi^{0.25}}{\sqrt{2\beta}}\hat{R}}\right)}{\frac{\pi^{0.25}}{\sqrt{2\beta}}\hat{R}}.
\end{equation}
At this moment, combine Eq.\ref{eq:rf} with Eq.\ref{eq:rfq}, and the only variable to be iterated is the $\hat{R}$ in the numerator of Eq.\ref{eq:rfq}:
\begin{equation}\label{eq:qr}
q(\kappa)=\frac{\hat{F}}{\hat{R}^2}=\frac{\rm{sinh}\left({\frac{\pi^{0.25}}{\sqrt{2\beta}}\breve{R}}\right)}{\frac{\pi^{0.25}}{\sqrt{2\beta}}\hat{R}}
\Rightarrow \breve{R}=\frac{1}{\pi^{0.25}/\sqrt{2\beta}}\rm{sinh}^{-1}\left({\frac{\pi^{0.25}}{\sqrt{2\beta}}\frac{\hat{F}}{\hat{R}}}\right).
\end{equation}
Actually, this is a gentle way to calculate $q(\kappa)$. The procedure of MFPI method is summarized as follows:
\begin{description}
    \item[Step (1)] Set the results of NS equations to be initial $\mathbf{m}$ (Eq.\ref{eq:mm}).
    \item[Step (2)] Calculate $\hat{R}$ and $\hat{F}$ through Eq.\ref{eq:nccrmfpi}.
    \item[Step (3)] Use Eq.\ref{eq:qr} to update $q(\kappa)$.
    \item[Step (4)] Use the updated $q(\kappa)$ to update $\mathbf{m}$ through Eq.\ref{eq:nccr1}.
    \item[Step (5)] If $\mathbf{m}$ does not change any more, take it as the result, or go to step $2$.
\end{description}

Different from the classical fixed point iteration method, in the MPFI method, two kinds of iteration curves are merged together. One is as Eq.\ref{eq:nccr1}, in which $\mathbf{m}$ is to be iterated and $q(\kappa)$ is const. The other is as Eq.\ref{eq:qr}, in which $q(\kappa)$ is to be iterated and $\mathbf{m}$ is constant. This explains why it is difficult to analyze the accuracy and convergency of MFPI method. But qualitatively, $q(\kappa)$ is calculated very gently in this method. So its robustness is sufficient usually. From Fig.\ref{fig1a} and Fig.\ref{fig3c}, it is interpreted in Sec.\ref{sec:only} that sometimes the result of MFPI method does not accord with the NCCR equation.

\section*{Appendix B: A method for calculating $\tilde{q}_{\rm{max}}$}\label{sec:qmax}
In this section, a method is introduced to calculate $\tilde{q}_{\rm{max}}$, and at $\tilde{q}>\tilde{q}_{\rm{max}}$, there will never be singularity any more. Firstly, $\mathbf{A}$ in Eq.\ref{eq:aa} is written into:
\begin{equation}\label{eq:qr}
\mathbf{A}=\mathbf{D}-q(\kappa)\mathbf{I}=\mathbf{K}^{-1}\left({\mathbf{D}_{\rm{J}}-q(\kappa)\mathbf{I}}\right)\mathbf{K},
\end{equation}
where $\mathbf{D}$ is the matrix after separating $q(\kappa)$ from $\mathbf{A}$. $\mathbf{K}$ is a identity orthogonal matrix and $\mathbf{D}_{\rm{J}}$ is the corresponding Jordan form of $\mathbf{D}$, or specifically a diagonal matrix. The diagonal values of $\mathbf{D}_{\rm{J}}$ are the eigenvalues of $\mathbf{D}$.

Obviously, singularity appears in $F_{\rm{ob}}(\tilde{q})$ (Eq.\ref{eq:fq}) because $\mathbf{A}$ is singularity, which means $\mid\mathbf{A}\mid=0$ and $\mid\mathbf{K}^{-1}\mid\mid\left({\mathbf{D}_{\rm{J}}-q(\kappa)\mathbf{I}}\right)\mid\mid\mathbf{K}\mid=0$. It is known that $\mid\mathbf{K}\mid=\mid\mathbf{K}^{-1}\mid=1$. Therefore, singularity appears only when $q(\kappa)$ equals to any one of diagonal values of $\mathbf{D}_{\rm{J}}$. As a result, actually, $\tilde{q}_{\rm{max},{\rm{1}}}$ is set to be 1.1 times of the largest eigenvalue of $\mathbf{D}$.

As follows, a power method is introduced to calculate the largest eigenvalue of $\mathbf{D}$.
\begin{description}
    \item[Step (1)] Set an initial vector $\mathbf{w}^{\rm{0}}=\left({1,1,...,1}\right)^{\rm{T}}$ and set the largest element of $\mathbf{w}^{\rm{0}}$ to be $w_{\rm{max}}^{\rm{0}}$.
    \item[Step (2)] Calculate $\mathbf{w}^{\rm{n+1}}=\mathbf{D}\cdot\mathbf{w}^{\rm{n}}$ and set the largest element of $\mathbf{w}^{\rm{n+1}}$ to be $w_{\rm{max}}^{\rm{n+1}}$.
    \item[Step (3)] If $\mid\frac{\mathbf{w}^{\rm{n+1}}}{w_{\rm{max}}^{\rm{n+1}}}-\frac{\mathbf{w}^{\rm{n}}}{w_{\rm{max}}^{\rm{n}}}\mid>10^{-4}$, go to step $2$.
    \item[Step (4)] The value of largest eigenvalue of $\mathbf{D}$ is $w_{\rm{max}}^{\rm{n+1}}/w_{\rm{max}}^{\rm{n}}$.
\end{description}

There is another situation, where there is no singularity. For this condition, a experience equation is suitable,  $\tilde{q}_{\rm{max},{\rm{2}}}=2\times\rm{max}\left({\mid A_{\rm{ij}}\mid}\right)+2$. And $\tilde{q}_{\rm{max}}=\rm{max}\left({\tilde{q}_{\rm{max},{\rm{1}}},\tilde{q}_{\rm{max},{\rm{2}}}}\right)$.

\section*{Acknowledgements}
This work is supported by National Natural Science Foundation of China (11902264, 11902266, 12072283, 12172301) and 111 project of China (B17037). And it is supported by the high performance computing power and technical support provided by Xi'an Future Artificial Intelligence Computing Center. Junzhe Cao thanks Mr. Rui Zhang at School of Aeronautics, Northwestern Polytechnical University for data of CDUGKS in Sec.\ref{sec:ssn2}.

\clearpage
\section*{References}
\bibliography{multisolution}

\end{document}